\documentclass[a4paper,12pt]{article}
\usepackage{amsmath}
\usepackage{amsfonts,amssymb}
\usepackage{preprint}
\begin{document}
\numberwithin{equation}{section}
\def\calO{{\cal O}}
\def\calH{{\cal H}}
\def\calV{{\cal V}}
\def\calA{{\cal A}}
\def\calS{{\cal S}}
\def\calB{{\cal B}}
\def\calI{{\cal I}}
\def\calJ{{\cal J}}
\def\calD{{\cal D}}
\def\calE{{\cal E}}
\def\boldC{{\mathbf{C}}}
\def\boldN{{\mathbf{N}}}
\def\boldR{{\mathbf{R}}}
\def\boldZ{{\mathbf{Z}}}
\def\bolda{\hbox{\boldmath$\mathit{a}$}}
\def\hati{\skew3\hat{i}}
\def\ti{\skew3\tilde{i}}
\def\hatj{\skew3\hat{j}}
\def\tj{\skew3\tilde{j}}
\def\tphi{\tilde{\varphi}}
%
\hrule height0pt depth0pt
\vspace*{-12pt}
\rightline{\bf RIMS-1362}
\vspace*{50pt}
\centerline{
\Huge
Recursive Fermion System in Cuntz Algebra. II }
\vskip5pt
\centerline{
\Large
--- Endomorphism, Automorphism and Branching of Representation --- }
\vskip50pt
\renewcommand{\thefootnote}{\alph{footnote}}
\centerline{\large
 Mitsuo Abe\footnote{E-mail address: \abemail}
 and Katsunori Kawamura\footnote{E-mail address: \kkmail}
}
\vskip10pt
\centerline{\it Research Institute for Mathematical Sciences,
Kyoto University, Kyoto 606-8502, Japan}
\vskip25pt
%
\vskip150pt
\centerline{\bf Abstract}
\vskip10pt
Based on an embedding formula of the CAR algebra into the Cuntz algebra 
${\mathcal O}_{2^p}$, properties of the CAR algebra are studied in detail
by restricting those of the Cuntz algebra.
Various $\ast$-endomorphisms of the Cuntz algebra are explicitly constructed,
and transcribed into those of the CAR algebra.
In particular, a set of $\ast$-endomorphisms of the CAR algebra into its even 
subalgebra are constructed.
According to branching formulae, which are obtained by composing 
representations and $\ast$-endomorphisms, it is shown that a KMS state of 
the CAR algebra is obtained through the above even-CAR endomorphisms
from the Fock representation. 
A $U(2^p)$ action on ${\mathcal O}_{2^p}$ induces $\ast$-automorphisms of 
the CAR algebra, which are given by nonlinear transformations 
expressed in terms of polynomials in generators.
It is shown that, among such $\ast$-automorphisms of the CAR algebra, 
there exists a family of one-parameter groups of $\ast$-automorphisms 
describing time evolutions of fermions, in which the particle number 
of the system changes by time while the Fock vacuum is kept invariant.   
\vfill\eject
%
%
\section{Introduction}
\par
In our previous papers,\cite{AK1,AK2} we have presented a recursive 
construction of the CAR (canonical anticommutation relation) algebra\cite{KR} 
for fermions in terms of the Cuntz algebra\cite{Cuntz} $\calO_{2^p}$ 
($p\in\boldN$), and shown that it may provide us a useful tool to study 
properties of fermion systems by using explicit expressions in terms of 
generators of the algebra.  As a concrete example of applications,
we have constructed an infinite-dimensional (outer) $\ast$-automorphism 
group of the CAR algebra, in which the transformations are expressed 
in terms of polynomials in creation/annihilation operators.\cite{AK2}
The basic ingredient necessary for this embedding is called a {\it recursive
fermion system\/} and denoted by RFS$_p$, where a subscript $p$ stands for 
$\calO_{2^p}$. As a special example, the {\it standard\/} RFS$_p$, which 
describes an embedding of $\calO_{2^p}$ onto its $U(1)$-invariant 
subalgebra $\calO_2^{U(1)}$, has been introduced, and it has been shown 
that a certain permutation representation\cite{BJ,DP} of $\calO_{2^p}$ 
reduces to the Fock representation of the CAR algebra. 
We have also shown\cite{AK3} that it is possible to generalize this 
recursive construction to the algebra for the FP ghost fermions in string 
theory by introducing a $\ast$-algebra called the pseudo Cuntz algebra 
suitable for actions on an indefinite-metric state vector space.
We have found that, according to embeddings of the FP ghost algebra
into the pseudo Cuntz algebra with a special attention to the zero-mode
operators, unitarily inequivalent representations for the FP ghost are 
obtained from a single representation of the pseudo Cuntz algebra.
\par
The purpose of this paper is to develop the study of the recursive fermion
system and to show concretely that it becomes to possible to manage some 
complicated properties of the CAR algebra 
as follows:
\begin{itemize}
\item[(1)] Systematic construction of proper (i.e., not surjective) 
$\ast$-endomorphisms which are not necessarily expressed in terms of 
linear transformations: The existence of proper $\ast$-endomorphisms 
is characteristic for the infinite dimensionality of the algebra.
\item[(2)] Description of branchings of representations induced by 
proper $\ast$-endomorphisms: By using branchings, various reducible 
representations or mixed states for fermions are obtained.
\item[(3)] Systematic construction of outer $\ast$-automorphisms which 
are not necessarily expressed in terms of Bogoliubov (linear) 
transformations:
Nonlinearity of transformations in one-parameter groups of (outer) 
$\ast$-automorphisms corresponding to time evolutions implies that 
the fermions under consideration are no longer (quasi-)free.
\end{itemize}
For this purpose, it is necessary to prepare beforehand some useful 
formulae for representations,\cite{Kawamura1} embeddings, and 
$\ast$-endomorphisms\cite{Kawamura2,Kawamura3} of the Cuntz algebra. 
As for embeddings, from a fundamental formula\cite{Cuntz} for embedding 
of $\calO_3$ into $\calO_2$, we can easily obtain some basic formulae 
for embeddings among the Cuntz algebras. 
Then, using an important relation between embeddings (of some 
$\calO_{d'}$'s into $\calO_d$) and $\ast$-endomorphisms (of $\calO_d$), 
various $\ast$-endomorphisms of the Cuntz algebra are explicitly 
constructed. 
Conversely, from a set of given $\ast$-endomorphisms, we may also obtain 
new embeddings. By composing irreducible permutation representations and
$\ast$-endomorphisms, some branching formulae are derived.\cite{Kawamura3}
Based on these properties of the Cuntz algebra, we study the recursive 
fermion systems in detail. First, by restricting the irreducible 
permutation representations of the Cuntz algebra, the corresponding 
representations of the CAR algebra are obtained in the form of direct 
sums of irreducible ones. On the other hand, it is also shown that, for 
a certain type of irreducible permutation representation of $\calO_2$, 
we can construct a RFS$_1$ such that the restricted representation is 
irreducible. Such a RFS$_1$ gives, in general, an embedding of the CAR 
algebra onto a subalgebra of $\calO_2$ which is not $U(1)$-invariant. 
Furthermore, it is shown that a certain RFS$_1$ similar to the above one 
yields a direct sum of an infinite number of irreducible representations 
of the CAR algebra from any irreducible permutation representation of 
$\calO_2$. 
Next, from some $\ast$-endomorphisms of $\calO_2$, we explicitly 
construct $\ast$-endomorphisms of the CAR algebra, especially, a set of 
those giving $\ast$-homomorphism to its even 
subalgebra.\cite{Stormer,Binnenhei} 
It is shown that, by composing the Fock representation and
the above even-CAR endomorphisms, we may obtain a KMS state\cite{BR} 
of the CAR algebra with respect to a one-parameter group of 
$\ast$-automorphisms describing the time evolution of a (quasi-)free 
fermion system. In contrast with some KMS states of the Cuntz 
algebra,\cite{OP,Evans,BJO} the inverse temperature is not unique 
since the KMS condition is satisfied only by the induced state of the 
CAR algebra, but not by that of the Cuntz algebra. We also give some 
discussions on the relation to the Araki-Woods classification of factors 
for the CAR algebra.\cite{AW}
Finally, we apply the induced $\ast$-automorphisms of the CAR 
algebra\cite{AK2} to construct one-parameter groups of $\ast$-automorphisms 
describing nontrivial time evolutions of fermions. Since it is possible 
to describe nonlinear transformations of the CAR algebra by these 
$\ast$-automorphisms, the time evolutions are not restricted to those 
for (quasi-)free fermions.
We explicitly construct some examples for such one-parameter groups of
$\ast$-automorphisms of the CAR algebra, in which the particle number 
changes by time with keeping the Fock vacuum invariant.
\par
The present paper is organized as follows. 
In Sec.\,2 and Sec.\,3, we summarize various properties of the Cuntz 
algebra and obtain some convenient formulae necessary for our discussions.
In Sec.\,4, after reviewing the construction of the recursive fermion
system, we show the relation between RFS$_1$ and RFS$_p$ $(p\geqq2)$.
In Sec.\,5, we study the restriction of the permutation representations 
of the Cuntz algebra.
In Sec.\,6, various $\ast$-endomorphisms of the CAR algebra are explicitly 
obtained from those of the Cuntz algebra.
In Sec.\,7, based on some formulae constructed in the previous 
sections, it is shown that a KMS state is obtained from the Fock
representation through a certain $\ast$-endomorphism.
In Sec.\,8, we summarize $\ast$-automorphisms of the CAR algebra induced 
by a $U(2^p)$ action on $\calO_{2^p}$, and obtain one-parameter groups of
$\ast$-automorphisms describing nontrivial time evolutions for fermions.
The final section is devoted to discussion.
\vskip50pt
\section{Properties of Cuntz Algebra: Embedding and Endomorphism}
In this section and the next, we summarize some properties of the Cuntz 
algebra\footnote{Throughout this paper, we restrict ourself to consider 
the dense subset of the Cuntz algebra.}
necessary for our discussions in the succeeding sections. 
\par
First, let us recall that the Cuntz algebra\cite{Cuntz} $\calO_d$ $(d\geqq2)$ 
is a simple $C^*$-algebra generated by $s_1,\ s_2,\ \ldots,\ s_d$ satisfying 
the following relations:
\begin{alignat}{1}
s_i^* \, s_j &= \delta_{i,j} I,  \label{CR1}\\
\sum_{i=1}^d s_i \, s_i^* &= I, \label{CR2}
\end{alignat}
where $^*$ is a $\ast$-involution (or an adjoint operation), 
$I$ being the unit (or the identity operator). 
We often use brief descriptions as follows:
\begin{alignat}{1}
s_{i_1,i_2,\ldots,i_m}&\equiv s_{i_1}s_{i_2}\cdots s_{i_m}, \\
s^*_{i_1,i_2,\ldots,i_m}&\equiv s^*_{i_m}\cdots s^*_{i_2} s^*_{i_1}, \\
s_{i_1,\ldots,i_m;\,j_n,\ldots,j_1}&\equiv
s_{i_1}\cdots s_{i_m} s_{j_n}^* \cdots s_{j_1}^*.
\end{alignat}
From the relation \eqref{CR1}, $\calO_d$ is a linear space generated by 
monomials of the form $s_{i_1,\ldots,i_m;\,j_n,\ldots,j_1}$ with 
$m+n\geqq1$.
\par
From \eqref{CR1} and \eqref{CR2}, it is obvious that there is a 
$\ast$-automorphism $\alpha$ on $\calO_d$ defined by a $U(d)$ action as follows:
\begin{alignat}{1}
& \alpha_u(s_i) = \sum_{j=1}^d s_j \, u_{j,i}, \quad i=1,\ldots, d\,;\quad 
   u=(u_{j,i}) \in U(d).
 \label{U(d)-action}
\end{alignat}
Especially, we consider a $U(1)$ action $\gamma$ defined by
\begin{alignat}{1}
&\gamma_z(s_i) = z\,s_i, \quad i=1,\ldots, d\,;\quad z\in\boldC,\ |z|=1.
 \label{U(1)-action}
\end{alignat}
Then, the $U(1)$ invariant subalgebra $\calO_d^{U(1)}$ of $\calO_d$ is 
a linear subspace generated by monomials of the form 
$s_{i_1,\ldots, i_m;\,j_m,\ldots, j_1}$ with $m\geqq1$.
\vskip20pt
\subsection{Embedding}
If there exists an injective unital $\ast$-homomorphism $\psi$ from 
$\calO_{d'}$ to $\calO_{d}$, which is defined by a mapping 
$\psi\,:\,\calO_{d'} \to \calO_{d}$ satisfying 
\begin{alignat}{2}
\psi( \alpha X + \beta Y ) &= \alpha \psi(X) + \beta \psi(Y), &\quad 
  &\alpha, \, \beta \in \boldC,\ X,\,Y \in \calO_{d'},      \label{embed-1}\\
\psi( X Y ) &= \psi(X)\psi(Y) &\quad &X,\,Y \in \calO_{d'}, \label{embed-2}\\
\psi( X^* ) &= \psi(X)^*, &\quad &X \in \calO_{d'},         \label{embed-3}\\
\psi( I_{d'}) &= I_{d},   &&                                \label{embed-4}
\end{alignat}
with $I_{d'}$ and $I_{d}$ being unit of $\calO_{d'}$ and that of $\calO_{d}$, 
respectively, we say that $\calO_{d'}$ is {\it embedded into\/} $\calO_{d}$, 
and call $\psi$ an {\it embedding of $\calO_{d'}$ into $\calO_{d}$.}
We also denote an embedding as $\psi : \calO_{d'} \hookrightarrow \calO_{d}$.
In this paper, we always assume the condition \eqref{embed-4} for embeddings.
To define an embedding of $\calO_{d'}$ into $\calO_{d}$, it is sufficient 
and necessary to give a correspondence of generators between these two Cuntz 
algebras because of the following reason.
Let $\{s'_i\mid i=1,\,\ldots,\,d'\}$ be generators of $\calO_{d'}$.
If $\calO_{d'}$ is embedded into $\calO_{d}$, define $S_i\equiv\psi(s'_i) 
\in \calO_d$ $(i=1,\,\ldots,\,d')$  by the above unital $\ast$-homomorphism 
$\psi$.  Then, it is straightforward to show that 
$\{S_i\mid i=1,\,\ldots,\,d'\}$ satisfy \eqref{CR1} and \eqref{CR2} 
by using \eqref{embed-1}--\eqref{embed-4}.
Conversely, if there exists a set of elements $\{ S_i \in \calO_{d} \mid 
i=1,\,\ldots,\,d'\}$ satisfying \eqref{CR1} and \eqref{CR2}, it is also 
straightforward to construct the $\ast$-homomorphism 
$\psi:\calO_{d'}\to\calO_{d}$ 
by defining  $\psi(s'_i)\equiv S_i$ $(i=1,\,\ldots,\,d')$ and by uniquely 
extending its domain to the whole $\calO_{d'}$ in such a way that it 
satisfies \eqref{embed-1}--\eqref{embed-4}. 
Therefore, we also denote an embedding by giving a set of generators as
$\{S_1, \,\ldots,\, S_{d'}\}: \calO_{d'} \hookrightarrow \calO_{d}$.
\par
In the following, we present some fundamental formulae for embeddings 
among the Cuntz algebras.
\begin{enumerate}
\item[(1)] {Fundamental embedding by Cuntz: }
It is remarkable that $\calO_d$ with arbitrary $d$ $(d\geqq2)$ can be 
embedded into $\calO_2$. For example, by setting\cite{Cuntz}
\begin{gather}
S_1 = s_1, \ \ S_2 = s_2 s_1, \ \ S_3 = (s_2)^{2} s_1, \,\ldots, \ \ 
S_{d-1} = (s_2)^{d-2}s_1, \ \ S_d = (s_2)^{d-1}, \label{embedding1}
\end{gather}
where $s_1$ and $s_2$ are the generators of $\calO_2$, 
it is straightforward to show that $S_i$'s satisfy the relations 
\eqref{CR1} and \eqref{CR2}. This is called {\it Cuntz embedding}.
\item[(2)] {Inductive construction: }
From any embedding $\{S_1,\ \ldots,\ S_d\}:\calO_d\hookrightarrow 
\calO_2$,  we can obtain an embedding of $\calO_{d+1}$ into $\calO_2$ 
as follows:
\begin{alignat}{1}
&\{S_1,\ \ldots,\ S_{d-1},\ S_d\,s_1,\ S_d\,s_2\}:\calO_{d+1}
\hookrightarrow\calO_2. 
\end{alignat}
\item[(3)] {Generalized Cuntz embedding: }
It is straightforward to generalize \eqref{embedding1} for embedding 
of $\calO_{(d-1)n+1}$ into $\calO_{d}$ with $n \geqq1$ as follows
\begin{alignat}{2}
S_i &= s_i &\quad &\text{for }\  1\leqq i \leqq d-1,   \nonumber\\
S_{(d-1)k+i} &= (s_{d})^{k} s_i &\quad 
 &\text{for }\ 1\leqq k \leqq n-1,\ 1\leqq i \leqq d-1, \label{generic-embed}\\
S_{(d-1)n+1} &= (s_{d})^{n}, & &                  \nonumber
\end{alignat}
where $\{s_i\mid i=1,\ldots,d\}$ is the generators of $\calO_{d}$.
\item[(4)] {Generalized inductive construction: } 
From any embedding 
$\{S_1, \, \ldots, \, S_{(d-1)n+1} \}:\calO_{(d-1)n+1}
\hookrightarrow\calO_{d}$, 
we can obtain an embedding of $\calO_{(d-1)(n+1)+1}$ into $\calO_{d}$ 
as follows:
\begin{alignat}{1}
&\{S_1, \, \ldots, \, S_{(d-1)n}, \, S_{(d-1)n+1}s_1, \,\ldots, \, 
S_{(d-1)n+1}s_{d} \}:
\calO_{(d-1)(n+1)+1}\hookrightarrow\calO_{d}. \label{inductive-embed}
\end{alignat}
\item[(5)] {Homogeneous embedding: }
For $\calO_{d^p}$, we have its embedding into $\calO_d$ in which all
generators of $\calO_{d^p}$ are mapped homogeneously to elements of 
$\calO_d$ as follows:
\begin{gather}
\varPsi_p: \calO_{d^p} \hookrightarrow \calO_d, \nonumber\\
\varPsi_p(s'_i)\equiv S^{(p)}_i \equiv  
s_{i_1,i_2,\ldots,i_p},
\qquad i-1=\sum\limits_{k=1}^p(i_k-1)d^{k-1}    \label{hom-embed}\\
i=1,\,2,\,\ldots,\,d^p; \quad i_1,\,i_2,\,\ldots,\,i_p=1,\,2,\,\ldots,\,d,
\nonumber
\end{gather}
where the correspondence of $i-1$ and $(i_p-1,\,\ldots,\,i_1-1)$ is the 
same as that of a decimal number and its $d$-ary expression.
The embedding $\varPsi_p$ \eqref{hom-embed} for 
$\calO_{d^p}\hookrightarrow\calO_d$ is constructed inductively with 
respect to $p$ as follows:
\begin{alignat}{2}
S^{(p+1)}_{(i-1)d^p+j}&=S^{(p)}_j\,s_i &\quad
&\text{for} \ i=1,\,2,\,\ldots,\,d;\ j=1,\,2,\,\ldots,\,d^p, 
     \label{hom-embed-rec}
\end{alignat}
or
\begin{alignat}{2}
S^{(p+1)}_{(j-1)d+i}&=s_i S^{(p)}_j &\quad
&\text{for} \ i=1,\,2,\,\ldots,\,d;\ j=1,\,2,\,\ldots,\,d^p.
     \label{hom-embed-rec2}
\end{alignat}
\end{enumerate}
\vskip10pt
\par
From \eqref{hom-embed}, it is obvious that any monomial 
$s_{i_1,\ldots,i_n}\in\calO_d$ is one of homogeneously embedded generators
$\{S^{(n)}_1,\,\ldots,\,S^{(n)}_{d^n}\}:\calO_{d^n}\hookrightarrow\calO_d$.
It should be noted that, from \eqref{generic-embed} and \eqref{inductive-embed}, 
there is also an embedding of $\calO_{(d-1)n+1}$ into $\calO_{d}$ for
any monomial $s_{i_1,\ldots,i_n}\in\calO_d$ such that $s_{i_1,\ldots,i_n}$ 
can be set on one of embedded generators $\{S_1,\,\ldots,\,S_{(d-1)n+1}\}$. 
Such an example is given by the following: 
\begin{alignat}{1}
S_{j} &=
 \begin{cases}
   \displaystyle 
   s_{j}  &\mbox{for $1\leqq j \leqq i_1-1$,}\\[5pt]
   \displaystyle 
   s_{j+1}&\mbox{for $i_1\leqq j \leqq d-1$,}\\[5pt]
   \displaystyle 
   s_{i_1,\ldots, i_{k} \tj}  
          &\mbox{for $(d-1)k+1\leqq j \leqq (d-1)k+i_{k+1}-1,
      \quad 1\leqq k \leqq n-1$,}\\[5pt]
   \displaystyle 
   s_{i_1,\ldots, i_{k}, \tj+1}
          &\mbox{for $(d-1)k+i_{k+1}\leqq j \leqq (d-1)(k+1),
      \quad\ \ \   1\leqq k \leqq n-1$,}\\[5pt]
   \displaystyle 
   s_{i_1,\ldots, i_{n-1}, i_n} &\mbox{for $j=(d-1)n+1$},
 \end{cases}
\label{mono-embed}
\end{alignat}
where $\tj\equiv j-(d-1)k$.  
%
%
\vskip20pt
\subsection{Endomorphism}
An embedding of $\calO_d$ into itself is a {\it unital $\ast$-endomorphism\/} 
of $\calO_d$.
A typical \hbox{$\ast$-endomorphism} of $\calO_d$ is the {\it canonical 
endomorphism\/} $\rho$ defined by\footnote{We always use the symbol $\rho$
for the canonical endomorphism.}
\begin{equation}
\rho(X) = \sum_{i=1}^d s_i X s_i^*, \quad X \in \calO_d. 
   \label{c-endo}
\end{equation}
Indeed, from \eqref{CR1}, $\rho$ satisfies $\rho(X)\rho(Y)=\rho(XY)$ for
$X,\,Y\in\calO_d$.
From \eqref{CR2}, $\rho$ is unital, that is, $\rho(I)=I$, hence
$S_i\equiv\rho(s_i)$ satisfy the relations \eqref{CR1} and \eqref{CR2}.
\par
Let $U(k,\calO_d)$ $(k\in\boldN)$ be a set of all $k\times k$ unitary 
matrices in which each entry is an element of $\calO_d$. 
Then, any unital $\ast$-endomorphism $\varphi$ has a one-to-one 
correspondence with a unitary $u \in U(1,\,\calO_d)$ given by
\begin{alignat}{2}
\varphi(s_i) &= u\, s_i,     &\quad &i=1,\ldots,d,     \label{lr-unitary-1}\\
           u &= \sum_{i=1}^d \varphi(s_i) \,s_i^*. &&  \label{lr-unitary-2}
\end{alignat}
Likewise, there is a one-to-one correspondence between any unital 
$\ast$-endomorphism $\varphi$ and a $d\times d$ unitary 
$v=(v_{j,i}) \in U(d,\,\calO_d)$ as follows:
\begin{alignat}{2}
\varphi(s_i) &= \sum_{j=1}^d s_j \, v_{j,i}, &\quad i&=1,\ldots,d, 
                                                       \label{lr-unitary-3}\\
      v_{j,i} &= s_j^* \,\varphi(s_i), &\quad i,j&=1,\ldots,d.     
                                                       \label{lr-unitary-4}
\end{alignat}
It should be noted that, if it is possible to embed $\calO_{d'}$ into 
$\calO_d$ for certain $d'$ and $d$, then any unitary $u \in U(1,\,\calO_d)$ 
is expressed in the following form:\footnote{The symbols $S^{[k]}_i$ and 
$S^{(k)}_i$ should not be confused. The former is used just for distinguishing
one from some others, while the latter denotes the homogeneous embedding.} 
\begin{equation}
u = \sum_{i=1}^{d'}S^{[2]}_i S^{[1]\,\ast}_i,  \label{hom-hom}
\end{equation}
where $\{ S^{[k]}_1,\,\ldots,\,S^{[k]}_{d'} \}$  $(k=1,\,2)$ are
embeddings of $\calO_{d'}$ into $\calO_d$.
Indeed, it is straightforward to show that $u$ defined by \eqref{hom-hom}
satisfies $u\,u^* = u^*\, u = I$ by using that 
$\{ S^{[k]}_1,\,\ldots,\,S^{[k]}_{d'}\}$ satisfy \eqref{CR1} and \eqref{CR2} 
for each $k=1,\,2$.
Conversely, for an arbitrary unitary $u$ and an arbitrary embedding
$\{ S^{[1]}_1,\,\ldots,\,S^{[1]}_{d'}\}\,:\,\calO_{d'} \hookrightarrow 
\calO_d$, it is possible to obtain another embedding 
$\{ S^{[2]}_1,\,\ldots,\,S^{[2]}_{d'}\}\,:\,\calO_{d'} 
\hookrightarrow \calO_d$ 
in \eqref{hom-hom} as follows
\begin{equation}
S^{[2]}_i = u\, S^{[1]}_i \quad i=1,\,\ldots,\,d'.
\end{equation}
Using this formula, we obtain a new $\ast$-endomorphism from two known
embeddings, and conversely, a new embedding from a known $\ast$-endomorphism
and a known embedding.
For example, for $d=2, \ d'=3, \ \varphi=\rho, \ 
\{ S_1^{[1]}\equiv s_1, \, S_2^{[1]}\equiv s_{2,1}, \,S_3^{[1]}\equiv s_{2,2}\}$,
we obtain a new embedding of $\calO_3$ into $\calO_2$ by
\begin{equation}
S_1^{[2]} = \rho(s_1), \quad S_2^{[2]}=s_{1,2}, \quad S_3^{[2]}= s_{2,2}.
\end{equation}
\par
Although \eqref{lr-unitary-1}--\eqref{hom-hom} are general formulae, they are 
not convenient to construct various $\ast$-endomorphisms explicitly.
Next, we present a more effective way to express a generic $\ast$-endomorphism 
of $\calO_d$ in terms of some embeddings of $\calO_{d'}$ into $\calO_d$ without 
recourse to unitaries.
For this purpose, we need the following $d+1$ embeddings:
\begin{gather}
\{S^{[i]}_1,\ \ldots,\ S^{[i]}_{d_i}\}: \ 
   \calO_{d_i} \hookrightarrow \calO_d,
\quad i=1,\ \ldots,\ d, \label{embend1}\\
\{S^{[d+1]}_1,\ \ldots,\ S^{[d+1]}_D\}: \ 
   \calO_D \hookrightarrow \calO_d, 
\quad D\equiv\sum_{i=1}^d d_i, \label{embend2}
\end{gather}
where $d_i\equiv (d-1)n_i+1$ $(i=1,\ldots,d)$ with $\{n_1,\ldots,n_d\}$ 
being nonnegative integers. For $d_i=d$, a trivial embedding 
(i.e., $S^{[i]}_j\equiv s_j$) is used, while for $d_i=1$, we define 
$S^{[i]}_1\equiv I$.  
It should be noted that we have 
$D=(d-1)\big(\sum\limits_{i=1}^d n_i + 1\big) + 1$. 
Thus, for any $\{d_1,\ldots,d_d\}$, there exists an embedding of $\calO_D$ 
into $\calO_d$ from \eqref{generic-embed}.
Given the above $d+1$ embeddings, we can define a $\ast$-endomorphism 
$\varphi$ of $\calO_d$ as follows:
\begin{alignat}{1}
&\varphi(s_i) \equiv
 \sum_{j=1}^{d_i}S^{[d+1]}_{D_{i-1} + j}\,S^{[i]\,\ast}_j, \qquad
  D_i\equiv \sum_{j=1}^id_j. \label{embend3}
\end{alignat}
Indeed, it is straightforward to show that  $\{ \varphi(s_1),\,\ldots,\,
\varphi(s_d)\}$ satisfy \eqref{CR1} and \eqref{CR2} by using \eqref{embend1}
and \eqref{embend2}. 
Conversely, for an arbitrary $\ast$-endomorphism $\varphi$ of $\calO_d$ and $d$ 
arbitrary embeddings 
$\{S^{[i]}_1,\,\ldots,\,S^{[i]}_{d_i}\}:\calO_{d_i}\hookrightarrow\calO_d$\  
$(i=1,\,\ldots,\,d)$, we obtain an embedding
$\{ S_{D_{i-1}+j}\equiv\varphi(s_i)S^{[i]}_j \mid 
i=1,\ldots,d\,;\ j=1,\ldots,d_i\} : \calO_D\hookrightarrow\calO_d$ 
with $D_i=\sum\limits_{j=1}^i d_j, \ D=D_d$, which reproduces $\varphi(s_i)$
itself when substituted into $S^{[d+1]}_{D_{i-1}+j}$ in \eqref{embend3}.
Therefore, any $\ast$-endomorphism of $\calO_d$ is expressed in the form of 
\eqref{embend3}.
\par
From a $\ast$-endomorphism $\varphi$ in the form of \eqref{embend3}, 
we obtain various $\ast$-endomorphisms by using the $U(D)$ action on 
$\calO_{D}$ given by \eqref{U(d)-action} as follows:  
\begin{alignat}{1}
&\begin{array}{l}
\displaystyle
\varphi_u(s_i) \equiv
 \sum_{j=1}^{d_i}S^{[d+1]\,\prime}_{D_{i-1} + j}\,S^{[i]\,\ast}_j, \\[5pt]
\displaystyle
  S^{[d+1]\,\prime}_k\equiv \sum_{\ell=1}^D S^{[d+1]}_\ell\,u_{\ell,k},
  \quad k=1,\ldots,D,\quad u\in U(D).
\end{array}  \label{embend-u}
\end{alignat}
Especially, by using permutations given by 
$S^{[d+1]}_i \mapsto S^{[d+1]}_{\sigma(i)}$ 
($\sigma\in\mathfrak{S}_D\subset U(D)$), we obtain $D!$ $\ast$-endomorphisms 
for a given set of $d+1$ embeddings. 
In the case $D=d^{p+1}$ $(p\in\boldN)$, we may adopt the homogeneous 
embedding defined by \eqref{hom-embed} for the embedding of $\calO_D$ 
in to $\calO_d$, $S^{[d+1]}_i=s_{i_1,\ldots,i_{p+1}}$ 
$(i=1,\ldots,D;\ i_1,\ldots,i_{p+1}=1,\ldots,d)$. 
Then, each permutation of the indices $i\in\{1,\ldots,d^{p+1}\}$ 
induces a permutation of the multi indices 
$(i_1,\ldots,i_{p+1})\in\{1,\ldots,d\}^{p+1}$ according to the 
one-to-one correspondence between them given by \eqref{hom-embed}. 
For simplicity of description, we denote this induced permutation of 
the multi indices by the same symbol $\sigma$ as for the single indices, 
that is, $S^{[d+1]}_{\sigma(i)}=s_{\sigma(i_1,\ldots,i_{p+1})}
=s_{i^\sigma_1,\ldots,i^\sigma_{p+1}}$.
\par
Hereafter, we assume that $\ast$-endomorphisms of $\calO_d$ are 
expressed in terms of a finite sum of monomials.
We, now, consider $\ast$-endomorphisms $\varphi$ of $\calO_d$ 
which commute with the $U(1)$ action $\gamma$ defined by 
\eqref{U(1)-action}. Then, it satisfies the following:
\begin{alignat}{1}
\gamma_z\big(\varphi(s_i)\big)
 &=\varphi\big(\gamma_z(s_i)\big) \nonumber\\
 &=z\,\varphi(s_i),\quad i=1,\ldots,d,\quad z\in\boldC,\quad |z|=1.
\end{alignat}
Hence, from $\gamma_z(s_{i_1,\ldots,i_m;\,j_n,\ldots,j_1})
=z^{m-n}\,s_{i_1,\ldots,i_m;\,j_n,\ldots,j_1}$, each term in 
$\varphi(s_i)$ $(i=1,\ldots,d)$ is a monomial in the form of
$s_{i_{1},\ldots,i_{n+1};\,j_{n},\ldots,j_{1}}$ with 
$0\leqq n\leqq p_i$ $(i=1,\ldots,d)$, where 
$P\equiv\{p_i \mid i=1,\ldots,d\}$ is a set of nonnegative 
integers. Let $p$ be the maximum of $P$. 
By using \eqref{CR2}, we can rewrite $\varphi(s_i)$ $(i=1,\ldots,d)$ 
into a homogeneous polynomial of degree $(p+1,\,p)$, that is, a finite
sum of monomials in the form of $s_{i_{1},\ldots,i_{p+1};\,j_{p},\ldots,j_{1}}$.
Here, any monomial $s_{i_1,\ldots,i_p}$ 
(or $s_{i_1,\ldots,i_{p+1}}$) is one of the homogeneously embedded 
generators of $\calO_{d^p}$ (or $\calO_{d^{p+1}}$) into $\calO_d$ 
defined by \eqref{hom-embed}, hence $\varphi$ is written as
\begin{alignat}{1}
&\varphi(s_i) 
= \sum_{j=1}^{d^p} \sum_{k=1}^{d^{p+1}} 
    c_{k,j;i}\,S^{(p+1)}_k S^{(p)\,\ast}_j
\end{alignat}
with an appropriate set of coefficients $c_{k,j;i}\in\boldC$.
Since $\{\varphi(s_i)\mid i=1,\ldots,d\}$ satisfies \eqref{CR1} and 
\eqref{CR2}, the relations among the coefficients $c_{k,j;i}$'s are 
obtained as follows: 
\begin{alignat}{1}
&\sum_{k=1}^{d^{p+1}}\bar c_{k,j;i}\, c_{k,j';i'} 
  = \delta_{j,j'}\delta_{i,i'}, \qquad
\sum_{i=1}^d\sum_{j=1}^{d^p} c_{k,j;i}\, \bar c_{k',j;i} 
  = \delta_{k,k'},
\end{alignat}
hence $u_{k,\ell}\equiv c_{k,j;i}$ with $\ell\equiv (j-1)d+i$ is 
an element of $U(d^{p+1})$. Therefore, any $\ast$-endomorphism $\varphi$ of 
$\calO_d$, which is expressed in terms of a finite sum of monomials, 
commuting with the U(1) action $\gamma$ is written as follows:
\begin{alignat}{1}
&\begin{array}{l}
\displaystyle
\varphi(s_i) \equiv
 \sum_{j=1}^{d^p}S^{(p+1)\,\prime}_{(j-1)d + i}\,S^{(p)\,\ast}_j, 
  \quad i=1,\ldots,d,\\[5pt]
\displaystyle
  S^{(p+1)\,\prime}_\ell \equiv \sum_{k=1}^{d^{p+1}} S^{(p+1)}_k\,u_{k,\ell},
  \quad \ell=1,\ldots,d^{p+1},\quad u=(u_{k,\ell})\in U(d^{p+1}).
\end{array} \label{hom-endo}
\end{alignat}
We call this type of $\ast$-endomorphism the {\it $(p+1)$-th order homogeneous 
endomorphism}. Here, one should note that if $u_{k,\ell}=\delta_{k,\ell}$,
\eqref{hom-endo} becomes the identity map $\varphi(s_i)=s_i$ 
from \eqref{hom-embed-rec2} as follows:
\begin{alignat}{1}
\varphi(s_i) 
&= \sum_{j=1}^{d^p}S^{(p+1)}_{(j-1)d + i}\,S^{(p)\,\ast}_j \nonumber\\
&= s_i \sum_{j=1}^{d^p}S^{(p)}_{j}\,S^{(p)\,\ast}_j = s_i, \quad i=1,\ldots,d.
\end{alignat}
Especially, as for the second order homogeneous endomorphism, by setting
\begin{alignat}{1}
u&=\begin{pmatrix} v & 0 & \cdots &0\\ 
                   0 & v & \cdots &0\\
                     &   & \ddots & \\
                   0 & 0 & \cdots &v\\
\end{pmatrix}\in U(d^2),\quad v\in U(d),
\end{alignat}
so that we have
\begin{alignat}{1}
S^{(2)\,\prime}_{(j-1)d+i}
&=\sum_{k=1}^d S^{(2)}_{(j-1)d+k}\,v_{k,i}\nonumber\\ 
&=\sum_{k=1}^d s_{k,j}\,v_{k,i}, \qquad  v=(v_{k,i})\in U(d),
\end{alignat}
where use has been made of \eqref{hom-embed-rec2} for $p=1$,
the $\ast$-automorphism of $\calO_d$ by the $U(d)$ action \eqref{U(d)-action} 
is reproduced as follows:
\begin{alignat}{1}
\varphi(s_i) 
&= \sum_{j=1}^d \sum_{k=1}^d s_{k,j}\, s_j^*\, v_{k,i}\nonumber\\
&= \sum_{k=1}^d s_k\, v_{k,i}, \quad i=1,\ldots,d\,;\quad v=(v_{k,i})\in U(d).
\end{alignat}
\par
In the case that $u$ in \eqref{hom-endo} is a permutation 
$\sigma\in\mathfrak{S}_{d^{p+1}}\subset U(2^{p+1})$, $\varphi$ is called the 
{\it $(p+1)$-th order permutation endomorphism}\cite{Kawamura2}\footnote{As 
far as the present authors know, the first nontrivial example of the 
permutation endomorphisms other than the canonical endomorphism is the 
second order one in $\calO_3$ presented by N.\ Nakanishi in private 
communication. The discussions in this subsection are based on the 
generalization of his result.} of $\calO_d$.
Explicitly, the permutation endomorphisms are written in the following:
\begin{enumerate}
\item[(1)] {The second order permutation endomorphism:}
\begin{alignat}{1}
&
 \begin{array}{ll}
  \displaystyle 
  \varphi(s_i)
    =\sum\limits_{j=1}^d S^{(2)}_{\sigma((j-1)d+i)}s^*_j
    =\sum\limits_{j=1}^d s_{\sigma(i,j)}\,s_j^*,
 \end{array}
\label{perm-endo}
\end{alignat}
where $\sigma(i_1,i_2)$ denotes 
a permutation of multi indices $(i_1,i_2)\in\{1,\ldots,d\}^2$ induced
from that of indices $i\in\{1,\ldots,d^2\}$ by the one-to-one 
correspondence defined by $i=\sum\limits_{k=1}^2(i_k-1)d^{k-1}+1$. 
\item[(2)] {The ${(p+1)}$-th order permutation endomorphism:}
\begin{alignat}{1}
&
 \begin{array}{ll}
   \displaystyle 
   \varphi(s_i)
   =\sum\limits_{j=1}^{d^p} S^{(p+1)}_{\sigma((j-1)d+i)}S^{(p)\,\ast}_j
   =\sum\limits_{j_1,\ldots,j_p=1}^{d} 
                 s_{\sigma(i,j_1,\ldots,j_p)}\,s_{j_1\cdots j_p}^*,
 \end{array}
\qquad
\label{ext-perm-endo}
\end{alignat}
where $\sigma(i_1,\ldots,i_{p+1})
$ denotes a permutation of multi indices 
$(i_1,\ldots,i_{p+1})\in\{1,\ldots,d\}^{p+1}$ induced
from that of indices $i\in\{1,\ldots,d^{p+1}\}$
by the one-to-one correspondence defined by
$i=\sum\limits_{k=1}^{p+1}(i_k-1)d^{k-1}+1$. 
\end{enumerate}
\par
Let PEnd$_{p+1}(\calO_d)$ be a set of all $(p+1)$-th order permutation 
endomorphisms of $\calO_d$.
Then, we have 
\begin{alignat}{1}
\mbox{PEnd}_{2}(\calO_d) \subset \mbox{PEnd}_{3}(\calO_d)\subset \cdots 
 \subset\mbox{PEnd}_{p+1}(\calO_d)\subset\cdots, 
\end{alignat}
since a subset of PEnd$_{p+1}(\calO_d)$ with 
$\mathfrak{S}_{d^{p+1}}\supset\mathfrak{S}_{d^p}\ni\sigma$ 
preserving $j_p$ in \eqref{ext-perm-endo} $(j_p^\sigma=j_p)$ is 
nothing but PEnd$_p(\calO_d)$ because of \eqref{CR2}.
The canonical endomorphism $\rho$ given by \eqref{c-endo} is the special 
case of the second order permutation endomorphism \eqref{perm-endo} 
with $\sigma \in \mathfrak{S}_{d^2}$ being a product of $d(d-1)/2$ 
transpositions $(i,j) \mapsto (j,i)$ with $i\not=j$.  
Likewise, $\rho^p$ is a special case of the \hbox{$(p+1)$-th} order 
permutation endomorphism with $\sigma: (i,j_1,\ldots,j_n)\mapsto
(j_1,\ldots,j_n,i)$.
\par
Of course, a generic $\ast$-endomorphism given by \eqref{embend3} does 
{\it not\/} necessarily commute with the $U(1)$ action $\gamma$. 
We call $\ast$-endomorphisms not commuting with $\gamma$ the {\it inhomogeneous
endomorphisms}.
A typical example of the inhomogeneous endomorphism in the form of 
\eqref{embend3} is obtained by setting $d_i=1$ $(i=1,\ldots,d-1)$
and $d_d=(d-1)(n-1)+1$ (hence $D=(d-1)n+1$) with $n-1\in\boldN$
as follows:
\begin{alignat}{1}
&\hspace*{-30pt}
 \begin{array}{l}
   \displaystyle \varphi(s_i)=S^{[d+1]}_i, \quad i=1,\ldots,d-1,\\[5pt]
   \displaystyle \varphi(s_d)=\sum_{j=1}^{d_d} S^{[d+1]}_{j+d-1}S^{[d]\,\ast}_j,
 \end{array}
\label{inhom-endo}
\end{alignat}
where 
$\{S^{[d]}_j\mid j=1,\ldots,d_d\}$ and $\{S^{[d+1]}_j\mid  
j=1,\ldots,D\}$ are embeddings of $\calO_{d_d}$ and $\calO_D$ into 
$\calO_d$, respectively, in which the order of $s_j$'s minus that of $s^*_k$'s
appearing in at least one of $S^{[d+1]}_i$ $(i=1,\ldots, d-1)$ is not equal to 1.
\par
In \eqref{inhom-endo}, we can assign an arbitrary $n$-th order monomial 
$s_{i_1,\ldots, i_n}$ $(i_1,\ldots,i_d=1,\ldots,d)$ to $\varphi(s_1)$ by
adjusting the embedding \eqref{mono-embed} of $\calO_{(d-1)n+1}$ into 
$\calO_d$. This fact will be applied later.
\vskip40pt
\section{Properties of Cuntz Algebra: Representation and Branching}
\subsection{Permutation representation}
A permutation representation\cite{BJ,DP} of $\calO_d$ on a countable 
infinite-dimensional Hilbert space $\calH$ is defined as follows. 
Let $\{e_n\mid n\in\boldN\}$ be a complete orthonormal basis of $\calH$. 
A {\it branching function system\/} $\{\mu_i\}_{i=1}^d$ on $\boldN$ 
is defined by
\begin{alignat}{2}
&\mu_i: \boldN \to \boldN \text{ is injective},&\ \  i&=1,2,\ldots,d,\\
&\mu_i(\boldN) \cap \mu_j(\boldN) = \emptyset \ 
     \text{ for } i\not=j, &\quad  i,j&=1,2,\ldots,d, \\
&\bigcup_{i=1}^d \mu_i(\boldN) = \boldN.&&
\end{alignat}
Given a branching function system $\{\mu_i\}_{i=1}^d$ and a set of 
complex numbers 
$\{z_{i,n}\in\boldC\,\mid\,|z_{i,n}|=1,\ i\!=\!1,\ldots,d;\ n\in\boldN\}$, 
the permutation representation $\pi$ of $\calO_d$ on $\calH$ is 
defined by\footnote{The original definition of the permutation 
representation of the Cuntz algebra in Ref.\,\citen{BJ}) is the case 
of $z_{i,n}=1$ $(i=1,\ldots,d,\ n\in\boldN)$. We have introduced 
a set of coefficients $z_{i,n}$ according to Ref.\,\citen{DP}).}
\begin{eqnarray}
\pi(s_i) e_n = z_{i,n}e_{\mu_i(n)}, \quad 
     i=1,\,\ldots,\,d,\ \ n\in\boldN. \label{PR1}
\end{eqnarray}
By this definition, $\pi(s_i)$ is defined on the whole $\calH$ 
linearly as a bounded operator. 
Then, the action of $\pi(s_i^*)=\pi(s_i)^*$ on $e_n$ is determined 
by the definition of the adjoint operation. 
Since for any $n\in\boldN$ there exists a pair $\{j,\ m\}$
which satisfy $\mu_j(m)=n$,  we consider $\pi(s_i)^*$ on $e_{\mu_j(m)}$: 
\begin{alignat}{1}
\langle \pi(s_i)^* e_{\mu_j(m)} | e_\ell \rangle 
&= \langle e_{\mu_j(m)} | \pi(s_i) e_{\ell} \rangle    
= z_{i,\ell}\langle e_{\mu_j(m)} | e_{\mu_i(\ell)} \rangle 
= z_{i,\ell}\delta_{i,j}\delta_{m,\ell}             \nonumber\\
&= z_{i,m}\delta_{i,j}\langle e_m | e_\ell \rangle, \qquad 
    \ell\in\boldN,\\
\noalign{\noindent hence we obtain}
\pi(s_i)^* e_{\mu_j(m)} &= \delta_{i,j} \bar z_{i,m}e_m. \label{PR2}
\end{alignat}
Here, $\langle\,\cdot\,|\,\cdot\,\rangle$ denotes the inner product 
on $\calH$. It is, now, straightforward to show that $\pi(s_i)$ and 
$\pi(s_i)^*$ defined by \eqref{PR1} and \eqref{PR2} satisfy the 
relation \eqref{CR1} and \eqref{CR2} on any $e_n$. 
\par
We can classify permutation representations into two types as 
follows:\cite{BJ,DP,Kawamura1}
\begin{itemize}
\item[(1)] Permutation representation with a {\it central cycle\/}: 
There exists a monomial $\pi(s_{i_0,\ldots, i_{\kappa-1}})$ having 
an eigenvalue $z$ with $z\in\boldC$, $|z|=1$.
This representation is denoted by Rep$(i_0,\ldots,i_{\kappa-1};\,z)$ 
and a positive integer $\kappa$ is called the {\it length\/} of the 
central cycle.  For the special case $z=1$, we denote 
Rep$(i_0,\ldots,i_{\kappa-1})\equiv\,$Rep$(i_0,\ldots,i_{\kappa-1};\,1)$.
\item[(2)] Permutation representation with a {\it chain\/}: 
There is no eigenvector for any monomial in $s_i$'s and there exists 
a vector $v\in\calH$ satisfying 
$\|\pi(s^*_{i_0,\ldots, i_N})v\|=1$, $(N\in\boldN)$ for a certain 
sequence $\{i_k\}_{k=0}^\infty$ $(i_k=1,\ldots,d)$.
This representation is denoted by Rep$(\{i_k\})$.
\end{itemize}
Any of other permutation representations is expressed as a direct 
sum and a direct integral of (1) and (2) with multiplicity.
For Rep$(i_0,\ldots,i_{\kappa-1};\,z)$, a {\it label\/} 
$(i_0,\ldots,i_{\kappa-1})$ is called to be {\it periodic\/}, 
if $i_k=i_{M+k}$ $(k=0,1,\ldots, \kappa-1)$ is satisfied for a certain 
positive integer $M (<\kappa)$ under understanding that the subscripts 
of $i_k$'s take values in $\boldZ_\kappa$. The integer $M$ (if there 
are more than one, the minimum of such $M$'s) is called the 
{\it period\/} of the label $(i_0,\ldots,i_{\kappa-1})$.  
By definition, $M$ is a divisor of $\kappa$ smaller than $\kappa$.
If $\pi(s_{i_0,\ldots, i_{\kappa-1}})$ has an eigenvalue $z$, so does 
any of its cyclic permutations $\pi(s_{i'_0,\ldots, i'_{\kappa-1}})$.  
Hence all of $\kappa$ Rep$(i'_0,\ldots,i'_{\kappa-1};\,z)$'s obtained 
by cyclic permutations of a label $(i_0,\ldots,i_{\kappa-1})$ are 
identified.
Likewise for Rep$(\{i_k\})$, a label $\{i_k\}_{k=0}^\infty$ is called 
to be {\it eventually periodic\/} if there exist a positive integer $M$ 
satisfying $i_{k+M}=i_k$ for $k\geqq N$ with a nonnegative integer $N$. 
Rep$(\{i_k\})$ and Rep$(\{j_k\})$ are called to be {\it tail equivalent} 
if there exist nonnegative integers $M$ and $M'$ such that 
$i_{k+M}=j_{k+M'}$ $(k\in\boldN)$. Two tail equivalent permutation 
representations with chains are unitarily equivalent to each other.\cite{DP} 
It is known\cite{BJ,DP} that a permutation representation of $\calO_d$ is
irreducible if and only if it is cyclic and its label is not (eventually) 
periodic. 
\par
In the following, we give explicit realizations of permutation
representations. According to each type, it is convenient to rearrange 
the basis of $\calH$ in an appropriate form.
\par 
Rep$(i_0,\ldots,i_{\kappa-1};\,z)$ of $\calO_d$: The complete orthonormal 
basis of $\calH$ is denoted by 
$\{e_{\lambda,\,m}\mid \lambda\in\boldZ_\kappa,\,m\in\boldN\}$.  
The previous basis $\{e_n\}_{n=1}^\infty$ is recovered by such an 
identification as $e_{\kappa(m-1)+\lambda+1}\equiv e_{\lambda,\,m}$. 
We define the action of $\pi(s_i)$ on $\calH$ by
\begin{alignat}{1}
&\pi(s_i)\,e_{\lambda,\,1}
 =\begin{cases} 
      e_{\lambda-1,\,i+1} & \mbox{for $1\leqq i \leqq i_{\lambda-1} -1$,}\\
      z^{1/\kappa}\,e_{\lambda-1,\,1}   & \mbox{for $i=i_{\lambda-1}$,}\\
      e_{\lambda-1,\,i}   & \mbox{for $i_{\lambda-1}+1\leqq i \leqq d$,}
    \end{cases} \label{perm-rep-cycle-1}  \\ 
&\pi(s_i)\,e_{\lambda,\,m}
 = e_{\lambda-1,\,d(m-1)+i} \quad \hbox{for } m\geqq2. 
                \label{perm-rep-cycle-2}
\end{alignat}
Then, $e_{\lambda,\,1}$'s become eigenvectors of operators
$\{\pi(s_{i_{\lambda},\ldots,i_{\kappa-1},i_0,\ldots,i_{\lambda-1}}) \mid
\lambda\in\boldZ_\kappa\}$ as follows:
\begin{alignat}{1}
& \pi(s_{i_{\lambda},\ldots,i_{\kappa-1},i_0,\ldots,i_{\lambda-1}})\,
     e_{\lambda,\,1} 
   = z\, e_{\lambda,\,1}, \quad \lambda\in\boldZ_\kappa. 
    \label{perm-rep-eigenvector}
\end{alignat}
The set of eigenvectors $\{ e_{\lambda,\,1} \mid \lambda\in\boldZ_\kappa\}$ 
is called the {\it central cycle\/} of Rep$(i_0,\ldots,i_{\kappa-1};\,z)$.
Here, one should note that the subspace 
spanned by $\{e_{\lambda,\,m}\}_{m=1}^\infty$ for a fixed $\lambda$ is 
generated by the action of $n\kappa$-th monomials 
$\{ s_{j_1,\ldots,j_{n\kappa}} \mid n\geqq0;\, 
j_1,\ldots,j_{n\kappa}=1,\ldots,d \}$ on $e_{\lambda,\,1}$.
The special case Rep$(1)\equiv\,$Rep$(1;\,1)$ is called the 
{\it standard representation\/} and denoted by $\pi_s$ in 
Ref.\,\citen{AK1}):
\begin{alignat}{1}
&\pi_s(s_i)\,e_n = e_{d(n-1)+i},\qquad 
    i=1,2,\ldots,d\,;\ n\in\boldN,
 \label{rep1-formula1}
\end{alignat}
where $e_n\equiv e_{0,\,n}$. From \eqref{rep1-formula1}, it is 
straightforward to obtain the following formula:
\begin{alignat}{1}
 \begin{array}{l}
   \displaystyle
      \pi_s(s_{i_1,\ldots,i_k})\,e_n = e_{N(i_1,\ldots,i_k;n)}, \\ 
   \displaystyle\qquad 
      N(i_1,\ldots,i_k;n)\equiv 
        (n-1)d^k + \sum_{j=1}^k(i_j-1)d^{j-1} + 1, 
 \end{array}
 \label{rep1-formula2} 
\end{alignat}
for $i_1,\cdots,i_k=1,\ldots,d\,;\ n\in\boldN$.
\par
Rep$(\{i_k\})$ of $\calO_d$: The complete orthonormal basis of $\calH$ 
is denoted by $\{e_{\lambda,\,m}\mid \lambda\in\boldZ,\,m\in\boldN\}$. 
We define the action of $\pi(s_i)$ on $\calH$ by
\begin{alignat}{1}
&\pi(s_i)\,e_{\lambda,\,1}
 =\begin{cases}
    e_{\lambda-1,\,i+1} & \mbox{for $1\leqq i \leqq i_{\lambda-1}-1$,}\\
    e_{\lambda-1,\,1}   & \mbox{for $i = i_{\lambda-1}$,}\\
    e_{\lambda-1,\,i}   & \mbox{for $i_{\lambda-1}+1\leqq i \leqq d$,}
  \end{cases} \qquad 
     \hbox{for } \lambda\geqq 1, \label{perm-rep-chain-1}\\[5pt]
&\pi(s_i)\, e_{\lambda,\,m}
  = e_{\lambda-1,\, d(m-1)+i} \qquad 
    \hbox{for } \lambda\leqq0 \ \ \hbox{or}\ \ m\geqq2. 
                                 \label{perm-rep-chain-2}
\end{alignat}
Then, we obtain
\begin{alignat}{1}
&\pi(s^*_{i_0,\ldots,i_N})\,e_{0,\,1}=e_{N+1,\,1}, \qquad 
    N\in\boldN.      \label{perm-rep-chain-3}
\end{alignat}
The set of vectors $\{e_{\lambda,\,1}\mid \lambda\in\boldZ\}$ is 
called the {\it chain\/} of Rep$(\{i_k\})$.
The subspace spanned by $\{e_{\lambda,\,m}\}_{m=1}^\infty$ for a fixed 
$\lambda$ is generated by the action of $\calO_d^{U(1)}$ on 
$e_{\lambda,\,1}$.
It should be noted that from an equality 
\begin{equation}
\pi(s^*_{i_{M},\ldots, i_{N}})\,e_{M,\,1}=e_{N+1,\,1}, \qquad 
   0\leqq M\leqq N,             \label{perm-rep-chain-4}
\end{equation}
it is obvious that Rep$(\{j_k\})$ with $\{j_k\equiv i_{k+M}\}_{k=0}^\infty$, 
which is tail equivalent with $\{i_k\}$, is obtained from Rep$(\{i_k\})$ 
by rearranging the basis of $\calH$, hence Rep$(\{j_k\})$ is unitarily 
equivalent with Rep$(\{i_k\})$.
\par
In concluding this subsection, we remark on an important property of the 
standard representation. 
By using the homogeneous embedding $\varPsi_q$ of $\calO_{d^q}$ $(q\geqq2)$ 
into $\calO_d$ defined by \eqref{hom-embed}, the standard representation 
$\pi_s^{(q)}$ of $\calO_{d^q}$ is obtained from $\pi_s^{(1)}$ of $\calO_d$ 
as follows:
\begin{equation}
\pi_s^{(q)}=\pi_s^{(1)}\circ\varPsi_q.  \label{pi_s-reduction}
\end{equation}
Indeed, from \eqref{hom-embed}, \eqref{rep1-formula1} and 
\eqref{rep1-formula2}, we obtain
\begin{alignat}{1}
(\pi_s^{(1)}\circ\varPsi_q)(s'_i)e_n
&= \pi_s^{(1)}(s_{i_1,\ldots, i_q})e_n,  \qquad 
    i=\sum_{k=1}^q (i_k-1) d^{k-1} + 1   \nonumber\\
&= e_{(n-1)d^q+i}                        \nonumber\\
&= \pi_s^{(q)}(s'_i) e_n,
\end{alignat}
where $\{s'_1,\,\ldots,\,s'_{d^q}\}$ and 
$\{s_1,\,\ldots,\,s_d\}$ are the generators of $\calO_{d^q}$ 
and $\calO_d$, respectively.
On the other hand, as for other permutation representations with 
central cycles of length 1 and eigenvalue 1, we have
\begin{equation}
\pi_{i_0}^{(1)}\circ\varPsi_q \cong \pi_{\ti_0}^{(q)},
 \qquad \ti_0\equiv \frac{d^q-1}{d-1}(i_0-1)+1, \quad
 i_0=2,\ldots,d,    \label{pi_i-reduction}
\end{equation}
where $\pi_{i_0}^{(1)}$ and $\pi_{\ti_0}^{(q)}$ stand for Rep$(i_0)$ 
of $\calO_d$ and Rep$(\ti_0)$ of $\calO_{d^q}$, respectively, and we 
have used the symbol ``$\cong$'' to denote the unitary equivalence 
with taking into account that realizations of the representations 
are different from those given by \eqref{perm-rep-cycle-1}.
Here, it should be noted that $\varPsi_q(s'_{\ti_0})=(s_{i_0})^{d}$, 
and \eqref{pi_i-reduction} is obvious since there are no other 
monomials in $s_i$'s having eigenvector except for those only in 
$s_{i_0}$.
Generally, for an irreducible permutation representation with a label
$L=(i_0,\ldots,i_{\kappa-1})$ $(2\leqq \kappa <\infty)$, we obtain
\begin{alignat}{1}
& \pi_L^{(1)}\circ\varPsi_q \cong 
  \bigoplus_{j=1}^{\kappa q / r} \pi_{\tilde L_j}^{(q)},
\end{alignat}
where $r$ is the least common multiple of $\kappa$ and $q$, and   
$\{\tilde L_j\}_{j=1}^{\kappa q / r}$ is a certain set of nonperiodic
labels with length $r/q$ in $\calO_{d^q}$ determined by $L$ and $q$.
\vskip20pt
\subsection{Branching of permutation representations}
Let $\calA$ and $\calB$ be algebras on $\boldC$.
From a representation $\pi$ of $\calA$ and a homomorphism 
$\varphi:\calB\to\calA$, we have a representation $\pi\circ\varphi$ 
of $\calB$ by composing $\pi$ and $\varphi$.
Even if $\pi$ is irreducible (or indecomposable), $\pi\circ\varphi$ 
is not necessarily so. 
If it is possible to decompose $\pi\circ\varphi$ into a direct sum 
of a family $\{\pi_\lambda\}_{\lambda\in\Lambda}$ of representations
of $\calB$, which are representatives of the unitary equivalence class 
of representations of $\calB$, we write
\begin{alignat}{1}
&\pi\circ\varphi \cong \bigoplus_{\lambda\in\Lambda}\pi_\lambda,
\end{alignat}
and call it the {\it branching\/} of $\pi$ by $\varphi$.
It should be noted that the symbol $\!$``$\cong$''$\!$ denotes unitary 
equivalence. Likewise, for an endomorphism $\varphi:\calA\!\to\!\calA$, 
we have a similar branching.
\par
In general, for a given irreducible permutation representation 
$\pi$ of the Cuntz algebra, a branching of $\pi$ by a kind of 
$\ast$-endomorphism $\varphi$ is given by\cite{Kawamura3}
\begin{equation}
\pi\circ\varphi \cong \bigoplus_{L\in\calS}\pi_L, \label{branch}
\end{equation}
where $\calS$ denotes a certain set of irreducible permutation 
representations determined by $\pi$ and $\varphi$.
Here, we give a few examples in $\calO_2$ necessary in later discussions.
For more examples and detailed discussions, see Ref.\,\citen{Kawamura3}).
\par
We consider a family of particular $(p+1)$-th order permutation 
endomorphisms $\{\varphi_{\sigma_p}\}_{p\geqq1}$ defined by 
\eqref{ext-perm-endo} with 
$\sigma_p\!\in\!{\mathfrak S}_{2^{p+1}}$ being the transposition 
$\sigma_p(1,j_1,\ldots, j_p)\equiv(1,j_1,\ldots,j_p)$, 
$\sigma_p(2,j_1,\ldots, j_{p-1},j_p)\equiv(2,j_1,\ldots, j_{p-1},\hatj_p)$, 
$\hatj_p\equiv 3-j_p$:
\begin{eqnarray}
&& 
\begin{cases}
\displaystyle \varphi_{\sigma_p}(s_1) = s_1, \\
\displaystyle \varphi_{\sigma_p}(s_2) = s_2\,\rho^{p-1}(J), \quad 
   J\equiv s_{2;1} + s_{1;2} = J^* = J^{-1},
\end{cases} \label{p-branch-endo}
\end{eqnarray}
where $\rho$ is the canonical endomorphism of $\calO_2$ and 
$\rho^0(X)\equiv X$. Then, it is shown that
\begin{alignat}{2}
\varphi_{\sigma_p}\circ\varphi_{\sigma_q} 
 &=\varphi_{\sigma_q}\circ\varphi_{\sigma_p},
 &\quad  &p,\,q\in\boldN, \label{p-branch-endo-1}\\
(\varphi_{\sigma_p})^2 &=\varphi_{\sigma_{2p}}, 
 &\quad &p\in\boldN.\label{p-branch-endo-2}
\end{alignat}
Indeed, from the equality 
\begin{alignat}{1}
 J \rho(X) &= \rho(X) J, \quad X\in\calO_2,
\end{alignat}
we have
\begin{alignat}{1}
 \varphi_{\sigma_p}(\rho^n(J)) &= \rho^n(J)\,\rho^{p+n}(J), \qquad 
  n+1\in\boldN. 
\end{alignat}
Hence we obtain
\begin{alignat}{1}
(\varphi_{\sigma_p}\circ\varphi_{\sigma_q})(s_1) &= s_1\nonumber\\
 &= (\varphi_{\sigma_q}\circ\varphi_{\sigma_p})(s_1), \\
(\varphi_{\sigma_p}\circ\varphi_{\sigma_q})(s_2) 
 &= s_2\, \rho^{p-1}(J)\, \rho^{q-1}(J)\, \rho^{p+q-1}(J) \nonumber\\
 &= (\varphi_{\sigma_q}\circ\varphi_{\sigma_p})(s_2).
\end{alignat}
\par
Now, we consider eigenvectors of operators in the range of 
$\pi_s\circ\varphi_{\sigma_p}$.
First, one should note that \eqref{p-branch-endo} is rewritten 
as follows:
\begin{alignat}{1}
&\begin{array}{l}
\varphi_{\sigma_p}(s_i) = s_i\, u_i, \quad i=1,\,2,\\
\displaystyle \quad 
  u_1 \equiv I, \quad 
  u_2 \equiv 
 \sum_{j=1}^{2^{p-1}}(S^{(p)}_{j+2^{p-1};\,j}+S^{(p)}_{j;\,j+2^{p-1}}) 
       = u_2^* = u_2^{-1}, 
\end{array}
\end{alignat}
where $\{S^{(p)}_j\mid j=1,2,\ldots,2^p\}$ denote the homogeneously 
embedded generators of $\calO_{2^p}$ into $\calO_2$ defined by 
\eqref{hom-embed}.  By using equalities
\begin{alignat}{1}
&u_2\, S^{(p)}_j = 
\begin{cases}
 S^{(p)}_{j+2^{p-1}} & \mbox{for $1\leqq j\leqq 2^{p-1}$,}\\[5pt]
 S^{(p)}_{j-2^{p-1}} & \mbox{for $2^{p-1}+1\leqq j\leqq 2^{p}$,}
\end{cases} \label{u_2-property-1} \\
&u_{j_1}s_{j_2,\ldots, j_p, 1} = s_{j_2,\ldots, j_p, j_1}, \qquad
 j_1,\ldots,j_p = 1,2 \label{u_2-property-2}
\end{alignat}
and $\pi_s(s_1)\,e_1=e_1$, we obtain
\begin{alignat}{1}
\pi_s\big(\varphi_{\sigma_p}(S^{(p)\,\ast}_k)\,S^{(p)}_j\big)\,e_1 
&=\pi_s(u_{k_p}s^*_{k_p}\cdots u_{k_1}s^*_{k_1}\,s_{j_1,\ldots,j_p})\,
     e_1 \nonumber\\
&=\delta_{k_1,j_1}
     \pi_s(u_{k_p}s^*_{k_p}\cdots u_{j_1}\,s_{j_2,\ldots,j_p,1})\,
     e_1     \nonumber\\
&=\delta_{k_1,j_1}\pi_s(u_{k_p}s^*_{k_p}\cdots u_{k_2}s^*_{k_2}\,
     s_{j_2,\ldots,j_p,j_1})\,e_1 \nonumber\\
&=\cdots\nonumber\\
&=\delta_{k_1,j_1}\cdots\delta_{k_p,j_p}\pi_s(s_{j_1,\ldots,j_p})\,
     e_1 \nonumber\\
&= \delta_{k,j}\pi_s\big(S^{(p)}_j\big)\,e_1,
   \qquad j, k = 1, 2, \ldots, 2^p.   \label{phi_p(S^*)S}
\end{alignat}
Making $\pi_s(\varphi_{\sigma_p}(S^{(p)}_k))$ act on 
\eqref{phi_p(S^*)S} and summing up with respect to $k=1,\ldots,2^p$, 
we obtain
\begin{alignat}{1}
&\pi_s\big(\varphi_{\sigma_p}(S^{(p)}_j\big)\,S^{(p)}_j)\,e_1 
   = \pi_s(S^{(p)}_j)\,e_1,  \quad    j=1,2,\ldots,2^p,
   \label{phi_p-eigenvector}
\end{alignat}
hence $\pi_s(S^{(p)}_j)\,e_1$ is the eigenvector
of $(\pi_s\circ\varphi_{\sigma_p})(S^{(p)}_j)$.
Furthermore, from \eqref{u_2-property-1} and \eqref{u_2-property-2}, 
for any set of indices $j_1,\ldots,j_{mp+k} = 1,2$ with $m\in\boldN$ 
and $1\leqq k \leqq p$, we can show that there is a unique set of 
indices $j'_1(=j_1),j'_2,\ldots,j'_{(m-1)p+k}=1,2$ and $j=1,\ldots,2^p$
such that\begin{alignat}{1}
s_{j_1, j_2,\ldots,j_{mp+k}}
&=\varphi_{\sigma_p}(s_{j'_1, j'_2, \ldots, j'_{(m-1)p+k}})\,S^{(p)}_j,
\quad m\in\boldN,\ k=1,\ldots,p.
\end{alignat}
As for $s_{j_1,\ldots,j_k}$ with $k=1,\ldots,p$, from $\pi_s(s_1)\,
e_1=e_1$, we have
\begin{alignat}{1}
\pi_s(s_{j_1,\ldots,j_k})\,e_1
&=\pi_s(s_{j_1,\ldots,j_k})\,\pi_s\big((s_1)^{p-k}\big)\,e_1 \nonumber\\
&=\pi_s\big(s_{j_1,\ldots,j_k}(s_1)^{p-k}\big)\,e_1 \nonumber\\
&=\pi_s\big(S^{(p)}_j\big)\,e_1, \qquad 
  j \equiv \sum_{\ell=1}^k (j_\ell-1)2^{\ell-1}+1.
\end{alignat}
Therefore, any of the basis $\{e_n\}_{n=1}^\infty$, which 
satisfies \eqref{rep1-formula2} with $d=2$ and $n=1$, is given by 
an action of $(\pi_s\circ\varphi_{\sigma_p})(s_{j_1,\ldots,j_m})$ 
$(j_1,\ldots,j_m=1,2,\ m\in\boldN)$ 
on one of the $2^p$ eigenvectors in \eqref{phi_p-eigenvector} .
\par
For any divisor $\kappa$ of $p$, we can rewrite $u_2$ as
\begin{alignat}{1}
&u_2 = \rho^{(\kappa)\,\frac{p}{\kappa}-1}\Big(
\sum_{i=1}^{2^{\kappa-1}}\big(S^{(\kappa)}_{i+2^{\kappa-1};\,i}
+S^{(\kappa)}_{i;\,i+2^{\kappa-1}}\big)\Big),\\
&\rho^{(\kappa)}(X)\equiv 
 \sum_{k=1}^{2^\kappa} S^{(\kappa)}_k\, X\, S^{(\kappa)\,\ast}_k 
= \rho^\kappa(X), \quad X\in\calO_2,
\end{alignat}
where $\{S^{(\kappa)}_i \mid i=1,2,\ldots,2^\kappa\}$ denote the 
homogeneously embedded generators of $\calO_{2^\kappa}$ into 
$\calO_2$.  Then, in the same way as above, it is shown that
\begin{alignat}{1}
&\pi_s\big(\varphi_{\sigma_p}(S^{(\kappa)}_i)\,
                             (S^{(\kappa)}_i)^{\frac{p}{\kappa}}
      \big)\,e_1 
 = \pi_s\big((S^{(\kappa)}_i)^{\frac{p}{\kappa}}\big)\,e_1, \quad
   i=1,2,\ldots,2^\kappa.
   \label{phi_p-eigenvector-2}
\end{alignat}
From \eqref{hom-embed}, we have
\begin{alignat}{1}
&(S^{(\kappa)}_i)^{\frac{p}{\kappa}}=S^{(p)}_{\ti}, 
\quad \ti\equiv \frac{2^p-1}{2^\kappa-1}(i-1)+1,
\quad i=1,\ldots,2^\kappa,
\end{alignat}
hence some eigenvectors in \eqref{phi_p-eigenvector} are reduced to 
those in \eqref{phi_p-eigenvector-2}.
By writing $S^{(\kappa)}_i$ $(i=1,\ldots,2^\kappa)$ explicitly as
\begin{alignat}{1}
S^{(\kappa)}_i &= s_{i_0,\ldots,i_{\kappa-1}}, \quad 
i= \sum_{k=1}^{\kappa} (i_{k-1}-1)2^{k-1}+1
\end{alignat}
with $i_0,\ldots,i_{\kappa-1}=1,2$, we obtain
\begin{alignat}{1}
&\pi_s\big(\varphi_{\sigma_p}(s_{i_{\kappa-1}})\,
(S^{(\kappa)}_i)^{\frac{p}{\kappa}}\big)\,e_1 
 = \pi_s\big((S^{(\kappa)}_{i'})^{\frac{p}{\kappa}}\big)\,e_1, \qquad
S^{(\kappa)}_{i'}\equiv s_{i_{\kappa-1},i_0,\ldots,i_{\kappa-2}}.
\end{alignat}
Therefore, if the set of indices $(i_0,\ldots,i_{\kappa-1})$ is not periodic
in the sense stated in Sec.\,3-1, we can see that 
$(\pi_s\circ\varphi_{\sigma_p})(s_i)$ $(i=1,2)$ 
act on a set of $\kappa$ vectors 
\begin{alignat}{1}
&
\Big\{
\pi_s\big((s_{i_0,\ldots,i_{\kappa-1}})^{\frac{p}{\kappa}}\big)\,e_1,\ 
\pi_s\big((s_{i_1,\ldots,i_{\kappa-1},i_0})^{\frac{p}{\kappa}}\big)\,e_1,\
\ldots,\ 
\pi_s\big((s_{i_{\kappa-1},i_0,\ldots,i_{\kappa-2}})^{\frac{p}{\kappa}}\big)\,e_1\Big\}
\end{alignat}
in such way that they constitute a central cycle of length $\kappa$ 
of the irreducible permutation representation 
Rep$(i_0,i_1,\ldots,i_{\kappa-1})$ of $\calO_2$.
\par
From the above discussions, we can, now, show the branching formula of 
the standard representation $\pi_s$ by $\varphi_{\sigma_p}$ as follows:
\begin{equation}
\pi_s\circ\varphi_{\sigma_p} \cong \bigoplus_{L\in I\!P\!R_p}\pi_L, 
\label{branch-formula}
\end{equation}
where $I\!P\!R_p$ denotes a set of all irreducible permutation 
representations with central cycles and with eigenvalue 1, in which 
each length $\kappa$ of central cycles is a divisor of $p$.
Here, $\kappa$ eigenvectors in Rep$(L)$ with a nonperiodic label 
$L=(i_0,\ldots,i_{\kappa-1})$ $(1\leqq \kappa\leqq p)$ is given by
\begin{gather}
(\pi_s\circ\varphi_{\sigma_p})
 (s_{i_\lambda,\ldots,i_{\kappa-1},i_0,\ldots,i_{\lambda-1}})
 \,e_{N(L,\lambda)} = e_{N(L,\lambda)}, \quad   \lambda\in\boldZ_\kappa,\\
 N(L,\lambda)\equiv 
    \frac{2^p-1}{2^\kappa-1} 
    \sum_{\ell=1}^{\kappa} (i_{\lambda+\ell-1}-1)2^{\ell-1}+1, 
\end{gather}
\par
For better understanding, we explicitly write \eqref{branch-formula} 
for $p=1,2,3,4$ as follows:
\begin{alignat}{1}
\pi_s\circ\varphi_{\sigma_1} &\cong \pi_s \oplus \pi_2, \label{branch1} \\
\pi_s\circ\varphi_{\sigma_2} &\cong \pi_s \oplus \pi_2 \oplus \pi_{1,2}, \\
\pi_s\circ\varphi_{\sigma_3} &\cong \pi_s \oplus \pi_2  
                        \oplus \pi_{1,1,2} \oplus \pi_{1,2,2},\\
\pi_s\circ\varphi_{\sigma_4} &\cong \pi_s \oplus \pi_2  
                        \oplus \pi_{1,2} 
\oplus \pi_{1,1,1,2} \oplus \pi_{1,1,2,2} \oplus \pi_{1,2,2,2},
\end{alignat}
where $\pi_{i_0,\ldots,i_{\kappa-1}}$ $(\kappa=1,2,3,4)$ denotes 
Rep$(i_0,\ldots,i_{\kappa-1})$.
Since it is easy to reconfirm \eqref{branch1}, we show it concretely 
in the following.
\par
From $\pi_s(s_i)e_1=e_i$ $(i=1,2)$, which is obtained from 
\eqref{rep1-formula1} with $d=2$, and 
\begin{equation}
J s_j = s_\hatj,\quad \hatj\equiv 3-j,\quad j=1,2, \label{J s_i}
\end{equation}
we have
\begin{alignat}{1}
(\pi_s\circ\varphi_{\sigma_1})(s_1)\,e_{1}&=e_{1}, \\
(\pi_s\circ\varphi_{\sigma_1})(s_2)\,e_{2}&=e_{2}.
\end{alignat}
Furthermore, from \eqref{J s_i} and $u_2=J$, we obtain
\begin{alignat}{1}
s_{j_1,j_2,\ldots, j_{k+1}} 
  &= \varphi_{\sigma_1}(s_{j'_1,j'_2,\ldots, j'_{k}})\,s_{j'_{k+1}}, 
   \quad  j_1,j_2,\ldots,j_{k+1}=1,\,2,\\
   s_{j'_1} &\equiv s_{j_1}, \qquad
   s_{j'_\ell} \equiv u_{j'_{\ell-1}}s_{j_\ell}=
             \begin{cases}
               s_{j_\ell}     & \mbox{for $j'_{\ell-1}=1$,}\\[5pt]
               s_{\hatj_\ell} & \mbox{for $j'_{\ell-1}=2$,}
             \end{cases} \quad \ell=2,\ldots,k+1, \label{j'-j}
\end{alignat}
hence we have 
\begin{alignat}{1}
(-1)^{j'_{\ell}-1}&=(-1)^{j_{\ell}-1}(-1)^{j'_{\ell-1}-1}, \quad 
                      \ell=2,\ldots,k+1,\\
(-1)^{j'_{k+1}-1}&=\prod\limits_{\ell=1}^{k+1}(-1)^{j_\ell-1},
\end{alignat}
that is, we have $j'_{k+1}=1$ if the number of $2$ in 
$\{j_1,\ldots,j_{k+1}\}$ is even, and $j'_{k+1}=2$ otherwise.
Therefore, any of $\{e_n\}_{n=1}^\infty$, which is expressed as
$\pi_s(s_{i_1,\ldots,i_{k+1}})\,e_1$ with $i_1,\ldots,i_{k+1}=1,2$, 
is uniquely given by an action of 
$(\pi_s\circ\varphi_{\sigma_1})(s_{j_1,\ldots, j_k})$ on either $e_1$ 
or $e_2$ with an appropriate set of indices $\{j_1,\ldots,j_k = 1,\,2\}$.
Thus, we obtain \eqref{branch1}. 
\par
Next, we consider the branching number $B_p$ of $\pi_s$ by 
$\varphi_{\sigma_p}$, which is defined by the number of irreducible 
permutation representations appearing in the rhs of \eqref{branch-formula}.
We can obtain $B_p$ in the following way. 
First, let $C_n$ be the number of irreducible permutation representations 
Rep$(i_0,\ldots,i_{n-1})$.  
One should note the following:\ 
(1)\ 
Rep$(i_0,\ldots,i_{n-1})$ is defined up to cyclic permutations of the label; 
(2)\ 
if Rep$(i_0,\ldots,i_{n-1})$ is periodic, its periodicity is a divisor 
of $n$ except for $n$ itself;
(3)\ 
the total number of Rep$(i_0,\ldots,i_{n-1})$'s involving reducible or
redundant ones is given by $2^n$.
Then, the recurrence formula for $C_n$ is given by
\begin{eqnarray}
&&
\begin{cases}
C_1 = 2, \\[5pt]
\displaystyle\sum_{k\in D_n} k\,C_k = 2^n,
\end{cases} \label{rec-C_n}
\end{eqnarray}
with $D_n$ being the whole set of divisors of $n$. 
In terms of  $C_n$, $B_p$ is given by
\begin{equation}
B_p = \sum_{n\in D_p} C_n
\end{equation}
with $D_p$ being the whole set of divisors of $p$. 
Since it is an elementary problem to solve \eqref{rec-C_n}, we give here
its solution without proof. Let $n=n_1^{m_1}\cdots n_r^{m_r}$
be the factorization of $n$ in prime numbers. Then, $C_n$ $(n\geqq2)$ is 
given by
\begin{equation}
C_n = \frac{1}{n}\left[ 2^n 
 + \sum_{\ell=1}^r (-1)^\ell 
   \sum_{k_1<\cdots<k_\ell} 2^{n/(n_{k_1}\,\cdots\,n_{k_\ell})}\right].
\end{equation}
In particular, for a prime number $n(\geqq2)$, we have
\begin{equation}
C_n = \frac{2^n -2}{n}.
\end{equation}
\par
In concluding this subsection, we show that, for any irreducible 
permutation representation with a central cycle, $\pi_L$, we can 
explicitly construct a $\ast$-endomorphism $\varphi$ so that it yields 
only the standard representation $\pi_s$ as follows:
\begin{eqnarray}
&& \pi_L\circ\varphi \cong \pi_s. \label{branch-to-pi_s} 
\end{eqnarray}
For example, in $\calO_2$, \eqref{branch-to-pi_s} for $L=(1,2)$ 
is satisfied by the $\ast$-endomorphism $\varphi$ as follows:
\begin{eqnarray}
&& \varphi(s_1) \equiv s_{1,2}, \qquad
   \varphi(s_2) \equiv s_{2;1} + s_{1,1;2}, \label{pi_{1,2}-endo}
\end{eqnarray}
which is one of the inhomogeneous endomorphisms defined by \eqref{inhom-endo}.
Indeed, it is straightforward to show
that $\varphi(s_{i_1,\ldots, i_n})$ for any index $(i_1,\ldots,i_n)$
involves none of monomials in the form of 
$\{ (s_{2,1})^m,\ s_{j_1,\ldots, j_k}(s_{2,1})^{m-1}s^*_{j_1,\ldots, j_k},\ 
s_{j_1,\ldots, j_k}(s_{1,2})^{m-1}s^*_{j_1,\ldots, j_k} \}$ with 
$j_1,\ldots, j_k =1,2;\ k\geqq1$, and $m\geqq1$,
hence there is no eigenvector except for $(\pi_{1,2}\circ\varphi)(s_1)$.
On the other hand, from \eqref{pi_{1,2}-endo}, we have
\begin{alignat}{2}
&\begin{array}{l}
\displaystyle
\varphi\big((s_2)^{2m-1}\big)=s_2(s_1)^{m-1}s_1^* + (s_1)^{m+1}s_2^*,\\[5pt]
\displaystyle
\varphi\big((s_2)^{2m}\big)  =s_2(s_1)^{m}s_2^* + (s_1)^{m+1}s_1^*,
\end{array} &\quad &m\geqq1,\\
\noalign{\noindent hence we obtain}
&\begin{array}{l}
\displaystyle
\varphi\big((s_2)^{2m+1}s_1\big) = s_2(s_1)^m s_2, \\[5pt]
\displaystyle
\varphi\big((s_2)^{2m}s_1\big)   = (s_1)^m s_{1,2}, 
\end{array} &\quad &m\geqq0. \label{pi_{1,2}-endo-2}
\end{alignat}
Since any monomial $s_{j_1,\ldots,j_k,1,2}$ is uniquely 
written as a product of the monomials appearing in the rhs of 
\eqref{pi_{1,2}-endo-2}, it is rewritten into 
$\varphi(s_{j'_1,\ldots,j'_\ell})$ with an appropriate set of 
indices $\{j'_1,\ldots,j'_\ell= 1,2\}$. 
Therefore, any of the basis of Rep$(1,2)$, 
$\{e_{\lambda,\,m}\mid \lambda=0,1;\ m\in\boldN\}$, which is expressed as
$\pi_{1,2}(s_{j_1,\ldots\,j_k})\,e_{0,\,1}
=\pi_{1,2}(s_{j_1,\ldots\,j_k,1,2})\,e_{0,\,1}$, is rewritten into  
$(\pi_{1,2}\circ\varphi)(s_{j'_1,\ldots,j'_\ell})\,e_{0,\,1}$.
Thus, from $(\pi_{1,2}\circ\varphi)(s_1)\,e_{0,\,1}=e_{0,\,1}$, 
we obtain \eqref{branch-to-pi_s} for $L=(1,2)$.
Besides \eqref{pi_{1,2}-endo}, \eqref{branch-to-pi_s} is satisfied 
also by $\varphi'(s_1)\equiv s_{2,1}$ and 
$\varphi'(s_2)\equiv s_{2,2;1}+s_{1;2}$.
However, it should be noted that the $\ast$-endomorphism defined by 
$\varphi''(s_1)\equiv s_{1,2}$ and 
$\varphi''(s_2)\equiv s_{1,1;1}+s_{2;2}$ 
(or $\varphi''(s_1)\equiv s_{2,1}$ and 
$\varphi''(s_2)\equiv s_{1;1}+s_{2,2;2}$)
does not satisfy \eqref{branch-to-pi_s}, that is, 
$\pi_{1,2}\circ\varphi''$ yields a direct sum of $\pi_s$ 
and an infinite number of $\pi_2$. 
\par
In general, in $\calO_2$, for any $\pi_L$ with a nonperiodic 
label $L=(i_0,\ldots,i_{\kappa-1};\,z)$, \eqref{branch-to-pi_s} is 
satisfied by the $\ast$-endomorphism $\varphi$ defined by
\begin{alignat}{2}
\varphi(s_1) \equiv &\; S_{\kappa+1}, 
 &\quad&\varphi(s_2) \equiv \sum_{j=1}^{\kappa} S_{j}T^*_j, 
    \label{pi_L-endo-1}\\ 
\quad\left\{ \hspace*{-5pt}
  \begin{array}{c} 
    S_1\\ 
    S_j\\ 
	S_{\kappa+1} 
  \end{array}\right.\hspace*{-5pt}
&\hspace*{-5pt}
  \begin{array}{l} 
   \equiv s_{\hati_0},\\
   \equiv s_{i_0,\ldots, i_{j-2}, \hati_{j-1}},\\
   \equiv \bar{z}\,s_{i_0,\ldots, i_{\kappa-2}, i_{\kappa-1}},
  \end{array}
 &\quad & 2\leqq j \leqq \kappa, \label{pi_L-endo-2}\\
\quad\left\{
  \begin{array}{c} 
    T_1\\ 
    T_j\\ 
    T_\kappa
  \end{array}\right.\hspace*{1pt}
 &\hspace*{-5pt}
  \begin{array}{l} 
   \equiv s_{i_0},\\
   \equiv s_{\hati_0, \ldots, \hati_{j-2}, i_{j-1}},\\
   \equiv s_{\hati_0, \ldots, \hati_{\kappa-3}, \hati_{\kappa-2}},
  \end{array}
 &\quad & 2\leqq j \leqq \kappa-1  \label{pi_L-endo-3}
\end{alignat}
with $\hati\equiv 3-i$, where $\{S_1,\ldots,S_{\kappa+1}\}$ and 
$\{T_1,\ldots,T_\kappa\}$ are specific generators of 
$\calO_{\kappa+1}$ and $\calO_{\kappa}$ embedded into $\calO_2$, 
respectively. 
\par
It is possible to obtain such a $\ast$-endomorphism also for $\calO_d$ 
$(d\geqq3)$, but it is rather complicated to construct a general 
formula similar to \eqref{pi_L-endo-1}--\eqref{pi_L-endo-3}. 
We give here an example in $\calO_3$: \eqref{branch-to-pi_s} for 
$\pi_L$ with $L=(1,2,1,3)$ is satisfied by the $\ast$-endomorphism 
$\varphi$ defined by
\begin{alignat}{1}
&\begin{array}{l}
\displaystyle
\varphi(s_1)\equiv s_{1,2,1,3}, \\[5pt]
\displaystyle
\varphi(s_2)\equiv s_2, \\[5pt]
\displaystyle
\varphi(s_3)\equiv s_{3;\,1} + s_{1,3;\,2,2} + s_{1,1;\,3,2} + s_{1,2,2;\,1,2} 
                 + s_{1,2,3;\,2,3} + s_{1,2,1,1;\,3,3} + s_{1,2,1,2;\,1,3}.
\end{array}
\end{alignat}
\vskip40pt
\section{Recursive Fermion System}
In this section, we summarize the construction of  the recursive 
fermion system (RFS$_p$),\cite{AK1} which gives embeddings of the 
CAR algebra into $\calO_{2^p}$ $(p\in\boldN)$.
We denote the generators of the CAR algebra by 
$\{a_n \mid n\in\boldN \}$ which satisfy
\begin{equation}
\{ a_m, \, a_n \} =0, \quad \{ a_m, \, a_n^* \} = \delta_{m,n}I, 
 \quad m,\,n\in\boldN.\label{car}
\end{equation}
%
%
\vskip20pt
\subsection{Definition of RFS$_p$ in $\boldsymbol{\calO_{2^p}}$} 
Let $\bolda_1,\,\bolda_2,\,\ldots,\,\bolda_p\in\calO_{2^p}$, 
$\zeta_p: \calO_{2^p} \to \calO_{2^p}$ be a linear mapping, 
and $\varphi_p$ a unital $\ast$-endomorphism of $\calO_{2^p}$, 
respectively.
A set 
$R_p=(\bolda_1,\,\bolda_2,\,\ldots,\,\bolda_p\,;\,\zeta_p,\,\varphi_p)$ 
is called a {\it recursive fermion system of order $p$ (RFS$_p$) 
in $\calO_{2^p}$,} if it satisfies the following conditions 
\begin{alignat}{4}
&\phantom{ii}\mbox{(i) seed condition: }
&&\{\bolda_j,\,\bolda_k\}=0,\quad 
     \{\bolda_j,\,\bolda_k^*\}\!=\!\delta_{j,k}I, 
     &\quad &j,k\!=\!1,\ldots,p,  \\[3pt]
&\phantom{i}\mbox{(ii) recursive condition: }
&&\{\bolda_i,\,\zeta_p(X)\}\!=\!0,\quad 
     \zeta_p(X)^*\!=\!\zeta_p(X^*),  
    &&X \in \calO_{2^p},                 \label{RFS_p-a-phi}\\[3pt]
&\mbox{(iii) normalization condition: }
&&\zeta_p(X)\zeta_p(Y)=\varphi_p(XY),   
    &&X,\,Y \in \calO_{2^p}              \label{RFS_p-zeta}
\end{alignat}
and none of $\{\bolda_1,\,\ldots,\bolda_p\}$ is expressed as 
$\zeta_p(X)$ with $X\in\calO_{2^p}$. 
We call $\bolda_j$ $(j=1,\ldots,p)$ and $\zeta_p$ the {\it seeds\/} 
and the {\it recursive map\/} of RFS$_p$, respectively.
The embedding $\varPhi_{R_p}$ of the CAR algebra into 
$\calO_{2^p}$ associated with $R_p$ is defined by
\begin{equation}
\begin{array}{c}
   \displaystyle
   \varPhi_{R_p}: \mbox{CAR} \hookrightarrow \calO_{2^p},  \\[10pt]
   \displaystyle
   \varPhi_{R_p}(a_{p(m-1)+j}) \equiv \zeta_p^{n-1}(\bolda_j) 
      \quad  j=1,\,\ldots,p\,;\ m\in\boldN.  
\end{array}  \label{RFSp-embed}
\end{equation}
We denote $\calA_{R_p}\equiv\varPhi_{R_p}(\mbox{CAR})$ and call it
the {\it CAR subalgebra of $\calO_{2^p}$ associated with $R_p$}.
\par
The simplest example of RFS$_p$ is given by the {\it standard\/} 
RFS$_{p}$
$SR_p=(\bolda_1,\,\ldots,\,\bolda_p\,;\, \zeta_p, \, \varphi_p)$, 
which is defined by
\begin{alignat}{1}
& \bolda_j = \sum_{k=1}^{2^{p-j}}\sum_{\ell=1}^{2^{j-1}}
         (-1)^{\sum\limits_{m=1}^{j-1}\left[\frac{\ell-1}{2^{m-1}}\right]}
         s_{2^j(k-1)+\ell}s_{2^{j-1}(2k-1)+\ell}^*, 
   \quad j=1,\,\ldots,\,p, \label{RFSp-1}\\
& \zeta_p(X) = \sum_{i=1}^{2^p}
             (-1)^{\sum\limits_{m=1}^{p}\left[\frac{i-1}{2^{m-1}}\right]}
             s_i X s_i^*, \quad X \in \calO_{2^p}, \label{RFSp-2}  \\
& \varphi_p(X) = \rho_{2^p}(X) \equiv \sum_{i=1}^{2^p} s_i X s_i^*,
           \quad X \in \calO_{2^p}, \label{RFSp-3}
\end{alignat}
where $[x]$ denotes the largest integer not greater than $x$, and 
$\rho_{2^p}$ being the canonical endomorphism \eqref{c-endo} of 
$\calO_{2^p}$.
It is shown that $\calA_{SR_p}=\calO_{2^p}^{U(1)}$ by mathematical 
induction, that is, any 
$s_{i_1,\ldots, i_k;\,j_k,\ldots, j_1} \in \calO_{2^p}^{U(1)}$ 
is expressed in terms of $a_n \ (n\leqq kp)$.
\par
Especially, $SR_p$ for $p=1,\,2,\,3,\,4$ are given by
\begin{alignat}{1}
&SR_1\ \ 
\begin{cases}
\bolda_1 \equiv s_{1;2},\\[3pt]
\zeta_1(X) \equiv s_1 X s_1^* - s_2 X s_2^*, \quad X \in \calO_2,\\[3pt]
\varphi_1(X) \equiv \rho_2(X) = \sum\limits_{i=1}^2 s_i X s_i^*, 
   \quad X \in \calO_2.
\end{cases} \label{srfs_1} \\[5pt]
&SR_2\ \ 
\begin{cases}
\bolda_1 \equiv s_{1;2} + s_{3;4}, \\[3pt]
\bolda_2 \equiv s_{1;3} - s_{2;4}, \\[3pt]
\zeta_2(X) \equiv s_1 X s_1^* - s_2 X x_2^* - s_3 X s_3^* + s_4 X s_4^*, 
   \quad X \in \calO_4, \\[3pt]
\varphi_2(X) \equiv \rho_4(X) = \sum\limits_{i=1}^4 s_i X s_i^*, 
   \quad X \in \calO_4.
\end{cases}  \label{srfs_2} \\[5pt]
&SR_3\ \ 
\begin{cases}
\bolda_1 \equiv s_{1;2} + s_{3;4} + s_{5;6} + s_{7;8}, \\[3pt]
\bolda_2 \equiv s_{1;3} - s_{2;4} + s_{5;7} - s_{6;8}, \\[3pt]
\bolda_3 \equiv s_{1;5} - s_{2;6} - s_{3;7} + s_{4;8}, \\[3pt]
\zeta_3(X) \equiv 
  s_1 X s_1^* - s_2 X x_2^* - s_3 X s_3^* + s_4 X s_4^* \\[3pt]
\phantom{\zeta_2(X) = } 
  - s_5 X s_5^* + s_6 X x_6^* + s_7 X s_7^* - s_8 X s_8^*, 
  \quad X \in \calO_8, \\[3pt]
\varphi_3(X) \equiv \rho_8(X) 
     = \sum\limits_{i=1}^8 s_i X s_i^*, \quad X \in \calO_8.
\end{cases} \label{srfs_3} \\[5pt]
\noalign{\eject}
&SR_4\ \ 
\begin{cases}
\bolda_1 \equiv  s_{1;2} + s_{3;4} + s_{5;6} + s_{7;8}
               + s_{9;10} + s_{11;12} + s_{13;14} + s_{15;16}, \\[3pt]
\bolda_2 \equiv s_{1;3} - s_{2;4} + s_{5;7} - s_{6;8}
               + s_{9;11} - s_{10;12} + s_{13;15} - s_{14;16}, \\[3pt]
\bolda_3 \equiv s_{1;5} - s_{2;6} - s_{3;7} + s_{4;8}
               + s_{9;13} - s_{10;14} - s_{11;15} + s_{12;16}, \\[3pt]
\bolda_4 \equiv s_{1;9} - s_{2;10} - s_{3;11} + s_{4;12}
               - s_{5;13} + s_{6;14} + s_{7;15} - s_{8;16},    \\[3pt]
\zeta_3(X) \equiv 
 s_1 X s_1^* - s_2 X x_2^* - s_3 X s_3^* + s_4 X s_4^* \\[3pt]
\phantom{\zeta_2(X) = } 
 - s_5 X s_5^* + s_6 X x_6^* + s_7 X s_7^* - s_8 X s_8^* \\[3pt]
\phantom{\zeta_2(X) = }
 - s_9 X s_9^* + s_{10} X x_{10}^* + s_{11} X s_{11}^* - s_{12} X s_{12}^*    
    \\[3pt]
\phantom{\zeta_2(X) = } 
+s_{13}Xs_{13}^* - s_{14} X x_{14}^* - s_{15} X s_{15}^* + s_{16} X s_{16}^*,  
\quad X \in \calO_{16}, \\[3pt]
\varphi_3(X) \equiv \rho_{16}(X) 
     = \sum\limits_{i=1}^{16} s_i X s_i^*, \quad X \in \calO_8.
\end{cases} \label{srfs_4}
\end{alignat}
\par
As for the standard RFS$_1$, it is easy to write down 
$s_{i_1,\ldots, i_k;\,j_k,\ldots, j_1} \in \calO_2^{U(1)}$, $k\geqq1$,
in terms of $\varPhi_{SR_1}(a_n) \ (n\leqq k)$ explicitly as follows:
\begin{alignat}{1}
& s_{i_1,\ldots,i_k;\,j_k,\ldots,j_1}
  = (-1)^{\sum\limits_{m=1}^{k-1}(j_m-1)N_m}A_1 A_2 \cdots A_k, \\ 
&A_m \equiv
 \begin{cases}
   \varPhi_{SR_1}(a_m)\varPhi_{SR_1}(a_m)^* & \mbox{for $(i_m,\,j_m)=(1,\,1)$,}\\[5pt]
   \varPhi_{SR_1}(a_m)                      & \mbox{for $(i_m,\,j_m)=(1,\,2)$,}\\[5pt]
   \varPhi_{SR_1}(a_m)^*                    & \mbox{for $(i_m,\,j_m)=(2,\,1)$,}\\[5pt]
   \varPhi_{SR_1}(a_m)^*\varPhi_{SR_1}(a_m) & \mbox{for $(i_m,\,j_m)=(2,\,2)$,}
 \end{cases} \quad m=1,\,2,\,\ldots,\,k,\qquad \\
& \quad N_m 
   \equiv \sum_{\ell=m+1}^k(i_\ell+j_\ell-2)=
   \sharp\Big\{ i\in\{i_{m+1}, \ldots, i_k, j_{m+1},\ldots,j_k\}
                \; \Big| \; i=2\Big\}.
\end{alignat}
\par
Besides the above standard RFS $SR_p$ satisfying 
$\calA_{SR_p}=\calO_{2^p}^{U(1)}$, we can construct a RFS $R_p$ 
so as to obtain $\calA_{R_p}\not\subset\calO_{2^p}^{U(1)}$.
Let $\varphi$ be an arbitrary inhomogeneous endomorphism of $\calO_{2^p}$
in the form of \eqref{inhom-endo}, and define a RFS$_p$ by 
$R_p=(\bolda_1,\ldots,\bolda_p;\ \zeta_p,\,\varphi_p)$ which is obtained 
from \eqref{RFSp-1}--\eqref{RFSp-3} by replacing $s_i$ by $\varphi(s_i)$ 
$(i=1,\ldots,2^p)$. Then, the embedding $\varPhi_{R_p}$ of the CAR algebra 
into $\calO_{2^p}$ associated with $R_p$ is given by
\begin{alignat}{1}
&\varPhi_{R_p}\equiv\varphi\circ\varPhi_{SR_p}. \label{R_p-SR_p}
\end{alignat}
Since $\varphi$ does not commute with the $U(1)$ action $\gamma$ defined 
by \eqref{U(1)-action}, we have indeed
$\varPhi_{R_p}(\mbox{CAR})=(\varphi\circ\varPhi_{SR_p})(\mbox{CAR})
=\varphi(\calO_{2^p}^{U(1)})\not\subset\calO_{2^p}^{U(1)}$.
For later use, we give two such examples in the case of $p=1$ :
\begin{alignat}{1}
&\begin{cases}
\displaystyle
\varphi(s_1)\equiv s_{1,2}, \qquad 
\varphi(s_2)\equiv s_{2;\,1}+s_{1,1;\,2},&\\[5pt]
\displaystyle \quad
\bolda_1\equiv s_{1,2,1;\,2} + s_{1,2,2;\,1,1}, &\\[5pt]
\displaystyle \quad
\zeta_1(X) \equiv s_{1,2}Xs_{1,2}^* 
- (s_{2;1}+s_{1,1;\,2}) X (s_{1;2}+s_{2;\,1,1}),&
\end{cases}
\label{U(1)-variant-RFS-1}
\\[5pt]
&\begin{cases}
\displaystyle
\varphi(s_1)\equiv s_{1;\,1} + s_{2,1;\,2}, \qquad \quad
\varphi(s_2)\equiv s_{2,2},&\\[5pt]
\displaystyle \quad
\bolda_1 \equiv  s_{1;\,1,2,2} + s_{2,1;\,2,2,2}, &\\[5pt]
\displaystyle \quad
\zeta_1(X) \equiv 
   (s_{1;1} + s_{2,1;\,2}) X (s_{1;1} + s_{2;\,1,2}) - s_{2,2} X s_{2,2}^*.&
\end{cases}
\label{U(1)-variant-RFS-2}
\end{alignat}
Here, the $\ast$-endomorphism $\varphi$ in \eqref{U(1)-variant-RFS-1} is 
the same as \eqref{pi_{1,2}-endo}.
\vskip20pt
\subsection{Reduction of the standard RFS$\boldsymbol{_p}$ $\boldsymbol{(p\geqq2)}$ 
to the standard RFS$\boldsymbol{_1}$}
Using the homogeneous embedding $\varPsi_p$ of $\calO_{2^p}$ $(p\geqq2)$ 
into $\calO_2$ defined by \eqref{hom-embed} with $d=2$, it is shown that 
the standard RFS$_p$ reduces to the standard RFS$_1$.
In this subsection, we denote the seeds of the standard RFS$_p$ $(p\geqq1)$
in $\calO_{2^p}$ by $\bolda^{(p)}_j$ $(j=1,\ldots,p)$. 
\par
First, we show that the following equality is satisfied:
\begin{equation}
\varPsi_p(\bolda^{(p)}_j) = \zeta_{1}^{j-1}(\bolda^{(1)}_1), \qquad
 j=1,\,2,\,\ldots,\,p.
\end{equation}
Indeed, from \eqref{RFSp-1} and \eqref{hom-embed-rec2} with $d=2$, we have
\begin{alignat}{1}
\varPsi_p(\bolda^{(p)}_1)
&=\sum_{k=1}^{2^{p-1}}S^{(p)}_{2k-1}S^{(p)\,\ast}_{2k}
 = \sum_{k=1}^{2^{p-1}}s_1 S^{(p-1)}_k S^{(p-1)\,\ast}_k s_2^* 
 = s_{1;2} = \bolda^{(1)}_1, \\
\hspace*{-20pt}
\varPsi_p(\bolda^{(p)}_j)
&=\sum_{k=1}^{2^{p-j}}\sum_{\ell=1}^{2^{j-1}}
   (-1)^{\sum\limits_{m=1}^{j-1}\left[\frac{\ell-1}{2^{m-1}}\right]}
      S^{(p)}_{2^j(k-1)+\ell}S^{(p)\,\ast}_{2^{j-1}(2k-1)+\ell}
   \nonumber \\
&=\sum_{k=1}^{2^{p-j}}\sum_{\ell'=1}^{2^{j-2}}
\Big[ (-1)^{\sum\limits_{m=1}^{j-1}\left[\frac{2\ell'-2}{2^{m-1}}\right]}
      S^{(p)}_{2^j(k-1)+2\ell'-1}S^{(p)\,\prime}_{2^{j-1}(2k-1)+2\ell'-1}
  \nonumber\\
&\hspace*{120pt}
    + (-1)^{\sum\limits_{m=1}^{j-1}\left[\frac{2\ell'-1}{2^{m-1}}\right]}
      S^{(p)}_{2^j(k-1)+2\ell'}S^{(p)\,\prime}_{2^{j-1}(2k-1)+2\ell'}\Big]
 \nonumber \\
&=\sum_{k=1}^{2^{p-j}}\sum_{\ell'=1}^{2^{j-2}}
(-1)^{\sum\limits_{m=1}^{j-2}\left[\frac{\ell'-1}{2^{m-1}}\right]}
\Big[
 s_1S^{(p-1)}_{2^{j-1}(k-1)+\ell'}S^{(p-1)\,\ast}_{2^{j-2}(2k-1)+\ell'}s_1^*
  \nonumber\\
&\hspace*{120pt}
-s_2S^{(p-1)}_{2^{j-1}(k-1)+\ell'}S^{(p-1)\,\ast}_{2^{j-2}(2k-1)+\ell'}s_2^*
\Big]\nonumber \\
&=\zeta_1(\varPsi_{p-1}(\bolda^{(p-1)}_{j-1}))
=\cdots=\zeta_1^{j-1}(\varPsi_{p-j+1}(\bolda^{(p-j+1)}_1)) \nonumber\\
&=
\zeta_1^{j-1}(\bolda^{(1)}_1), \qquad j=2,\,3,\ldots,\,p,\quad 
\end{alignat}
where \eqref{CR2} for $\calO_{2^{p-1}}$ is used, and $\varPsi_1$ should
be understood as the identity map on $\calO_2$.
Likewise, for the recursive map $\zeta_p$, we have
\begin{alignat}{1}
\varPsi_p(\zeta_p(X))
&= \sum_{i=1}^{2^p}
   (-1)^{\sum\limits_{m=1}^{p}\left[\frac{i-1}{2^{m-1}}\right]}
        S^{(p)}_i \varPsi_p(X) S^{(p)\,\ast}_i \nonumber\\
&= \sum_{i'=1}^{2^{p-1}}
\Big[(-1)^{\sum\limits_{m=1}^{p}\left[\frac{2i'-2}{2^{m-1}}\right]}
        S^{(p)}_{2i'-1}\varPsi_p(X) S^{(p)\,\ast}_{2i'-1} 
    +(-1)^{\sum\limits_{m=1}^{p}\left[\frac{2i'-1}{2^{m-1}}\right]}
        S^{(p)}_{2i'}\varPsi_p(X) S^{(p)\,\ast}_{2i'}\Big] \nonumber\\
&= \sum_{i'=1}^{2^{p-1}}
(-1)^{\sum\limits_{m=1}^{p-1}\left[\frac{i'-1}{2^{m-1}}\right]}
\Big[   s_1S^{(p-1)}_{i'}\varPsi_p(X) S^{(p-1)\,\ast}_{i'}s_1^* 
 - s_2S^{(p-1)}_{i'}\varPsi_p(X) S^{(p-1)\,\ast}_{i'}s_2^* \Big] \nonumber\\
&= \zeta_1\Big(\sum_{i=1}^{2^{p-1}}
   (-1)^{\sum\limits_{m=1}^{p-1}\left[\frac{i-1}{2^{m-1}}\right]}
        S^{(p-1)}_i \varPsi_p(X) S^{(p-1)\,\ast}_i\Big) \nonumber\\
&=\cdots = \zeta_1^p(\varPsi_p(X)), \qquad X \in \calO_{2^p}.
\end{alignat}
Therefore, we obtain
\begin{alignat}{1}
(\varPsi_p\circ\varPhi_{SR_p})(a_{p(m-1)+j})
&= \varPsi_p(\zeta_p^{m-1}(\bolda^{(p)}_j)) \nonumber\\
&= \zeta_1^{p(m-1)+j-1}(\bolda^{(1)}_1) \nonumber\\
&= \varPhi_{SR_1}(a_{p(m-1)+j}),\quad j=1,2,\ldots,p\,;\quad m\in\boldN,
\end{alignat}
hence
\begin{equation}
(\varPsi_p\circ\varPhi_{SR_p})(a_n)=\varPhi_{SR_1}(a_n), \qquad
 n\in\boldN. \label{varPhi_SR-reduction}
\end{equation}
\par
From the above calculations, it is straightforward to generalize
\eqref{varPhi_SR-reduction} to the following form:
\begin{equation}
\varPsi_{r,p}\circ\varPhi_{SR_p} = \varPhi_{SR_r}, \label{varPhi_SR-reduction-2}
\end{equation}
where $\varPsi_{r,p}$ denotes the homogeneous embedding of 
$\calO_{2^p}$ into $\calO_{2^r}$ with $r$ being an arbitrary divisor 
of $p$.
\vskip40pt
\section{Restriction of Permutation Representations to CAR Subalgebra}
In the previous sections, we have discussed on some properties of embeddings,
$\ast$-endomorphisms, the permutation representations and branchings in 
the Cuntz algebra, and introduced the construction of the recursive 
fermion system.
Hereafter, we discuss on properties of the CAR algebra by restricting
those of the Cuntz algebra through the recursive fermion system. 
\subsection{Fock(-like) representation}
As shown in Ref.\,\citen{AK1}), the restriction of the standard representation
$\pi^{(p)}_s$ of $\calO_{2^p}$ to $\calA_{SR_p}$ for an arbitrary $p$ gives 
the Fock representation, which is denoted by Rep$[1]$, as follows:
\begin{alignat}{1}
&\pi^{(p)}_s(a_n)\,e_1 = 0, \quad n\in\boldN, \label{RFSvac}\\
&
\pi^{(p)}_s(a_{n_1}^*a_{n_2}^* \cdots a_{n_k}^*) \, e_1 
=e_{N(n_1,\ldots,n_k)}, \quad 1\leqq n_1 < n_2 <\cdots<n_k,\label{RFSfock}\\
&
\qquad
N(n_1,\ldots,n_k) \equiv 1+2^{n_1-1}+\cdots+2^{n_k-1}, \label{binary}
\end{alignat}
where we make an identification of $\varPhi_{SR_p}(a_n)$ with $a_n$
for simplicity of description.
Since it is obvious that any $n\in\boldN$ is expressible in the form of
$N(n_1,\ldots,n_k)-1$, $e_n$ $(n\in\boldN)$ is uniquely given 
in the form of the lhs of \eqref{RFSfock}, that is, 
$e_1$ is the unique vacuum and a cyclic vector of the representation. 
We can, now, see the fact that the above Fock representation is strictly 
common to all $p$ is nothing but a direct consequence of 
\eqref{pi_s-reduction} and \eqref{varPhi_SR-reduction}:
\begin{alignat}{1}
\pi_s^{(p)}\circ\varPhi_{SR_p}
&=(\pi_s\circ\varPsi_p)\circ\varPhi_{SR_p} \nonumber\\
&=\pi_s\circ(\varPsi_p\circ\varPhi_{SR_p})
=\pi_s\circ\varPhi_{SR_1},
\end{alignat}
where $\pi_s$ is the standard representation of $\calO_{2}$.
\par
As a straightforward generalization of the above, we consider the 
restriction of Rep$(i_0;\,z)$ $(i_0=1,\ldots,2^p)$ of $\calO_{2^p}$ to 
$\calA_{SR_p}$. From \eqref{RFSp-embed}--\eqref{RFSp-2} and 
\eqref{perm-rep-cycle-1} with $e_{0,\, m}\equiv e_m$, 
we have
\begin{alignat}{1}
\pi^{(p)}_{i_0}(a_{p(m-1)+j})\,e_1 &=
\delta_{i_{0,j},\,2}\,
(-1)^{N_{i_0,j,m}}\, z^{-m}\, \pi^{(p)}_{i_0}((s_{i_0})^{m-1}s_{i_0-2^{j-1}})\, e_1, 
\\
\pi^{(p)}_{i_0}(a^*_{p(m-1)+j})\,e_1 &=
\delta_{i_{0,j},\,1}\,
(-1)^{N'_{i_0,j,m}}\, z^{-m}\, 
\pi^{(p)}_{i_0}((s_{i_0})^{m-1}s_{i_0+2^{j-1}})\, e_1. 
\end{alignat}
where $m\in\boldN$, $j=1,2,\ldots,p$, and $i_{0,j}$ is obtained from 
$i_0\equiv\sum\limits_{j=1}^p(i_{0,j} - 1)2^{j-1} + 1$ 
with $i_{0,1},\ldots,i_{0,p}=1,2$; $N_{i_0,j,m}$ and $N'_{i_0,j,m}$ 
are certain integers determined by $i_0$, $j$, $m$.
Therefore, using a Bogoliubov transformation 
$\phi_{i_0}(=\phi_{i_0}^{-1})$ defined by
\begin{eqnarray}
\phi_{i_0}(a_{p(m-1)+j})\equiv
\begin{cases}
a_{p(m-1)+j} &\mbox{for $i_{0,j}=1$,}\\[5pt]
a^*_{p(m-1)+j} &\mbox{for $i_{0,j}=2$,}
\end{cases}\quad m\in\boldN, \ \ j=1,\ldots,p, \label{pi_j-Bogoliubov}
\end{eqnarray}
we obtain
\begin{equation}
\pi^{(p)}_{i_0}(a^{(i_0)}_n)\,e_1 = 0, \quad
a^{(i_0)}_n \equiv \phi_{i_0}(a_n), \qquad n\in\boldN.
\end{equation}
It is shown in the same way as \eqref{RFSfock} that any of
$\{e_n\}_{n=1}^\infty$ is given by an action of 
$\pi^{(p)}_{i_0}(a^{(i_0)\,\ast}_{n_1}\cdots a^{(i_0)\,\ast}_{n_k})$ 
$(n_1<\cdots<n_k)$ on $e_1$.
We call this (irreducible) Fock-like representation of the CAR algebra 
the {\it $\phi_{i_0}$-Fock representation}, and denote it by 
Rep$^{(p)}[i_0]$. We have the following relations:
\begin{alignat}{1} 
\mbox{Rep}^{(p)}[i_0] 
&\equiv \mbox{Rep}^{(p)}(i_0;\,z)\circ \varPhi_{SR_p} \nonumber\\ 
&= \mbox{Rep}^{(p)}(1)\circ\varPhi_{SR_p}\circ\phi_{i_0}  \nonumber\\
&=\mbox{Rep}[1] \circ \phi_{i_0}   \nonumber\\
&=\mbox{Fock} \circ \phi_{i_0},
\quad  i_0=1,2,\ldots,2^p. \label{Fock-like rep}
\end{alignat}
\vskip20pt
\subsection{Restriction of permutation representation with central cycle to CAR}
Now, we consider a generic irreducible permutation representation
with a central cycle and with an eigenvalue $z$ $(|z|=1)$, 
Rep$(L;\,z)$ of $\calO_2$, where $L=(i_0,i_1,\ldots,i_{\kappa-1})$ denotes 
the label of the representation. 
First, let us recall the $\kappa$ eigenvectors 
$e_{\lambda,\,1}$ $(\lambda=0,1,\ldots, \kappa-1)$
in Rep$(L;\,z)$ given by \eqref{perm-rep-eigenvector}.  
Then, for $n=\kappa(m-1)+\ell$ with $m\in\boldN$ and $\ell=1,2,\ldots,\kappa$, 
we have 
\begin{alignat}{1}
&\hspace*{-10pt}\pi_L(a_n)\,e_{\lambda,\,1} \nonumber\\
&=\pi_L\big(\zeta_1^{\kappa(m-1)+\ell-1}(s_{1;2})\big)\,e_{\lambda,\,1}\nonumber\\
&= \bar z^{m} \pi_L\big(\zeta_1^{\kappa(m-1)+\ell-1}(s_{1;2})\big)\,
    \pi_L\big((s_{i_{\lambda},\ldots,i_{\kappa-1}, i_0,\ldots, i_{\lambda-1}})^{m}\big)
    \,e_{\lambda,\,1}     \nonumber\\
&= (-1)^{(m-1)N_{\lambda,\kappa}+N_{\lambda,\ell-1}}\bar z^{m}
   \pi_L\big((s_{i_{\lambda},\ldots,i_{\lambda-1}})^{m-1}
    s_{i_{\lambda},\ldots,i_{\lambda+\ell-2}, 1;\,2} 
    s_{i_{\lambda+\ell-1}, i_{\lambda+\ell},\ldots,i_{\lambda-1}}\big)\,
    e_{\lambda,\,1}    \nonumber\\
&=
\delta_{i_{\lambda+\ell-1},\,2}\,
    (-1)^{N_{\lambda,n-1}}\bar z^{m}
       \pi_L\big((s_{i_{\lambda},\ldots,i_{\lambda-1}})^{m-1}
              s_{i_{\lambda},\ldots,i_{\lambda+\ell-2},1, 
                 i_{\lambda+\ell},\ldots,i_{\lambda-1}}\big)\,e_{\lambda,\,1},
\\[5pt]
&\hspace*{-10pt}\pi_L(a^*_n)\,e_{\lambda,\,1} \nonumber\\
&=\pi_L\big(\zeta_1^{\kappa(m-1)+\ell-1}(s_{2;1})\big)\,e_{\lambda,\,1}\nonumber\\
&= \bar z^{m} \pi_L\big(\zeta_1^{(m-1)N+\ell-1}(s_{2;1})\big)\,
    \pi_L\big((s_{i_{\lambda},\ldots,i_{\kappa-1}, i_0,\ldots,i_{\lambda-1}})^{m}\big)
     \,e_{\lambda,\,1}      \nonumber\\
&= (-1)^{(m-1)N_{\lambda,\kappa}+N_{\lambda,\ell-1}} \bar z^{m}
    \pi_L\big((s_{i_{\lambda},\ldots,i_{\lambda-1}})^{m-1}
           s_{i_{\lambda},\ldots,i_{\lambda+\ell-2}, 2;\,1} 
           s_{i_{\lambda+\ell-1}, i_{\lambda+\ell},\ldots,i_{\lambda-1}}\big)
           \,e_{\lambda,\,1} \nonumber\\
&=\delta_{i_{\lambda+\ell-1},\,1}\,
(-1)^{N_{\lambda,n-1}}\bar z^{m}
  \pi_L\big((s_{i_{\lambda},\ldots,i_{\lambda-1}})^{m-1}
         s_{i_{\lambda},\ldots,i_{\lambda+\ell-2},2,
            i_{\lambda+\ell},\ldots,i_{\lambda-1}}\big)\,e_{\lambda,\,1}.
\end{alignat}
where $N_{\lambda,j}\equiv\sum\limits_{r=0}^{j-1}(i_{\lambda+r}-1)$ 
$(N_{\lambda,0}\equiv0)$ 
is the number of $2$ in $\{i_{\lambda},\,\ldots,\,i_{\lambda+j-1}\}$. 
One should note that the subscripts of indices $i_k$'s take values in
$\boldZ_\kappa$. Therefore, using a Bogoliubov transformation 
$\phi_{L,\lambda}$ defined by 
\begin{equation}
\phi_{L,\lambda}(a_{\kappa(m-1)+\ell}) \equiv
\begin{cases}
(-1)^{N_{\lambda,n-1}} a_{\kappa(m-1)+\ell} 
   &\mbox{for $i_{\lambda + \ell - 1}=1$}, \\[5pt]
(-1)^{N_{\lambda,n-1}} a^*_{\kappa(m-1)+\ell} 
   &\mbox{for $i_{\lambda + \ell - 1}=2$},
\end{cases} \quad m\in\boldN,\ \ \ell=1,\ldots,\kappa,
\label{pi_L-Bogoliubov}
\end{equation}
we obtain
\begin{equation}
\pi_L(a^{(\lambda)}_n)\, e_{\lambda,\,1} = 0, \quad 
a^{(\lambda)}_n \equiv \phi_{L,\lambda}(a_n), \qquad n\in\boldN,
\end{equation}
hence $e_{\lambda,\,1}$ $(\lambda=0,1,\ldots, \kappa-1)$ is 
a vacuum for the annihilation operators 
$\{a^{(\lambda)}_n\,\mid\,n\in\boldN\}$, and the corresponding 
Fock space $\calH^{[\lambda]}$ is generated by 
$\pi_L(a^{(\lambda)\,\ast}_{n_1}a^{(\lambda)\,\ast}_{n_2}\cdots
a^{(\lambda)\,\ast}_{n_r})\,e_{\lambda,\,1}$ with $n_1<n_2<\cdots<n_r$, 
$r\in\boldN$.
In the special case $n_i=\kappa(m-1)+\ell_i$ 
$(m\in\boldN,\ 1\leqq\ell_1<\cdots<\ell_r\leqq \kappa,\ 1\leqq r\leqq k)$, 
we have
\begin{alignat}{1}
\begin{array}{c}
\displaystyle
\pi_L\big(a^{(\lambda)\,\ast}_{n_1}a^{(\lambda)\,\ast}_{n_2}\,\cdots\,
 a^{(\lambda)\,\ast}_{n_r}\big)
 \,e_{\lambda,\,1} =
\bar z^{m}\,
 \pi_L\big((s_{i_{\lambda+1},\ldots,i_\lambda})^{m-1} s_J\big)\,
    e_{\lambda,\,1},\\[5pt]
\displaystyle
  s_J\equiv s_{i_{\lambda},\ldots,
               i_{\lambda+\ell_1-2},\hati_{\lambda+\ell_1-1},i_{\lambda+\ell_1},\ldots,
			   i_{\lambda+\ell_2-2},\hati_{\lambda+\ell_2-1},i_{\lambda+\ell_2},\ldots,
               i_{\lambda+\ell_r-2},\hati_{\lambda+\ell_r-1},i_{\lambda+\ell_r},\ldots,
	           i_{\lambda-1}} 
\end{array}  \label{multi-index1}
\end{alignat}
with $\hati_\ell \equiv 3 - i_\ell$. 
Here, one should note that
$s_J$ in \eqref{multi-index1} takes any $\kappa$-th order monomial 
of $s_i$ $(i=1,2)$ other than $s_{i_{\lambda},\ldots, i_{\lambda-1}}$.
On the other hand, in the case $n_i=\kappa(m_i -1) + \ell_i$ 
$(m_1<\cdots<m_r,\,1\leqq \ell_i\leqq \kappa)$, we have
\begin{alignat}{1}
&\hspace*{-10pt}
\begin{array}{c}
\displaystyle
\pi_L\big(a^{(\lambda)\,\ast}_{n_1}a^{(\lambda)\,\ast}_{n_2}\cdots 
    a^{(\lambda)\,\ast}_{n_r}\big) \,e_{\lambda,\,1} 
=\bar z^{m_r}\,
 \pi_L\big((s_{i_{\lambda},\ldots,i_{\lambda-1}})^{m_1-1} 
        s_{J_1}
       (s_{i_{\lambda},\ldots,i_{\lambda-1}})^{m_2-m_1-1} 
        s_{J_2}
       \cdots\\[5pt]
\displaystyle
\hspace*{160pt}     
       \times
       \cdots 
       (s_{i_{\lambda},\ldots,i_{\lambda-1}})^{m_r-m_{r-1}-1} 
        s_{J_{r}}
        \big)\, e_{\lambda,\,1}, \\[5pt]
\displaystyle
\hspace*{20pt}
s_{J_k}\equiv s_{i_{\lambda},\ldots,
                 i_{\lambda+\ell_k-2},\hati_{\lambda+\ell_k-1},i_{\lambda+\ell_k},
		         \ldots,i_{\lambda-1}},
 \qquad k=1,\ldots,r.
\end{array}
\end{alignat}
Taking \eqref{perm-rep-eigenvector} into account, it is, now, easy 
to infer that 
any $n\kappa$-th $(n=0,\,1,\,\ldots)$ order monomial of $s_i$ $(i=1,\,2)$ 
acting on $e_{\lambda,\,1}$ is uniquely given by 
$\pi_L\big(a^{(\lambda)\,\ast}_{n_1}a^{(\lambda)\,\ast}_{n_2}\cdots 
a^{(\lambda)\,\ast}_{n_r}\big) \,e_{\lambda,\,1}$ up to a $U(1)$ 
factor with a suitable set of $\{n_1<n_2<\cdots<n_r\}$. 
Therefore, we have\footnote{It should be understood that the 
completion of the Hilbert space is carried out.}
\begin{alignat}{1}
\calH^{[\lambda]}
&=\mbox{Lin}\langle\,\{\ e_{\lambda,\,1},\ 
 \pi_L\big(a^{(\lambda)\,\ast}_{n_1}\,\cdots\,a^{(\lambda)\,\ast}_{n_r}\big)
    \,e_{\lambda,\,1},\ 
  1\leqq n_1<\cdots<n_r;\,r\geqq1\ \}\,\rangle \nonumber\\ 
&=\mbox{Lin}\langle\,\{\ \pi_L(s_{j_1,\ldots,j_{n\kappa}})\,e_{\lambda,\,1} \ \mid \  
  j_i=1,\,2;\ i=1,\ldots,n\kappa;\ n\geqq0\ \}\,\rangle \nonumber\\
&=\mbox{Lin}\langle\,\{\ e_{\lambda,\,m}\, \mid \, m\in\boldN\ \}\,\rangle.
  \label{Fockspace}
\end{alignat}
It is obvious that a direct sum of $\calH^{[\lambda]}$ 
$(\lambda=0,1,\ldots,\kappa-1)$ gives the total Hilbert space $\calH$:
\begin{alignat}{1}
&
\bigoplus_{\lambda=0}^{\kappa-1} \calH^{[\lambda]} 
=\mbox{Lin}\langle\,\{\ e_{\lambda,\,m}\ \mid \ 
   \lambda\in\boldZ_\kappa,\,m\in\boldN\ \}\,\rangle 
=\calH.
\end{alignat}
Therefore, the restriction of the irreducible permutation representation 
Rep$(L;\,z)$ of $\calO_2$ to $\calA_{SR_1}$ gives 
a direct sum of $\kappa$ $\phi_{L,\lambda}$-Fock representations as follows:
\begin{alignat}{1}
&\mbox{Rep}(L;\,z)\,\Big|_{\calA_{SR_1}}  
\cong \bigoplus_{\lambda=0}^{\kappa-1}
 \Big(\mbox{Fock} \circ \phi_{L,\lambda}\Big).
\label{rep-decompose}
\end{alignat}
This result is nothing but an explicit realization of the general theory 
for restriction of the permutation representation with a central cycle of 
$\calO_d$ to $\calO_d^{U(1)}$ discussed in Ref.\,\citen{BJ}).
\par
It should be noted that it is also possible to derive the above formula 
\eqref{rep-decompose} by using the homogeneous embedding $\varPsi_\kappa$ 
of $\calO_{2^\kappa}$ into $\calO_2$. From \eqref{hom-embed}, we have
\begin{equation}
\varPsi_\kappa^{-1}(s_{i_\lambda,\ldots,i_{\kappa-1}, i_0,\ldots,i_{\lambda-1}})
 \!=\!s'_{i(\lambda)},
\quad i(\lambda)\!\equiv\! 
    \sum_{\ell=1}^\kappa (i_{\lambda+\ell-1}-1)2^{\ell-1} +1,
\quad \lambda\!=\!0,1,\ldots,\kappa-1, 
\end{equation}
where the generators of $\calO_{2^\kappa}$ are denoted by 
$\{s'_i \mid i=1,2,\ldots,2^\kappa\}$.
Hence we can rewrite \eqref{perm-rep-eigenvector} as
\begin{eqnarray}
&&(\pi_L\circ\varPsi_\kappa)(s'_{i(\lambda)})\,e_{\lambda,\,1} 
  = z\,e_{\lambda,\,1}, 
\quad \lambda=0,1,\ldots,\kappa-1,
\end{eqnarray}
which shows that $\pi_L\circ\varPsi_\kappa$ is a reducible permutation
representation of $\calO_{2^\kappa}$ consisting of a direct sum of 
$\kappa$ irreducible ones, i.e., Rep$^{(\kappa)}(i(\lambda);\,z)$ 
$(\lambda=0,1,\ldots,\kappa-1)$. 
Therefore, we obtain
\begin{alignat}{1}
\mbox{Rep}(L;\,z)\,\Big|_{\calA_{SR_1}}  
&=\pi_L\circ\varPhi_{SR_1}
=\pi_L\circ\varPsi_\kappa\circ\varPhi_{SR_\kappa}\nonumber\\
&\cong\bigoplus_{\lambda=0}^{\kappa-1}  
  \mbox{Rep}^{(\kappa)}(i(\lambda);\,z)\circ\varPhi_{SR_\kappa}
\cong\bigoplus_{\lambda=0}^{\kappa-1} \Big(\mbox{Fock} \circ \phi_{i(\lambda)}\Big)
\label{perm-rep-restrict-embed}
\end{alignat}
where use has been made of \eqref{Fock-like rep} with $p=\kappa$.
Here, from \eqref{pi_j-Bogoliubov} with $p=\kappa$ and \eqref{pi_L-Bogoliubov}, 
$\phi_{i(\lambda)}(a_n)$ for each $n\in\boldN$ is identical with 
$\phi_{L,\lambda}(a_n)$ up to sign. 
Hence the rhs of \eqref{perm-rep-restrict-embed} is unitarily equivalent 
with the rhs of \eqref{rep-decompose}.
%
%
\vskip20pt
\subsection{Restriction of permutation representation with chain to CAR}
In the same way as above, it is straightforward to obtain the restriction of 
the permutation representation with a chain of $\calO_2$, Rep$(L_\infty)$ 
with $L_\infty=\{i_k\}_{k=1}^\infty$, to $\calA_{SR_1}$. 
By direct calculations using \eqref{perm-rep-chain-1}--\eqref{perm-rep-chain-4},
we have
\begin{alignat}{1}
&\pi_{L_\infty}(a_n)\,e_{\lambda,\,1}
 =\delta_{i_{\lambda+n-1},\,2}\,
    (-1)^{N_{\lambda,n}}s_{i_\lambda,\ldots,i_{\lambda+n-2},1}\,e_{\lambda+n,\,1},
\\
&\pi_{L_\infty}(a_n^*)\,e_{\lambda,\,1}
 =\delta_{i_{\lambda+n-1},\,1}\,
    (-1)^{N_{\lambda,n}}s_{i_\lambda,\ldots,i_{\lambda+n-2},2}\,e_{\lambda+n,\,1},
\end{alignat}
where $N_{\lambda,n}\equiv \sum\limits_{j=0}^{n-2}(i_{\lambda+j}-1)$ 
$(n\geqq2,\ N_{\lambda,1}\equiv0)$ is the number of 2
in $\{i_{\lambda},\,i_{\lambda+1},\,\ldots,\,i_{\lambda+n-2}\}$, and we set 
$i_k\equiv 1$ for $k<0$.  
By using a Bogoliubov transformation
$\phi_{L_\infty,\lambda}$ defined by
\begin{equation}
\phi_{L_\infty,\lambda}(a_n)\equiv
\begin{cases}
 (-1)^{N_{\lambda,n}} a_n & \mbox{for $i_{\lambda + n - 1}=1$}, \\[5pt]
 (-1)^{N_{\lambda,n}} a^*_n & \mbox{for $i_{\lambda + n - 1}=2$},
\end{cases}
\end{equation}
we obtain
\begin{equation}
\pi_{L_\infty}(a^{(\lambda)}_n)\, e_{\lambda,\,1} = 0, \quad
a^{(\lambda)}_n\equiv\phi_{L_\infty,\lambda}(a_n),\quad n\in\boldN,
\end{equation}
hence $e_{\lambda,\,1}$ $(\lambda\in\boldZ)$ is a vacuum for the 
annihilation operators $\{a^{(\lambda)}_n\,\mid\,n\in\boldN \}$, 
and the corresponding Fock space $\calH^{[\lambda]}$ is generated by 
$\pi_{L_\infty}\big(a^{(\lambda)\,\ast}_{n_1}a^{(\lambda)\,\ast}_{n_2}\cdots
a^{(\lambda)\,\ast}_{n_r}\big)\,e_{\lambda,\,1}$ with $n_1<n_2<\cdots<n_r$, 
$r\in\boldN$. 
From
\begin{alignat}{1}
&
\begin{array}{c}
\displaystyle
\pi_{L_\infty}\big(a^{(\lambda)\,\ast}_{n_1}\,\cdots\,
 a^{(\lambda)\,\ast}_{n_r}\big)\,e_{\lambda,\,1}
= \pi_{L_\infty}(s_J) \,e_{\lambda+n_r,\,1},\\
\displaystyle
  s_J\!\equiv\!s_{i_{\lambda},\ldots,
     i_{\lambda+n_1-2},\hati_{\lambda+n_1-1},i_{\lambda+n_1},\ldots,
	 i_{\lambda+n_2-2},\hati_{\lambda+n_2-1},i_{\lambda+n_2},\ldots,
     i_{\lambda+n_r-2},\hati_{\lambda+n_r-1}} 
\end{array}
\end{alignat}
with $\hati_j\equiv 3 - i_j$, and noting that
$s_J$ takes any $n_r$-th monomial except for 
$s_{i_\lambda,\ldots,i_{\lambda+n_r-1}}$, we obtain
\begin{alignat}{1}
\calH^{[\lambda]}
&=\mbox{Lin}\langle\,\{\ e_{\lambda,\,1},\ 
 \pi_L\big(a^{(\lambda)\,\ast}_{n_1}\,\cdots\,a^{(\lambda)\,\ast}_{n_r}\big)
      \,e_{\lambda,\,1},\ 
  1\leqq n_1<\cdots<n_r;\,r\geqq1\ \}\,\rangle \nonumber\\ 
&=\mbox{Lin}\langle\,\{\ e_{\lambda,\,m}\, \mid \, m\in\boldN\ \}\,\rangle.
  \label{Fockspace_infty}
\end{alignat}
Hence the total Hilbert space $\calH$ is a direct sum of an infinite number 
of the above Fock-like spaces:
\begin{equation}
\calH = \bigoplus_{\lambda\in\boldZ} \calH^{[\lambda]}. \label{infinite-decomp}
\end{equation}
Thus, the restriction of the permutation representation with a chain of 
$\calO_2$ to $\calA_{SR_1}$ gives a direct sum of an infinite number
of $\phi_{L_\infty,\lambda}$-Fock representations.
\vskip20pt
\subsection{$\boldsymbol{U(1)}$-variant RFS}
So far, we have studied the restriction of the permutation representations 
of $\calO_2$ (or $\calO_{2^p}$ with $p\geqq2$) to 
$\calA_{SR_1}=\calO_2^{U(1)}$ (or $\calA_{SR_p}=\calO_{2^p}^{U(1)}$), 
and found that the resultant representations are reducible in general 
except for the case of the permutation representation with a central 
cycle of length 1. 
However, for a RFS$_1$ $R_1$ with $\calA_{R_1}\not\subset\calO_2^{U(1)}$, 
the situation changes drastically. In the following, we briefly describe 
this feature.
\par
In $\calO_2$, for any irreducible permutation representation with 
a central cycle, $\pi_L$, there exists a RFS ${R_1}$ such that 
$\pi_L\circ\varPhi_{R_1}$ is an irreducible representation of the CAR 
algebra.
This fact is nothing but the result of the existence of the 
$\ast$-endomorphism $\varphi$ satisfying \eqref{branch-to-pi_s}, which 
is explicitly given by \eqref{pi_L-endo-1}--\eqref{pi_L-endo-3}. 
Indeed, if we define $R_1$ by \eqref{R_p-SR_p}, then we have
\begin{alignat}{1}
\pi_L\circ\varPhi_{R_1}
&=\pi_L\circ(\varphi\circ\varPhi_{SR_1})\nonumber\\
&=(\pi_L\circ\varphi)\circ\varPhi_{SR_1}\nonumber\\
&\cong\pi_s\circ\varPhi_{SR_1}\nonumber\\
&=\mbox{Fock}.
\end{alignat}
An example for the case $L=(1,2)$ is given by \eqref{U(1)-variant-RFS-1}.
In this case, the eigenvector $e_{0,1}$ of $\pi_{1,2}(s_{1,2})$ satisfies
\begin{alignat}{1}
& (\pi_{1,2}\circ\varPhi_{R_1})(a_n)\,e_{0,1} = 0, \quad n\in\boldN,
\end{alignat}
and any vector in $\{e_{0,m},\ e_{1,n}\}$ $(m\geqq2,\ n\geqq1)$ 
is uniquely given by
\begin{alignat}{1}
& (\pi_{1,2}\circ\varPhi_{R_1})(a^*_{n_1}\cdots a^*_{n_k})\,e_{0,1}
\end{alignat}
with an appropriate set of positive integers $n_1<\cdots<n_k$, $k\geqq1$.
\par
On the other hand, there also exists a RFS$_1$ $R_1$ with 
$\calA_{R_1}\not\subset\calO_1^{U(1)}$ such that $\pi_s\circ\varPhi_{R_1}$ 
(more generally, $\pi\circ\varPhi_{R_1}$ with an arbitrary irreducible
representation $\pi$) gives an infinitely decomposable representation 
just like $\pi_{L_\infty}\circ\varPhi_{SR_1}$.
An example is given by \eqref{U(1)-variant-RFS-2}.
Since the term involving $s_{1;1}$ at the right of $X$ in $\zeta_1(X)$ 
defined by \eqref{U(1)-variant-RFS-2} vanishes if $X$ includes $s_2^*$ 
at its right end, it is obvious that any of $\varPhi_{R_1}(a_n)$ 
$(n\in\boldN)$ involves $s_2^*$ at its right end. 
Therefore, from $\pi_s(s_1)\,e_{m}=e_{2m-1}$ $(m\in\boldN)$, 
we obtain
\begin{equation}
(\pi_s\circ\varPhi_{R_1})(a_n) e_{2m-1} = 0, \quad n \in\boldN
\end{equation}
for each $m\in\boldN$, which means that there exist an infinite number 
of vacuums $\{e_{2m-1}\}$ $(m\in\boldN)$.
Hence the restriction of Rep(1) to the above RFS is a direct sum 
of an infinite number of Fock representations as follows:
\begin{alignat}{1}
\pi_s\circ\varPhi_{R_1}
&\cong (\pi_s\circ\varPhi_{SR_1})^{\oplus\infty}\nonumber\\
&=(\mbox{Fock})^{\oplus\infty}.
\end{alignat}
\vskip40pt
\section{Restriction of Permutation Endomorphisms of $\boldsymbol{\calO_2}$ 
to $\boldsymbol{\calA_{SR_1}}$}
We consider the restriction of the permutation endomorphism $\varphi_\sigma$ 
of $\calO_2$ defined by \eqref{perm-endo} and \eqref{ext-perm-endo}
to the CAR subalgebra $\calA_{SR_1}$ associated with the standard RFS$_1$
defined by \eqref{RFSp-1}--\eqref{RFSp-3}. 
Since $\varphi_\sigma$ commutes with the $U(1)$ action $\gamma$ defined by 
\eqref{U(1)-action}, we have 
\begin{equation}
\varphi_\sigma(\calO_2^{U(1)})\subset\calO_2^{U(1)}.
\end{equation}
Thus, the restriction of $\varphi_\sigma$ to 
$\calO_2^{U(1)}=\calA_{SR_1}$ yields a $\ast$-endomorphism  
\begin{alignat}{1}
&\begin{array}{c}
\displaystyle
 \tilde\varphi_\sigma : \mbox{CAR} \to \mbox{CAR}, \\[5pt]
\displaystyle
 \tilde\varphi_\sigma\equiv
    \varPhi_{SR_1}^{-1}\circ\varphi_\sigma\circ\varPhi_{SR_1}
\end{array}
\end{alignat}
of the CAR algebra.
In this section, we identify $\varPhi_{SR_1}(a_n)$ with $a_n$, hence
$\tphi_\sigma$ with $\varphi_\sigma$, for simplicity of description. 
First, we study all the second order permutation endomorphisms
and after that we consider some higher order ones.
\vskip20pt
\subsection{The second order permutation endomorphisms}
To specify each of the second order permutation endomorphisms of
$\calO_2$ defined by \eqref{perm-endo}, we denote it as follows:
\begin{alignat}{1}
\varphi_\sigma (s_i) 
&
\equiv \sum_{j=1}^2 S^{(2)}_{\sigma((j-1)2+i)}s^*_j
= \sum_{j=1}^2 s_{\sigma(i,j)}\,s^*_j, \quad i=1,\,2,
\quad \sigma\in\mathfrak{S}_{4},
\end{alignat}
where $S^{(2)}_1=s_{1,1}$, $S^{(2)}_1=s_{2,1}$, $S^{(2)}_3=s_{1,2}$, 
$S^{(2)}_4=s_{2,2}$.
Here, from the one-to-one correspondence between $S^{(2)}_i$ 
and $s_{i_1,i_2}$,
a natural action of $\sigma\in\mathfrak{S}_{4}$ on $(i_1,i_2)\in\{1,2\}^2$
is induced. For example, $\sigma=[1,3]$ and $\sigma=[1,2,4]$ denote 
the transposition of $(1,1)\leftrightarrow(1,2)$, 
and the cyclic permutation of $(1,1)\to(2,1)\to(2,2)\to(1,1)$, respectively.
\par
Let $\alpha$ be the $\ast$-automorphism of $\calO_2$ defined by $\alpha(s_1)=s_2$, 
$\alpha(s_2)=s_1$.
Then, all the second order permutation endomorphisms of $\calO_2$ are 
given by\cite{Kawamura2}
\begin{alignat}{3}
\varphi_{id} &\!=\! \mathit{id},                       &&&&\label{phi_id}\\
\varphi_{[1,2]}(s_1)&\!=\!s_{2,1;1}\!+\!s_{1,2;2}      &\quad 
\varphi_{[1,2]}(s_2)&\!=\!s_{1,1;1}\!+\!s_{2,2;2},     &&\label{phi_[13]}\\
\varphi_{[1,3]}(s_1) &\!=\! s_{1,2;1}\!+\!s_{1,1;2},   &\quad 
\varphi_{[1,3]}(s_2) &\!=\! s_2,                       &\quad  
\varphi_{[1,3]}&\!=\!\alpha\circ\varphi_{[2,4]}\circ\alpha, \label{phi_[12]}\\
\varphi_{[1,4]}(s_1)&\!=\!s_{2,2;1}\!+\!s_{1,2;2}      &\quad 
\varphi_{[1,4]}(s_2)&\!=\!s_{2,1;1}\!+\!s_{1,1;2},     &&\label{phi_[14]}\\
\varphi_{[2,3]}(s_1)&\!=\!s_{1,1;1}\!+\!s_{2,1;2},     &\quad
\varphi_{[2,3]}(s_2)&\!=\!s_{1,2;1}\!+\!s_{2,2;2},     &\quad 
\varphi_{[2,3]}&\!=\!\rho,                              \label{phi_[23]}\\
\varphi_{[2,4]}(s_1)&\!=\!s_1,                         &\quad
\varphi_{[2,4]}(s_2)&\!=\!s_{2,2;1}\!+\!s_{2,1;2},     &&\label{phi_[34]}\\
\varphi_{[3,4]}(s_1)&\!=\!s_{1,1;1}\!+\!s_{2,2;2}      &\quad 
\varphi_{[3,4]}(s_2)&\!=\!s_{2,1;1}\!+\!s_{1,2;2},     &\quad
\varphi_{[3,4]}&\!=\!\varphi_{[1,2]}\circ\alpha,        \label{phi_[24]}\\
\varphi_{[1,2][3,4]}(s_1)&\!=\!s_2,                    &\quad
\varphi_{[1,2][3,4]}(s_2)&\!=\!s_1,                    &\quad
\varphi_{[1,2][3,4]}&\!=\!\alpha,                       \label{phi_[13][24]}\\
\noalign{\vfill\eject}
\varphi_{[1,3][2,4]}(s_1)&\!=\!s_{1,2;1}\!+\!s_{1,1;2}, &\quad
\varphi_{[1,3][2,4]}(s_2)&\!=\!s_{2,2;1}\!+\!s_{2,1;2}, &\quad 
\varphi_{[1,3][2,4]}&\!=\!\varphi_{[1,4][2,3]}\!\circ\!\alpha,\label{phi_[12][34]}\\
\varphi_{[1,4][2,3]}(s_1)&\!=\!s_{2,2;1}\!+\!s_{2,1;2}, &\quad
\varphi_{[1,4][2,3]}(s_2)&\!=\!s_{1,2;1}\!+\!s_{1,1;2},&&\label{phi_[14][23]}\\
\varphi_{[1,2,3]}(s_1)&\!=\!s_{2,1;1}\!+\!s_{1,1;2},    &\quad
\varphi_{[1,2,3]}(s_2)&\!=\!s_{1,2;1}\!+\!s_{2,2;2},    &\quad
 &   \label{phi_[132]}\\
\varphi_{[1,2,4]}(s_1)&\!=\!s_{2,1;1}\!+\!s_{1,2;2},    &\quad
\varphi_{[1,2,4]}(s_2)&\!=\!s_{2,2;1}\!+\!s_{1,1;2},   &&\label{phi_[134]}\\
\varphi_{[1,3,2]}(s_1)&\!=\!s_{1,2;1}\!+\!s_{2,1;2},    &\quad
\varphi_{[1,3,2]}(s_2)&\!=\!s_{1,1;1}\!+\!s_{2,2;2},    &\quad 
\varphi_{[1,3,2]}&\!=\!\varphi_{[2,3,4]}\circ\alpha,\label{phi_[123]}\\
\varphi_{[1,3,4]}(s_1)&\!=\!s_{1,2;1}\!+\!s_{2,2;2},    &\quad
\varphi_{[1,3,4]}(s_2)&\!=\!s_{2,1;1}\!+\!s_{1,1;2},    &\quad
\varphi_{[1,3,4]}&=\varphi_{[1,2,3]}\!\circ\!\alpha,\label{phi_[124]}\\
\varphi_{[1,4,2]}(s_1)&\!=\!s_{2,2;1}\!+\!s_{1,2;2},    &\quad
\varphi_{[1,4,2]}(s_2)&\!=\!s_{1,1;1}\!+\!s_{2,1;2},    &\qquad
\varphi_{[1,4,2]}&\!=\!\varphi_{[2,4,3]}\circ\alpha,\label{phi_[143]}\\
\varphi_{[1,4,3]}(s_1)&\!=\!s_{2,2;1}\!+\!s_{1,1;2},    &\quad
\varphi_{[1,4,3]}(s_2)&\!=\!s_{2,1;1}\!+\!s_{1,2;2},    &\qquad
\varphi_{[1,4,3]}&\!=\!\varphi_{[1,2,4]}\circ\alpha,\label{phi_[142]}\\
\hspace*{-15pt}
\varphi_{[2,3,4]}(s_1)&\!=\!s_{1,1;1}\!+\!s_{2,2;2},    &\quad
\varphi_{[2,3,4]}(s_2)&\!=\!s_{1,2;1}\!+\!s_{2,1;2},    &&\label{phi_[243]}\\
\varphi_{[2,4,3]}(s_1)&\!=\!s_{1,1;1}\!+\!s_{2,1;2},    &\quad
\varphi_{[2,4,3]}(s_2)&\!=\!s_{2,2;1}\!+\!s_{1,2;2},    &\qquad
\varphi_{[2,4,3]}&\!=\!\alpha\!\circ\!\varphi_{[1,2,3]}\!\circ\!\alpha,
                \label{phi_[234]}\\
\varphi_{[1,2,3,4]}(s_1)&\!=\!s_{2},                    &\quad
\varphi_{[1,2,3,4]}(s_2)&\!=\!s_{1,2;1}\!+\!s_{1,1;2},  &\qquad
\varphi_{[1,2,3,4]}&=\varphi_{[1,3]}\circ\alpha,\label{phi_[1324]}\\
\varphi_{[1,2,4,3]}(s_1)&\!=\!s_{2,1;1}\!+\!s_{1,1;2},  &\quad
\varphi_{[1,2,4,3]}(s_2)&\!=\!s_{2,2;1}\!+\!s_{1,2;2},  &\qquad
\varphi_{[1,2,4,3]}&=\varphi_{[1,4]}\circ\alpha,\label{phi_[1342]}\\
\varphi_{[1,3,2,4]}(s_1)&\!=\!s_{1,2;1}\!+\!s_{2,1;2},  &\quad
\varphi_{[1,3,2,4]}(s_2)&\!=\!s_{2,2;1}\!+\!s_{1,1;2},  &&\label{phi_[1234]}\\
\varphi_{[1,3,4,2]}(s_1)&\!=\!s_{1,2;1}\!+\!s_{2,2;2},  &\quad
\varphi_{[1,3,4,2]}(s_2)&\!=\!s_{1,1;1}\!+\!s_{2,1;2},  &\qquad
\varphi_{[1,3,4,2]}&=\varphi_{[2,3]}\circ\alpha,\label{phi_[1243]}\\
\varphi_{[1,4,2,3]}(s_1)&\!=\!s_{2,2;1}\!+\!s_{1,1;2}, &\quad
\varphi_{[1,4,2,3]}(s_2)&\!=\!s_{1,2;1}\!+\!s_{2,1;2}, &\qquad
\varphi_{[1,4,2,3]}&\!=\!\varphi_{[1,3,2,4]}\circ\alpha,\label{phi_[1432]}\\
\varphi_{[1,4,3,2]}(s_1)&\!=\!s_{2,2;1}\!+\!s_{2,1;2}, &\quad
\varphi_{[1,4,3,2]}(s_2)&\!=\!s_{1},                   &\quad
\varphi_{[1,4,3,2]}&=\varphi_{[2,4]}\circ\alpha,\label{phi_[1423]}
\end{alignat}
where $\rho$ is the canonical endomorphism of $\calO_2$.
\par
First, we note that there are four $\ast$-automorphisms in the above 
$\ast$-endomorphisms: \hbox{$\varphi_{id}=\mathit{id}$,} 
$\varphi_{[1,2][3,4]}=\alpha$, $\varphi_{[1,3][2,4]}$ and 
$\varphi_{[1,4][2,3]}$. 
As for $\varphi_{[1,4][2,3]}$, we can rewrite it as follows:
\begin{eqnarray}
&&\varphi_{[1,4][2,3]}(s_1) =  s_2 J = J s_1 J^*, \\
&&\varphi_{[1,4][2,3]}(s_2) =  s_1 J = J s_2 J^*, \\
&&\quad  J \equiv  s_{2;1} + s_{1;2}, \quad J^*=J,\quad \ J^2=I,
\end{eqnarray}
where $J$ satisfies $J s_1 = s_2$ and $J s_2 = s_1$.
Thus, $\varphi_{[1,4][2,3]}$ is an inner $\ast$-automorphism:
$s_1$ and $s_2$ are expressed in terms of 
$t_i\equiv\varphi_{[1,4][2,3]}(s_i)$, $(i=1,\,2)$ owing to 
the identity $t_{2;1}+t_{1;2}=JJJ^*=J$.
On the other hand, since $\alpha$ is an outer $\ast$-automorphism, 
so is $\varphi_{[1,3][2,4]}(=\varphi_{[1,4][2,3]}\circ\alpha)$.
\par
It should be noted that the restriction of the $\ast$-automorphism 
$\alpha=\varphi_{[1,2][3,4]}$ to $\calA_{SR_1}$ gives the following 
Bogoliubov transformation:
\begin{equation}
\alpha(a_n)=\alpha(\zeta_1^{n-1}(s_{1;2}))=(-1)^{n-1}a_n^*, \qquad 
    n\in\boldN. \label{outer Bog}
\end{equation}
Therefore, we obtain
$(\varphi_\sigma\circ\alpha)(a_n)=(-1)^{n-1}\varphi_\sigma(a_n)^*$.
Hence we hereafter restrict ourselves to consider only the eleven 
$\ast$-endomorphisms in the following:
$\varphi_{[1,2]}$, $\varphi_{[1,3]}$, 
$\varphi_{[1,4]}$, $\varphi_{[2,3]}(=\rho)$, 
$\varphi_{[2,4]}$, $\varphi_{[1,4][2,3]}$, 
$\varphi_{[1,2,3]}$, $\varphi_{[1,2,4]}$, 
$\varphi_{[2,3,4]}$, $\varphi_{[2,4,3]}$,
$\varphi_{[1,3,2,4]}$.
\par
Since $\varphi_{[1,4][2,3]}$ is an inner $\ast$-automorphism, it is 
easy to obtain its restriction to $\calA_{SR_1}$: 
\begin{equation}
\varphi_{[1,4][2,3]}(a_n) = J a_n J^* 
 = (a_1 + a_1^*) a_n (a_1 + a_1^*)
 = \begin{cases}
       a_1^*  &\mbox{for $n=1$,}  \\[5pt]
      -a_n     &\mbox{for $n\geqq2$.}
     \end{cases}
\end{equation}
Hence this $\ast$-automorphism of $\calA_{SR_1}$ is nothing but 
a Bogoliubov transformation up to sign for a specific mode $a_1$. 
\par
Next, we consider the restrictions of $\varphi_{[1,2,3]}$ and 
$\varphi_{[2,4,3]}$.
As for $\varphi_{[1,2,3]}$, we have
\begin{alignat}{1}
\varphi_{[1,2,3]}(a_1)
&= s_{2,1;\,2,1}+ s_{1,1;\,2,2}
= s_2\, a_1\, s_1^* + s_1\, a_1\, s_2^*.
\end{alignat}
If $X\in\calO_2$ satisfies 
$\varphi_{[1,2,3]}(X)=s_2\, X\, s_1^*+s_1\, X\, s_2^*$,
we have
\begin{alignat}{1}
\varphi_{[1,2,3]}(\zeta_1(X))
=&\, (s_{2,1;\,1}+s_{1,1;\,2})(s_2\, X\, s_1^*+s_1\, X\, s_2^*)
     (s_{1;\,1,2}+s_{2;\,1,1})\nonumber\\
&\, -(s_{1,2;\,1}+s_{2,2;\,2})(s_2\, X\, s_1^*+s_1\, X\, s_2^*)
     (s_{1;\,2,1}+s_{2;\,2,2})\nonumber\\
=&\, s_2 \zeta_1(X) s_1^* + s_1 \zeta_1(X) s_2^*. 
\end{alignat}
Hence we obtain
\begin{alignat}{1}
\varphi_{[1,2,3]}(a_n) 
&= s_2\, a_n\, s_1^* + s_1\, a_n\, s_2^* = (s_{2;1}-s_{1;2})\zeta_1(a_n) 
 \nonumber\\
&=(a_1^* - a_1) a_{n+1}, \quad n\in\boldN.
\end{alignat}
Then, from \eqref{phi_[234]} and \eqref{outer Bog}, we have
\begin{alignat}{1}
\varphi_{[2,4,3]}(a_n)
&=(\alpha\circ\varphi_{[1,2,3]}\circ\alpha)(a_n) 
=(-1)^{n-1}(\alpha\circ\varphi_{[1,2,3]})(a_n^*) \nonumber\\
&=(-1)^{n-1}\alpha(a_{n+1}^*(a_1 - a_1^*) ) 
=(-1)^{n-1}(-1)^{n} a_{n+1} (a_1^* - a_1) \nonumber\\ 
&= (a_1^* - a_1) a_{n+1} = \varphi_{[1,2,3]}(a_n).
\end{alignat}
Therefore, $\varphi_{[1,2,3]}$ and $\varphi_{[2,4,3]}$ induce the 
$\ast$-endomorphisms of the CAR algebra which are expressed in terms 
of the second order binomials.
\par
As for $\varphi_{[2,3,4]}$, we have
\begin{alignat}{1}
\varphi_{[2,3,4]}(a_1)
&=s_{1,1;\,2,1}+s_{2,2;\,1,2}
=s_1\, a_1\, s_1^* + s_2\, a_1^*\, s_2^*.
\end{alignat}
If $X\in\calO_2$ satisfies 
$\varphi_{[2,3,4]}(X) = s_1\, X\, s_1^* \pm s_2\, X^*\, s_2^*$, we have
\begin{alignat}{1}
\varphi_{[2,3,4]}(\zeta_1(X))
=&\,(s_{1,1;\,1}+s_{2,2;\,2})(s_1\, X\, s_1^* \pm s_2\, X^*\, s_2^*)
      (s_{1;\,1,1}+s_{2;\,2,2})\nonumber\\
&\,-(s_{1,2;\,1}+s_{2,1;\,2})(s_1\, X\, s_1^* \pm s_2\, X^*\, s_2^*)
      (s_{1;\,2,1}+s_{2;\,1,2})\nonumber\\
=&\,s_1 \zeta_1(X) s_1^* \mp s_2 \zeta_1(X)^* s_2^*.
\end{alignat}
Hence we obtain
\begin{alignat}{1}
\varphi_{[2,3,4]}(a_n) 
&=s_1\, a_n\, s_1^* + (-1)^{n-1}\, s_2\, a_n^* s_2^*
  = s_{1;1}\zeta_1(a_n)+(-1)^n s_{2;2}\zeta_1(a_n)^* \nonumber\\
&=a_1a_1^* a_{n+1} + (-1)^n a_1^* a_1 a_{n+1}^*,\quad n\in\boldN.
\end{alignat}
In a similar way, we obtain
\begin{alignat}{1}
\varphi_{[1,2,4]}(a_n)
&=\begin{cases}
   (-1)^{\frac{n-1}{2}}\varphi_{[2,3,4]}(a_n)^* &\mbox{for odd $n$,}\\[3pt]
   (-1)^{\frac{n}{2}}\varphi_{[2,3,4]}(a_n)     &\mbox{for even $n$}.
\end{cases}
\end{alignat}
For the canonical endomorphism $\rho=\varphi_{[2,3]}$, by simple calculations, 
we obtain
\begin{alignat}{1}
\rho(a_n) 
 &= \zeta_1(I)\zeta_1(a_n) = (s_{1,1}-s_{2,2})a_{n+1} = K_1 a_{n+1}, 
  \qquad n\in\boldN, \label{rho|car} \\
K_1 &\equiv a_1a_1^*-a_1^*a_1 = I-2a_1^*a_1
  =\exp(\sqrt{-1}\,\pi\,a_1^*a_1), \label{Klein}
\end{alignat}
where use has been made of \eqref{RFS_p-zeta} and an identity 
$(a_1^*a_1)^2 = a_1^*a_1$.
Here, $K_1$ is the Klein-Jordan-Wigner operator 
anticommuting with $a_1$, hence we have
\begin{equation}
[\rho(a_n),\,a_1]\ =\ [\rho(a_n),\,a_1^*]\ =\ 0,\quad n\in\boldN.
\end{equation}
Hence $\rho(\calA_{SR_1})$ is the commutant of the subalgebra
generated by $a_1$ and $a_1^*$.
Likewise, for $\varphi_{[1,4]}$, we obtain
\begin{alignat}{1}
\varphi_{[1,4]}(a_n)
&=(-1)^{n-1}\rho(a_n)^* 
=(-1)^{n-1} K_1 a_{n+1}^*, \quad n\in\boldN.
\end{alignat}
Therefore, $\varphi_{[2,3,4]}$, $\varphi_{[1,2,4]}$, $\rho=\varphi_{[2,3]}$, 
and $\varphi_{[1,4]}$ induce the $\ast$-endomorphisms
of the CAR algebra which are expressed in terms of the third order polynomials.
\par
Next, for $\varphi_{[2,4]}$, we have
\begin{alignat}{1}
\hspace*{-15pt}
\varphi_{[2,4]}(a_1) 
 =&\, s_{1,1;\,2,2} + s_{1,2;\,1,2}                \nonumber \\
 =&\, - a_1 ( a_2 + a_2^* ), 
   \label{phi_{[3,4]}-recurrence-1}\\
\varphi_{[2,4]}(a_2) 
 =&\, s_{1,1,1;\,2,2,1} + s_{1,1,2;\,1,2,1}  
   -  s_{2,2,1;\,2,1,2} - s_{2,2,2;\,1,1,2} \nonumber\\
 =&\, -(a_1a_1^*a_2 + a_1^*a_1a_2^*)(a_3 + a_3^*), 
   \label{phi_{[3,4]}-recurrence-2}\\
\varphi_{[2,4]}(a_3) 
 =&\, s_{1,1,1,1;\,2,2,1,1} + s_{1,1,1,2;\,1,2,1,1} 
      - s_{1,2,2,1;\,2,1,2,1} - s_{1,2,2,2;\,1,1,2,1} \nonumber\\
  &\, - s_{2,2,1,1;\,2,2,2,2} - s_{2,2,1,2;\,1,2,2,2} 
      + s_{2,1,2,1;\,2,1,1,2} + s_{2,1,2,2;\,1,1,1,2} \nonumber\\
 =&\, -a_1a_1^*(a_2a_2^*a_3+a_2^*a_2a_3^*)(a_4+a_4^*)
     +a_1^*a_1(a_2^*a_2a_3+a_2a_2^*a_3^*)(a_4+a_4^*).
     \label{phi_{[3,4]}-recurrence-3}
\end{alignat}
In general, we obtain the recurrence formula as follows:
\begin{equation}
\varphi_{[2,4]}(a_n) = a_1 a_1^* \,b'_{n-1} - a_1^* a_1 \,b''_{n-1} 
    \quad n\geqq3,   \label{phi_{[3,4]}-recurrence}
\end{equation}
where $b_{n-1}'$ is obtained from $\varphi_{[2,4]}(a_{n-1})$ by replacing 
$a_k$ and $a_k^*$ $(k=1,\,\ldots,\,n)$ by $a_{k+1}$ and $a_{k+1}^*$, 
respectively, while $b''_{n-1}$ is obtained from $b'_{n-1}$ by exchanging 
$a_2$ and $a_2^*$.
It should be noted that $\varphi_{[2,4]}(a_n)$ is expressed in terms of
the $(2n)$-th order polynomials. 
Likewise, for $\varphi_{[1,3]}$, we obtain
\begin{alignat}{1}
&\varphi_{[1,3]}(a_n)=
\begin{cases}
 (-1)^{\frac{n-1}{2}}\varphi_{[2,4]}(a_n)    &\mbox{for odd $n$,}\\[3pt]
 (-1)^{\frac{n-2}{2}}\varphi_{[2,4]}(a_n)^*  &\mbox{for even $n$}.
\end{cases}
\end{alignat}
Therefore, $\varphi_{[2,4]}$ and $\varphi_{[1,3]}$ induce the 
$\ast$-endomorphisms of the CAR algebra which are expressed in terms 
of even polynomials.
\par
For $\varphi_{[1,2]}$, we have
\begin{alignat}{1}
\varphi_{[1,2]}(a_1)
=&\,s_{2,1;\,1,1}+s_{1,2;\,2,2} \nonumber\\
=&\,a_1^*a_2a_2^* + a_1a_2^*a_2,\\
\varphi_{[1,2]}(a_2)
=&\,s_{1,2,1;\,1,1,2}+s_{2,1,2;\,2,2,1}-s_{2,2,1;\,1,1,1}-s_{1,1,2;\,2,2,2}
  \nonumber\\
=&\,(a_1^*+a_1)(-a_2^*a_3a_3^* + a_2a_3^*a_3),\\
\varphi_{[1,2]}(a_3)
=&\,s_{2,1,2,1;\,1,1,2,1}+s_{1,2,1,2;\,2,2,1,2}
  -s_{1,2,2,1;\,1,1,1,2}-s_{2,1,1,2;\,2,2,2,1}\nonumber\\
& -s_{1,1,2,1;\,1,1,2,2}-s_{2,2,1,2;\,2,2,1,1}
   +s_{2,2,2,1;\,1,1,1,1}+s_{1,1,1,2;\,2,2,2,2}\nonumber\\
=&\,(a_1^*-a_1)(-a_2^*+a_2)(-a_3^*a_4a_4^* + a_3a_4^*a_4).
\end{alignat}
In general, we obtain the recurrence formula as follows:
\begin{equation}
\varphi_{[1,2]}(a_n)=(a_1^* + (-1)^n a_1)b_{n-1}, \quad n\geqq2,
\end{equation}
where $b_{n-1}$ is obtained from $\varphi_{[1,2]}(a_{n-1})$ by
replacing $a_1$, $a_1^*$, $a_k$, and $a_k^*$ $(k=2,\,\ldots,\,n)$ by 
$a_2$, $-a_2^*$, $a_{k+1}$, and $a_{k+1}^*$, respectively.
It should be noted that $\varphi_{[1,2]}(a_n)$ is expressed in terms of
the $(n+2)$-th order polynomials.
Likewise, for $\varphi_{[1,3,2,4]}$, we obtain
\begin{equation}
\varphi_{[1,3,2,4]}(a_n)=(-1)^n \varphi_{[1,2]}(a_n)^*, \quad
 n\in\boldN.
\end{equation}
\par
Thus, we have completed to clarify restrictions of all the 
second-order permutation endomorphisms of $\calO_2$ to $\calA_{SR_1}$.
\par
In summary, the $\ast$-endomorphisms of the CAR algebra induced by the second 
order permutation endomorphisms of $\calO_2$ are divided broadly into 
the following:
\begin{enumerate}
\item[(1)] $\ast$-automorphisms 
\begin{alignat}{1}
&\begin{array}{ll}
\mbox{Identity map: }\quad \varphi_{id}, &\\[2pt]
\mbox{Bogoliubov (inner) $\ast$-automorphism: }\quad  &\varphi_{[1,4][2,3]},\\[2pt]
\mbox{Bogoliubov (outer) $\ast$-automorphisms: }\quad &\varphi_{[1,2][3,4]},
\quad \varphi_{[1,3][2,4]}\,;
\end{array}
\end{alignat}
\item[(2)] $\ast$-endomorphisms expressed in terms of the second order binomials
\begin{alignat}{1}
\varphi_{[1,2,3]},\quad \varphi_{[1,3,4]}, \quad 
\varphi_{[1,4,2]}, \quad \varphi_{[2,4,3]}\,;
\end{alignat}
\item[(3)] $\ast$-endomorphisms expressed in terms of the third order polynomials
\begin{alignat}{1}
\begin{array}{llll}
\varphi_{[1,4]},\quad &\varphi_{[2,3]}, 
\quad &\varphi_{[1,2,4]}, \quad &\varphi_{[1,3,2]}, \\[2pt] 
\varphi_{[1,4,3]}, \quad &\varphi_{[2,3,4]}, \quad 
&\varphi_{[1,3,4,2]}, \quad &\varphi_{[1,2,4,3]}\,;
\end{array}
\end{alignat}
\item[(4)] $\ast$-endomorphisms expressed in terms of even polynomials
\begin{alignat}{1}
\varphi_{[1,3]},\quad \varphi_{[2,4]},\quad 
\varphi_{[1,2,3,4]},\quad \varphi_{[1,4,3,2]}\,;
\end{alignat}
\item[(5)] other $\ast$-endomorphisms expressed in terms of polynomials
\begin{alignat}{1}
\varphi_{[1,2]},\quad \varphi_{[3,4]},\quad 
\varphi_{[1,3,2,4]},\quad \varphi_{[1,4,2,3]}.
\end{alignat}
\end{enumerate}
\par
It should be noted that the above division of the induced 
$\ast$-endomorphisms is according to their apparent differences only, 
but not to their intrinsic properties. 
Indeed, all $\ast$-endomorphisms of Item (2) and half of Item (3) are 
expressed as composites of those in Items (4) and (5) as follows:
\begin{alignat}{1}
&\left\{\begin{array}{ll}
\displaystyle
\varphi_{[1,2,3]}=\varphi_{[1,3]}\circ\varphi_{[1,2]},&\quad
\varphi_{[1,3,4]}=\varphi_{[1,3]}\circ\varphi_{[3,4]},\\[2pt]
\displaystyle
\varphi_{[1,4,2]}=\varphi_{[2,4]}\circ\varphi_{[1,2]},&\quad
\varphi_{[2,4,3]}=\varphi_{[2,4]}\circ\varphi_{[3,4]},
\end{array}\right.\\
&\left\{\begin{array}{ll}
\displaystyle
\varphi_{[1,2,4]}=\varphi_{[1,2]}\circ\varphi_{[2,4]},&\quad
\varphi_{[1,3,2]}=\varphi_{[1,2]}\circ\varphi_{[1,3]},\\[2pt]
\displaystyle
\varphi_{[1,4,3]}=\varphi_{[3,4]}\circ\varphi_{[1,3]},&\quad
\varphi_{[2,3,4]}=\varphi_{[3,4]}\circ\varphi_{[2,4]}.
\end{array}\right.
\end{alignat}
On the other hand, the rest of Item 3 are not expressed as above.
\par
In this subsection, we have restricted ourselves to consider the second 
order permutation endomorphisms only.  If we consider more generally 
the second order homogeneous endomorphisms involving the $\ast$-automorphism
by the $U(2)$ action, we can find a relation between Items (4) 
and (5) above. Indeed, by using a $\ast$-automorphism of $\calO_2$ 
given by 
\begin{alignat}{1}
\begin{array}{l}
\alpha_\theta(s_1) = \cos\theta\, s_1 - \sin\theta\, s_2,\\[2pt]
\alpha_\theta(s_2) = \sin\theta\, s_1 + \cos\theta\, s_2,
\end{array} \quad 0\leqq\theta<2\pi,
\end{alignat}
we can rewrite $\varphi_{[1,2]}$ as follows:
\begin{alignat}{1}
\varphi_{[1,2]}&=\alpha_{-\pi/4}\circ\varphi_{[2,4]}\circ\alpha_{\pi/4},
\end{alignat}
where $\alpha_\theta$ induces an outer $\ast$-automorphism of the CAR algebra
expressed in terms of a nonlinear transformation as discussed later in 
Sec.\,8.
%
%
\vskip20pt
\subsection{Even-CAR endomorphisms}
As pointed out in Ref.\,\citen{AK1}), $\varphi_{[2,4]}$ induces the 
$\ast$-endomorphism $\tilde\varphi_{[2,4]}\equiv\varPhi_{SR_1}{}^{-1}
\circ\varphi_{[2,4]}\circ\varPhi_{SR_1}$ of the CAR algebra onto 
its even subalgebra\cite{Stormer,Binnenhei}. 
Since $\varphi_{[2,4]}$ is nothing but the special case of 
$\varphi_{\sigma_p}$ with $p=1$ defined by \eqref{p-branch-endo}, 
we consider it in more general.
\par
Let $\Gamma$ and $\tilde\Gamma$ be the $\ast$-automorphism of $\calO_2$ 
defined by $\Gamma(s_1)=s_1$, $\Gamma(s_2)=-s_2$ and its induced 
$\ast$-automorphism of the CAR algebra defined by 
$\tilde\Gamma\equiv\varPhi_{SR_1}^{-1}\circ\Gamma\circ\varPhi_{SR_1}$, 
respectively. 
Then, from \eqref{RFSp-embed} with \eqref{srfs_1}, we have
\begin{alignat}{1}
\tilde\Gamma(a_n)&=-a_n, 
 \quad   n\in\boldN.
\end{alignat}
Hence, from $\varPhi_{SR_1}(\mbox{CAR})=\calO_2^{U(1)}$, we obtain 
the following $\ast$-isomorphism:
\begin{alignat}{1}
&\mbox{CAR}_{\mbox{\scriptsize e}}
 \cong (\calO_2^{U(1)})_{\mbox{\scriptsize e}},\\
&\mbox{CAR}_{\mbox{\scriptsize e}}
  \equiv\{ X \in \mbox{CAR} \mid \tilde\Gamma(X)=X \},\\
  &(\calO_2^{U(1)})_{\mbox{\scriptsize e}}
  \equiv\{ X\in\calO_2^{U(1)} \mid \Gamma(X)=X \},
\end{alignat}
where CAR$_{\mbox{\scriptsize e}}$ is the even subalgebra of the CAR
algebra, and $(\calO_2^{U(1)})_{\mbox{\scriptsize e}}$ is the 
$\Gamma$-fixed point subalgebra of $\calO_2^{U(1)}$. 
Since it is obvious that $\Gamma\circ\varphi_{\sigma_p}=\varphi_{\sigma_p}$ 
from \eqref{p-branch-endo}, we have 
\begin{equation}
\varphi_{\sigma_p}(\calO_2^{U(1)})\subset(\calO_2^{U(1)})_{\mbox{\scriptsize e}},
\quad p\in\boldN.
\end{equation}
\par
In the case $p\!=\!1$, we also have
$\varphi_{\sigma_1}(\calO_2^{U(1)})\!\supset\!(\calO_2^{U(1)})_{\mbox{\scriptsize e}}$,
since it is shown inductively that 
\begin{alignat}{1}
\varphi_{\sigma_1}(\calD_k)
&\!=\!{\cal E}_{k},\quad k\in\boldN, \label{evencarind1} \\
\calD_k
&\!\equiv\!\{ s_{i_1,\ldots,i_{k-1},\ell;\,\ell,j_{k-1},\ldots,j_1}
\mid 
i_1,\ldots,i_{k-1},j_1,\ldots,j_{k-1},\ell=1,2\}, \\
\calE_{k}
&\!\equiv\! \{ X=s_{i_1,\ldots,i_{k};\,j_{k},\ldots,j_1} \mid 
\Gamma(X) \!=\! X,\ 
i_1,\ldots,i_{k},j_1,\ldots,j_{k}\!=\!1,2\},
\end{alignat} 
where $\{\calE_k\mid k\in\boldN \}$ generates (the dense subset of) 
$(\calO_2^{U(1)})_{\mbox{\scriptsize e}}$. 
Since $\varphi_{\sigma_1}$ is injective and 
$\sharp\,\calD_k = 2^{2k-1} =\sharp\,\calE_k$,
it is sufficient to show $\varphi_{\sigma_1}(\calD_k)\subset\calE_k$.
First, from \eqref{p-branch-endo}, we have
\begin{alignat}{1}
\varphi_{\sigma_1}(s_{1;\,1})=s_{1;\,1},\quad
\varphi_{\sigma_1}(s_{2;\,2})=s_2JJ^*\,s_2^*=s_{2;\,2},
\end{alignat}
hence \eqref{evencarind1} is satisfied for $k=1$.
Next, suppose that \eqref{evencarind1} is satisfied for $k=m$, 
and set
$\varphi_{\sigma_1}
(s_{i_2,\ldots,i_m,\ell;\,\ell,j_m,\ldots,j_2})
\equiv s_{i'_2,\ldots,i'_{m+1};\,j'_{m+1},\ldots,j'_2}
\in\calE_{m}$ for a fixed $\ell=1,2$.
Then, from \eqref{J s_i}, we have
\begin{alignat}{1}
&
\varphi_{\sigma_1}(s_{i_1,i_2,\ldots,i_m,\ell;\,\ell,j_m,\ldots,j_2,j_1})
= s_{i_1,i''_2,i'_3,\ldots,i'_{m+1};\,j'_{m+1},\ldots,j'_3,j''_2,j_1},
\label{evencarind2} \\
&
\quad
i''_2\equiv
\begin{cases}
  i'_2    &\mbox{for $i_1=1$,} \\
  3-i'_2  &\mbox{for $i_1=2$,}
\end{cases}
\quad
j''_2\equiv
\begin{cases}
  j'_2    &\mbox{for $j_1=1$,} \\
  3-j'_2  &\mbox{for $j_1=2$.}
\end{cases}
\end{alignat}
Since the number of 2 in $\{i_1,i''_2,j_1,j''_2\}$ and that in 
$\{i'_2,j'_2,\}$ are congruent modulo 2, 
we have
$\varphi_{\sigma_1}
(s_{i_1,i_2,\ldots,i_m,\ell;\,\ell,j_m,\ldots,j_2,j_1})
\in{\cal E}_{m+1}$. Thus, \eqref{evencarind1} is obtained.
Therefore, we have 
\begin{alignat}{1}
\varphi_{\sigma_1}(\calO_2^{U(1)})=(\calO_2^{U(1)})_{\mbox{\scriptsize e}}.
\end{alignat}
\par
On the other hand, for $p\geqq2$, $\varphi_{\sigma_p}(\calO_2^{U(1)})$ 
generates a proper subset of $(\calO_2^{U(1)})_{\mbox{\scriptsize e}}$.
In-deed, in this case, there exists a proper (i.e., not surjective) 
$\ast$-endomorphism $\varphi'_{\sigma_p}$ such that
\begin{alignat}{1}
\varphi_{\sigma_p}&=\varphi'_{\sigma_p}\circ \varphi_{\sigma_1},\\
 \varphi'_{\sigma_p}(s_1)&\equiv s_1, 
\quad
  \varphi'_{\sigma_p}(s_2)\equiv s_2 \prod_{k=0}^{p-2} \rho^{k}(J).
\end{alignat}
Since $\varphi'_{\sigma_p}$ commutes not only with the $U(1)$ 
action $\gamma$ but also with $\varphi_{\sigma_1}$, its restriction
to $\varphi_{\sigma_1}(\calO_2^{U(1)})\subset\calO_2^{U(1)}$ is
also proper. 
Consequently, we obtain
\begin{alignat}{1}
&\varphi_{\sigma_p}(\calO_2^{U(1)})
=\varphi'_{\sigma_p}\big(\varphi_{\sigma_1}(\calO_2^{U(1)})\big)
\subsetneqq\varphi_{\sigma_1}(\calO_2^{U(1)})
=(\calO_2^{U(1)})_{\mbox{\scriptsize e}}, \quad p\geqq2.
\end{alignat}
\par
Therefore, the restriction of $\varphi_{\sigma_p}$ to
$\calA_{SR_1}=\calO_2^{U(1)}$ generally induces a $\ast$-endomorphism 
$\tilde\varphi_{\sigma_p}\equiv\varPhi_{SR_1}{}^{-1}\circ\varphi_{\sigma_p}
\circ\varPhi_{SR_1}$ 
of the CAR algebra into its even subalgebra, and only for the case $p=1$, 
it gives the $\ast$-isomorphism between them. 
In general, the $\ast$-endomorphism of the CAR algebra whose range is 
a subset of its even subalgebra is called the {\it even-CAR endomorphism}.
Thus, $\tphi_{\sigma_p}$ $(p\in\boldN)$ are typical examples of the 
even-CAR endomorphisms.
\par
We write down the explicit expression for $\tphi_{\sigma_p}(a_n)$ in the 
form of a recurrence formula similar to \eqref{phi_{[3,4]}-recurrence} 
in the following:
\begin{alignat}{1}
\tphi_{\sigma_p}(a_n)
&\!=\!\begin{cases}
 \displaystyle
 a_n \prod_{\ell=1}^{n+p-1}K_\ell\,(a_{n+p}+a_{n+p}^*)
   &\mbox{for $1\leqq n\leqq p$,}  \\ 
 \displaystyle
 b_{m,\,n}\prod_{\ell=n-p}^{n+p-1}K_\ell\,(a_{n+p}+a_{n+p}^*)
   &\mbox{for $mp+1\leqq n\leqq (m+1)p,\quad m\in\boldN$,} 
 \end{cases} \label{phi_{p}-recurrence}  \\
b_{1,\,n}
&\equiv a_{n-p}a_{n-p}^*a_n + a_{n-p}^*a_{n-p}a_n^*, 
    \label{phi_{p}-recurrence-1} \\
b_{2,\,n}
&\equiv a_{n-2p}a_{n-2p}^* b_{1,\,n}
                       + a_{n-2p}^*a_{n-2p} b'_{1,\,n}, 
    \label{phi_{p}-recurrence-2} \\
b_{m,\,n}
&\equiv a_{n-mp}a_{n-mp}^* b_{m-1,\,n}
                       - a_{n-mp}^*a_{n-mp} b'_{m-1,\,n}, \qquad m\geqq3,
    \label{phi_{p}-recurrence-3} \\
K_\ell
&\equiv a_\ell a_\ell^*-a_\ell^*a_\ell
=1-2a_\ell^*a_\ell=\exp(\sqrt{-1}\,\pi\,a_\ell^*a_\ell),
    \label{phi_{p}-recurrence-4}
\end{alignat}
where $b'_{m,\,n}$ is obtained from $b_{m,\,n}$ by 
exchanging $a_{n-mp}$ and $a_{n-mp}^*$. It is straightforward to 
see that \eqref{phi_{[3,4]}-recurrence-1}--\eqref{phi_{[3,4]}-recurrence} 
is reproduced from \eqref{phi_{p}-recurrence}--\eqref{phi_{p}-recurrence-4}
by setting $p=1$.
\vskip20pt
\subsection{Simple examples of higher order permutation endomorphisms}
As for other higher order permutation endomorphisms, we give two simple 
examples in the following.
\par
It is straightforward to generalize the inner $\ast$-automorphism 
$\varphi_{[1,4][2,3]}$ so that it should yield the Bogoliubov transformation 
exchanging $a_k$ and $a_k^*$ for an arbitrary $k$. 
For that purpose, we define a linear mapping $\xi: \calO_2 \to \calO_2$  by
\begin{equation}
\xi(X)\equiv s_2 X s_1^* + s_1 X s_2^*,  \quad X \in \calO_2.
\end{equation}
Then, it satisfies the following:
\begin{alignat}{1}
\xi(X)^* &= \xi(X^*), \\
 \xi(X)\xi(Y)&=\rho(XY), \quad
 \xi(X)\rho(Y)=\rho(X)\xi(Y)=\xi(XY),
 \quad X, \, Y\in \calO_2, \label{f-bog1}\\
\xi(I)&=J, \quad J=s_{2;1}+s_{1;2}, \quad J=J^*=J^{-1}.
\end{alignat}
As a generalization of the operator $J$, we introduce $J_k$ $(k\in\boldN)$
as follows:
\begin{alignat}{1}
J_k &\equiv \xi^k(I), \quad  
   J_k=J^*_k=J^{-1}_k, \quad J_1=J,\quad  k\in\boldN,  \label{f-bog2}
\end{alignat}
which satisfies 
\begin{alignat}{1}
J_k s_i &= s_{3-i} J_{k-1}, \quad  i=1,\,2,\quad  k\geqq2, \label{f-bog3} \\
J_k J_\ell &= J_\ell J_k = \rho^k(J_{\ell-k}), \quad k<\ell,
                                                      \label{f-bog4}\\ 
J_k\,\zeta_1^m(X)\, J_k^*
    &= \,
     \begin{cases}
      (-1)^m \zeta_1^m(J_{k-m} X J_{k-m}^*) & \mbox{for $m < k$}, \\[5pt]
      (-1)^k \zeta_1^m( X )                 & \mbox{for $m \geqq k$},
     \end{cases} 
     \quad X \in \calO_2.\qquad                      \label{f-bog5}
\end{alignat}
In terms of $J_k$, we define an inner $\ast$-automorphism $\hat\varphi_k$ 
$(k=1,\,2,\,\ldots)$ of $\calO_2$ by
\begin{alignat}{1}
\hat\varphi_k(X) &\equiv 
  \begin{cases}
    J_1 X J_1^*                    & \mbox{for $k=1$,} \\[5pt]
    J_{k-1} J_k X J_k^* J_{k-1}^*  & \mbox{for $k\geqq2$,}
  \end{cases}
 \quad X \in \calO_2.
\end{alignat}
Then, it is shown that $\hat\varphi_k$ is one of the $(k+1)$-th order 
permutation endomorphisms of $\calO_2$. 
Since $\hat\varphi_1=\varphi_{[1,4][2,3]}$, we consider the case $k\geqq2$:
\begin{alignat}{1}
\hat\varphi_k(s_i) 
&= J_{k-1} J_k s_i J_k^* J_{k-1}^* = s_i\, J_{k-2} J_{k-1} J_k J_{k-1}
 =  s_i\, \rho^{k-2}(J_2) \nonumber\\
&= \sum_{j_1,\ldots,j_{k-2}=1}^2
    s_i s_{j_1,\ldots,j_{k-2}}
            (s_{2,2;\,1,1}+s_{1,2;\,1,2}+s_{2,1;\,2,1}+s_{1,1;\,2,2})
    s_{j_1,\ldots, j_{k-2}}^* \nonumber\\
&= \sum_{j_1,\ldots,j_k=1}^2
    s_{\sigma(i,j_1,\ldots,j_k)}s_{j_1,\ldots, j_k}^*,  \nonumber\\[2pt]
&  \hspace*{-10pt}
    \sigma : (i, j_1,\ldots,j_{k-2},j_{k-1},j_k) 
             \mapsto (i,j_1,\ldots,j_{k-2},\hatj_{k-1},\hatj_k), \quad 
	     \hatj\equiv 3-j.
\end{alignat}
From \eqref{f-bog3} and \eqref{f-bog5}, it is straightforward to 
show that $\hat\varphi_k$ gives the following Bogoliubov transformation 
if restricted to $\calA_{SR_1}$:
\begin{alignat}{1}
\hat\varphi_k(a_n) 
 &= \begin{cases}
     a_n   & \mbox{for $n<k$,}\\[2pt]
     a_k^* & \mbox{for $n=k$,}\\[2pt]
    -a_n   & \mbox{for $n>k$.}
   \end{cases}
\end{alignat}
\par
Another example is the $p$-th power of the canonical endomorphism of 
$\calO_2$, $\rho^p$ $(p\geqq1)$, which is one of the (p+1)-th order 
permutation endomorphism.  Restricting it to $\calA_{SR_1}$, we obtain
\begin{alignat}{2}
\hspace*{-15pt}
&\rho^p(a_n)
  =\zeta_1^p(I)\, \zeta_1^p(a_n) = \prod_{m=1}^p K_m\, a_{n+p}, 
      &\quad &K_m\equiv\exp(\sqrt{-1}\,\pi\,a_m^* a_m ),\quad n\in\boldN, \\
&[\rho^p(a_n), \, a_m]
 =[\rho^p(a_n), \, a_m^*]  = 0, 
    &\quad &n\in\boldN, \quad  m=1,\,2,\,\ldots,p.
\end{alignat} 
Therefore, $\rho^p(\calA_{SR_1})$ is the commutant of the 
$\ast$-subalgebra generated by \hbox{$\{a_m\mid 1\leqq m\leqq p\}$.}
%
%
%
\vfill\eject
\section{Branching of Fock Representation and KMS state}
As shown in the previous section, the $\ast$-endomorphism $\varphi_{\sigma_p}$ 
$(p\in\boldN)$, which is defined by \eqref{p-branch-endo}, of $\calO_2$ 
induces the $\ast$-endomorphism 
$\tphi_{\sigma_p}=\varPhi^{-1}_{SR_1}\circ\varphi_{\sigma_p}\circ\varPhi_{SR_1}$
of the CAR algebra into its even subalgebra. 
In this section, we consider a branching of the Fock representation of 
the CAR algebra by $\tphi_{\sigma_p}$, and show that a certain KMS 
state\cite{BR} of the CAR algebra is obtained.
\par
Let $\pi^{\mbox{\scriptsize even}}_{p}$ be a representation of the CAR algebra
obtained by composing the Fock representation and the even-CAR endomorphism
$\tphi_{\sigma_p}$ as follows:
\begin{alignat}{1}
\pi^{\mbox{\scriptsize even}}_p
&\equiv\mbox{Fock}\circ\tphi_{\sigma_p}
=(\pi_s\circ\varPhi_{SR_1})\circ
  (\varPhi^{-1}_{SR_1}\circ\varphi_{\sigma_p}\circ\varPhi_{SR_1})\nonumber\\
&=\pi_s\circ\varphi_{\sigma_p}\circ\varPhi_{SR_1}.
\end{alignat}
Then, from \eqref{branch-formula}, 
\eqref{perm-rep-restrict-embed} and \eqref{varPhi_SR-reduction-2}, 
it is straightforward to have
\begin{alignat}{1}
\pi^{\mbox{\scriptsize even}}_{p}
&\cong \bigoplus_{L\in I\!P\!R_p} \pi_L\circ\varPhi_{SR_1} \nonumber\\
&\cong \bigoplus_{L\in I\!P\!R_p} \bigoplus_{\lambda=0}^{\kappa-1} 
         \pi^{(\kappa)}_{j(\lambda)}\circ\varPhi_{SR_\kappa} \nonumber\\
&\cong \bigoplus_{L\in I\!P\!R_p} \bigoplus_{\lambda=0}^{\kappa-1} 
 \pi^{(\kappa)}_{j(\lambda)}\circ\varPsi_{\kappa,\,p}\circ\varPhi_{SR_p},
\label{rfs-branch-1}
\end{alignat}
where the label of $\pi_L$ is set by $L\equiv (j_0,\ldots,j_{\kappa-1})$,
$j(\lambda)\equiv 
\sum\limits_{\ell=1}^{\kappa}(j_{\lambda+\ell-1}-1)2^{\ell-1}+1$
with the subscript of $j_{\lambda+\ell-1}$ taking values in
$\boldZ_\kappa$, and 
$\varPsi_{p,\kappa}$ denoting the homogeneous embedding of 
$\calO_{2^p}$ into $\calO_{2^\kappa}$.
Substituting \eqref{pi_i-reduction} with $d=2^\kappa$ and $q=p/\kappa$ 
into \eqref{rfs-branch-1}, we obtain the branching formula as follows:
\begin{alignat}{1}
\pi^{\mbox{\scriptsize even}}_{p}
&\cong \bigoplus_{L\in I\!P\!R_p} \bigoplus_{\lambda=0}^{\kappa-1}
    \pi^{(p)}_{\tj(\lambda)}\circ\varPhi_{SR_p}, \qquad
    \tj(\lambda) \equiv \frac{2^p-1}{2^\kappa-1}( j(\lambda) - 1)+1 \nonumber\\
&\cong \bigoplus_{i_0=1}^{2^p} \pi^{(p)}_{i_0}\circ \varPhi_{SR_p} \nonumber\\
&\cong \bigoplus_{i_0=1}^{2^p} \Big(\mbox{Fock}\circ\phi_{i_0}\Big).
 \label{phi_p-rfs-branch}
\end{alignat}
Here, the vacuum $e^{(i_0)}_1$ of Fock$\,\circ\,\phi_{i_0}$ is given by 
\eqref{phi_p-eigenvector}, that is,
\begin{alignat}{2}
(\pi_{i_0}^{(p)}\circ\varPhi_{SR_p})(a^{({i_0})}_n)\,e^{({i_0})}_1 &=0, 
 &\qquad &a^{({i_0})}_n\equiv\phi_{i_0}(a_n),\qquad n\in\boldN, \\
e^{(i_0)}_1 
&\equiv \pi_s(s_{i_{0,1},\ldots, i_{0,p}})\,e_1 &&\nonumber\\
&= e_{i_0}, &\qquad &i_0=1,2,\ldots,2^p,
\end{alignat}
where use has been made of \eqref{rep1-formula2}.
We denote the $\phi_{i_0}$-Fock space 
by $\calH^{[i_0]}$: 
\begin{equation}
\calH^{[i_0]}\equiv \mbox{Lin}\langle\,\{\, e^{(i_0)}_1, \ 
 (\pi_{i_0}^{(p)}\circ\varPhi_{SR_p})(a^{({i_0})\,\ast}_{n_1}\cdots 
 a^{({i_0})\,\ast}_{n_r})\, e^{({i_0})}_1, \ 
 n_1<\cdots<n_r,\ r\geqq1 \,\}\,\rangle.
\end{equation}
Although it is rather complicated to express the basis of 
$\{ e^{({i_0})}_n \}_{n=1}^\infty$ of $\calH^{[{i_0}]}$ directly 
in terms of the basis of $\pi_s^{(p)}$, $\{ e_n \}_{n=1}^\infty$, of $\calH$ 
except for the vacuum $e^{({i_0})}_1$,
we can adopt the similar numbering in \eqref{rep1-formula2} by
an appropriate unitary transformation:
\begin{eqnarray}
&&e^{({i_0})}_n  
\equiv
(\pi_{i_0}^{(p)}\circ\varPhi_{SR_p})(a^{({i_0})\,\ast}_{n_1}\cdots
 a^{({i_0})\,\ast}_{n_r})\, e^{({i_0})}_1, \qquad
 n \equiv \sum_{k=1}^r 2^{n_k} + 1.
\end{eqnarray}
In any way, the total Hilbert space $\calH$, which is the representation
space of $\pi_s\circ\varPhi_{SR_1}(=\pi^{(p)}_s\circ\varPhi_{SR_p}$), 
is decomposed into a direct sum of mutually orthogonal subspaces:
\begin{equation}
\calH = \bigoplus_{i_0=1}^{2^p} \calH^{[{i_0}]}, \qquad
\calH^{[{i_0}]} \perp \calH^{[{i'_0}]}, \quad 
 {i_0}\not={i'_0}. \label{H-ortho-decomp}
\end{equation}
\par
Now, we consider a state $\omega$ of the CAR algebra defined 
by the representation $\pi^{\mbox{\scriptsize even}}_{p}$ 
with an appropriate unit vector $\Omega \in \calH$ as follows:
\begin{alignat}{1}
&\omega(X)
\equiv
\langle\,\Omega\,|\,\pi^{\mbox{\scriptsize even}}_{p}(X)\,
\Omega\,\rangle,\qquad X \in \mbox{CAR}.
\end{alignat}
If we set
\begin{equation}
\Omega \equiv \sum_{{i_0}=1}^{2^p} \sqrt{\Lambda_{i_0}}\,e^{({i_0})}_1
\end{equation}
with $\{\Lambda_{i_0}\}_{i_0=1}^{2^p}$ being a set of nonnegative constants 
satisfying $\sum\limits_{{i_0}=1}^{2^p} \Lambda_{i_0} =1$, 
then, from \eqref{phi_p-rfs-branch} and \eqref{H-ortho-decomp}, we can 
rewrite $\omega$ as a convex sum of pure states as follows:
\begin{alignat}{1}
&\omega(X) = \sum_{{i_0}=1}^{2^p}\Lambda_{i_0}\,\omega_{i_0}(X), \label{state1}\\
&\omega_{i_0}(X)\equiv
 \langle\,e^{({i_0})}_1\,|\,(\pi^{(p)}_{i_0}\circ\varPhi_{SR_p})(X)\,
 e^{(i_0)}_1\,\rangle,         \label{state2}
\end{alignat}
where $\omega_{i_0}$ is pure since 
$\pi^{(p)}_{i_0}\circ\varPhi_{SR_p}=\mbox{Fock}\circ\phi_{i_0}$
is irreducible. 
We parametrize $\{\Lambda_{i_0}\}$ by
\begin{alignat}{1}
&\Lambda_{i_0} = \prod_{j=1}^p\Lambda_{j,\,i_{0,j}},\\
&\Lambda_{j,\,i_{0,j}}\equiv
  \begin{cases}
    1-\lambda_j   & \mbox{for $i_{0,j}=1$,}\\[5pt]
      \lambda_j   & \mbox{for $i_{0,j}=2$}
  \end{cases}
\end{alignat}
with $i_0=\sum\limits_{j=1}^p (i_{0,j}-1)2^{j-1}+1$ 
$(i_{0,1},\ldots,i_{0,p}=1,2)$
and $0\leqq\lambda_i\leqq1$ $(j=1,\ldots,p)$, so that we have
\begin{equation}
\sum_{i_0\in A_{j,\,1}}\Lambda_{i_0} = 1-\lambda_j,
\quad
\sum_{i_0\in A_{j,\,2}}\Lambda_{i_0} = \lambda_j
\end{equation}
with $A_{j,\,1}$ and $A_{j,\,2}$ $(j=1,\ldots,p)$ being a set of all
indices $i_0$'s which satisfy $i_{0,j}=1$ and that $i_{0,j}=2$, 
respectively. Then, we obtain
\begin{alignat}{1}
&\omega(a_{p(m-1)+j}\,a^*_{p(n-1)+k})
  = \delta_{m,n}\delta_{j,k}\sum_{i_0\in A_{j,\,1}}\Lambda_{i_0}
  = \delta_{m,n}\delta_{j,k}(1-\lambda_j), \label{state-aa^*}\\
&\omega(a^*_{p(m-1)+j}\,a_{p(n-1)+k})
  = \delta_{m,n}\delta_{j,k}\sum_{i_0\in A_{j,\,2}}\Lambda_{i_0}
  = \delta_{m,n}\delta_{j,k}\lambda_j \label{state-a^*a}
\end{alignat}
with $m,n\in\boldN,\,j,k=1,\ldots,p$, and 
states for products of the same number of $a_m$'s and $a^*_n$'s
expressed in terms of \eqref{state-aa^*} and \eqref{state-a^*a},
while others vanish.
\par
If each of $\{\lambda_j\mid j=1,\cdots,p\}$ satisfies $0<\lambda_j<1/2$, 
we rewrite it as follows:
\begin{alignat}{1}
&\lambda_j = \frac{1}{1+\exp(\beta \varepsilon_j)}, \quad
 \varepsilon_j>0, \quad \beta>0.
\end{alignat}
Then, $\omega$ is identical with the KMS state of the CAR algebra with 
the inverse temperature $\beta$ with respect to the one-parameter group 
$\{\tau^{(0)}_t\mid t\in\boldR\}$ of $\ast$-automorphisms defined by
\begin{equation}
\tau^{(0)}_t(a_{p(m-1)+j}) 
 \equiv \exp(-\sqrt{-1}\,\varepsilon_j\,t)\,a_{p(m-1)+j},
\quad m\in\boldN,\ j=1,\ldots,p, \label{CAR-1para-auto}
\end{equation}
which describes the time evolution of a (quasi-)free fermion system.
Here, the superscript $(0)$ stands for the free fermions.
Indeed, it is shown that $\omega$ satisfies the KMS condition\cite{BR} 
given by
\begin{alignat}{1}
& \omega(X\,\tau^{(0)}_{\sqrt{-1}\beta}(Y))=\omega(YX), \quad X,\,Y\in\mbox{CAR}.
\end{alignat}
It is remarkable that we can induce $\{\tau^{(0)}_t\mid t\in\boldR\}$ 
from a one-parameter group $\{\alpha_t^{(0)}\mid t\in\boldR\}$ 
of $\ast$-automorphisms of $\calO_{2^p}$ by using the embedding 
$\varPhi_{SR_p}$ of the CAR algebra into $\calO_{2^p}$ associated with 
the standard RFS$_p$ as follows:
\begin{alignat}{1}
&\tau^{(0)}_t = \varPhi_{SR_p}^{-1}\circ \alpha^{(0)}_t \circ \varPhi_{SR_p},
  \label{CAR-1para-auto-by-O_{2^p}-1} \\
&\alpha^{(0)}_t(s_{i}) 
  \equiv \exp\left(\sqrt{-1}\,\varepsilon{(i)}\,t\right)s_i, 
  \quad
  \varepsilon{(i)}\equiv\sum_{j=1}^p(i_j-1)\varepsilon_j + \varepsilon
  \label{CAR-1para-auto-by-O_{2^p}-2}
\end{alignat}
with $i=\sum\limits_{j=1}^p(i_j-1)2^{j-1}+1$ $(i=1,\ldots,2^p;\ i_j=1,2)$,
$\varepsilon$ being an arbitrary real constant.
However, one should note that the above $\omega$ is {\it not\/} induced 
from the KMS state of $\calO_{2^p}$ with respect to the one-parameter group
$\{\alpha^{(0)}_t\}$. 
In contrast with the KMS state for the CAR algebra, as is well-known, 
the inverse temperature $\beta$ for the KMS state of the Cuntz algebra 
is uniquely determined for a one-parameter group of $\ast$-automorphisms 
such as $\{\alpha^{(0)}_t\}$.\cite{OP,Evans,BJO}
\par
If each of $\{\lambda_j\mid j=1,\cdots,p\}$ satisfies $1/2<\lambda_j<1$, 
using the Bogoliubov transformation $\phi_{2^p}(a_n)=a_n^*$ $(n\in\boldN)$, 
we can identify $\omega\circ\phi_{2^p}$ with the KMS state as above.
If $\lambda_j=1/2$ for any $j=1,\ldots,p$, or equivalently 
$\Lambda_{i_0}=1/2^p$ for any $i_0=1,\ldots,2^p$, then, $\omega$ is 
the normalized trace, that is, it satisfies
\begin{alignat}{1}
\omega(XY)&=\omega(YX), \quad X, Y \in \mbox{CAR}.
\end{alignat}  
If $\lambda_j=0,1$ for any $j=1,\ldots,p$, or equivalently $\Lambda_{i_0}=1$ 
for one $i_0$ and all the other $\Lambda_{i_0}$'s vanishing, then, 
$\omega$ is nothing but the pure state obtained from the $\phi_{i_0}$-Fock 
representation. 
\par
For the case $p=1$, the above classification with respect to $\lambda_1$ 
is complete, and reproduces the Araki-Woods classification of factors 
for the CAR algebra.\cite{AW} 
For the case $p\geqq2$, we decompose the set $\{1,\ldots,p\}$ 
into three disjoint subsets as follows:
\begin{alignat}{1}
&\{1,\ldots,p\}=J_1\cup J_2 \cup J_3, \\ %
&\begin{array}{l}
\quad
J_1\equiv\big\{ j\in\{1,\ldots,p\} \mid \lambda_j=0,\ 1 \big\}, \\[3pt]
\quad
J_2\equiv\big\{ j\in\{1,\ldots,p\} \mid \lambda_j={\textstyle\frac{1}{2}} \big\}, \\[3pt]
\quad
J_3\equiv J_{3,1}\cup J_{3,2},\quad
\begin{cases}
J_{3,1}\equiv\big\{ j\in\{1,\ldots,p\} \mid 0<\lambda_j<{\textstyle\frac{1}{2}} \big\},\\
J_{3,2}\equiv\big\{ j\in\{1,\ldots,p\} \mid {\textstyle\frac{1}{2}}<\lambda_j<1 \big\}. 
\end{cases}
\end{array}
\end{alignat}
Correspondingly, we decompose each monomial in the CAR algebra into a product of 
elements of the three $C^*$-subalgebras as follows:
\begin{alignat}{1}
&\calA_k \equiv C^*\langle\,\{ a_{p(m-1)+j} \mid m\in\boldN,\ j\in J_k \}\,\rangle,
 \quad k=1,2,3.
\end{alignat}
Then, we can factorize $\omega$ into the following form:
\begin{alignat}{1}
&\omega(X_1\, X_2\, X_3)=\prod_{k=1}^3\omega_k(X_k), \quad X_k \in \calA_k,
\end{alignat}
where $\omega_1$, $\omega_2$, and $\omega_3$ 
are the vector state associated with the Fock-like representation, 
the normalized trace, and the $\phi_i$-transformed KMS state for the 
one-parameter group $\{\tau_t^{(0)}\}$ of $\ast$-automorphisms, 
respectively, where $\phi_i$ is the Bogoliubov transformation defined by 
\eqref{pi_j-Bogoliubov} with $i\equiv \sum\limits_{j\in J_{3,2}}2^{j-1}+1$.
Therefore, the state $\omega$ given by \eqref{state1} and \eqref{state2} 
for a generic $p$ may be also understood according to the classification 
by Araki-Woods.
\vskip40pt
\section{Induced Automorphism of the CAR Algebra}
In this section, we summarize the induced $\ast$-automorphism\cite{AK2} of 
the CAR algebra from the $U(2^p)$ action on the Cuntz algebra $\calO_{2^p}$, 
and apply it to construct one-parameter groups of $\ast$-automorphisms of 
the CAR algebra describing nontrivial time evolutions of fermions.
\par
We consider a $\ast$-automorphism $\alpha_u$ of $\calO_{2^p}$ obtained from 
the action of $U(2^p)$ as follows:
\begin{alignat}{3}
&\alpha_u (s_i) \equiv \sum_{k=1}^{2^p} s_{k} u_{k, i}, 
&\quad 
&u = (u_{k,i}) \in U(2^p), \quad i=1,\,2,\,\ldots,\,2^p,  \label{O_{2^p}-auto}
\end{alignat}
Since $\alpha_u$ commutes with the $U(1)$ action $\gamma$ defined by 
\eqref{U(1)-action}, the restriction of $\alpha_u$ to 
$\calO_{2^p}^{U(1)}$ gives a $\ast$-automorphism of $\calO_{2^p}^{U(1)}$.  
Therefore, from $\calA_{SR_p}=\calO_{2^p}^{U(1)}$, $\alpha_u$ induces a
$\ast$-automorphism $\tau_u$ of the CAR algebra as follows:
\begin{alignat}{1}
&
\begin{array}{c}
\tau_u : U(2^p) \curvearrowright \, \mbox{CAR}, \quad u\in U(2^p), 
  \\[7pt]
\tau_u \equiv \varPhi_{SR_p}^{-1} \circ \alpha_u \circ \varPhi_{SR_p}.
\end{array}
\label{CAR-auto}
\end{alignat}
It is straightforward to show that $\tau_u(a_n)$ is expressed in terms of 
a polynomial in $a_k$ and $a_\ell^*$ with $k,\,\ell\leqq pm$ for 
$p(m-1)+1\leqq n \leqq pm$. Therefore, $\tau_u$ gives a nonlinear transformation
of the CAR algebra associated with $u\in U(2^p)$. The whole set of $\tau_u$ 
with $u\in U(2^p)$ denoted by 
\begin{alignat}{1}
& \mbox{Aut}_{U(2^p)}(\mbox{CAR})\equiv\{\, \tau_u\,\mid\,u\in U(2^p)\,\}
\end{alignat}
constitutes a nonlinear realization of $U(2^p)$ on the CAR algebra.
Using the homogeneous embedding $\varPsi_{p,q}$ of $\calO_{2^{q}}$ 
into $\calO_{2^{p}}$ with $q$ being a nontrivial multiple of $p$, which is 
defined by \eqref{hom-embed}, it is shown that, for any $u\in U(2^p)$, 
there exists $v\in U(2^{q})$ such that
\begin{alignat}{1}
&\alpha_u \circ \varPsi_{p,q} = \varPsi_{p,q} \circ \alpha_v.
\end{alignat}
From \eqref{varPhi_SR-reduction-2} and \eqref{CAR-auto}, we obtain
\begin{alignat}{1}
\tau_u 
&=\varPhi^{-1}_{SR_p}\circ\alpha_u\circ\varPhi_{SR_p} \nonumber\\
&=\varPhi^{-1}_{SR_p}\circ(\alpha_u\circ\varPsi_{p,q})\circ\varPhi_{SR_{q}}
=(\varPhi^{-1}_{SR_p}\circ\varPsi_{p,q})\circ\alpha_v\circ\varPhi_{SR_{q}}
    \nonumber\\
&=\varPhi^{-1}_{SR_{q}}\circ\alpha_v\circ\varPhi_{SR_{q}} 
     \nonumber\\
&=\tau_v,
\end{alignat}
hence we obtain
\begin{alignat}{1}
&\mbox{Aut}_{U(2^p)}(\mbox{CAR}) \subset \mbox{Aut}_{U(2^{rp})}(\mbox{CAR}),
\quad r=2,3,\ldots.
\end{alignat}
Now, it becomes possible to define a product 
$\tau_{u_1}\circ\tau_{u_2}$ even for $u_i\in U(2^{p_i})$ $(p_1\not=p_2)$
explicitly by $\tau_{v_1}\circ\tau_{v_2}=\tau_{v_1v_2}$, $\tau_{v_i}=\tau_{u_i}$, 
$v_i\in U(2^q)$ with $q$ being the least common multiple of $p_1$ and $p_2$.
Thus, we obtain an infinite-dimensional $\ast$-automorphism group of the 
CAR algebra defined by
\begin{alignat}{1}
&\mbox{Aut}_U(\mbox{CAR})\equiv\{ \, 
    \tau_{u}\, \mid \, u\in U(2^p), \ p\in\boldN \, \}.
\end{alignat}
\par
Next, we consider $\ast$-automorphism subgroups $A^{(i)}_p$ 
$(i\!=\!1,\ldots,2^p)$ of $\calO_{2^p}$ defined by
\begin{alignat}{1}
A^{(i)}_p \equiv 
 \{\, \alpha_u \,\mid\, u_{i,i}\in U(1),\ u_{j,i}=0,\ j\not=i\},
 \quad i=1,\,\ldots,\,2^p,
\end{alignat}
where $\alpha_u$ is defined by \eqref{O_{2^p}-auto}.
Then, it is obvious that $A^{(i)}_{p}\cong U(1)\times U(2^p-1)$.
Correspondingly, we introduce the subgroup 
$\mbox{Aut}^{(i)}_{U(2^p)}(\mbox{CAR})$ of 
$\mbox{Aut}_{U(2^p)}(\mbox{CAR})$ by
\begin{alignat}{1}
&\mbox{Aut}^{(i)}_{U(2^p)}(\mbox{CAR}) \equiv \{\, 
    \tau_u=\varPhi_{SR_p}^{-1} \circ \alpha_u \circ \varPhi_{SR_p} \,\mid\,
    \alpha_u\in A^{(i)}_p\,\}.
\end{alignat}
Since we have $\pi_L\circ\alpha_u\cong\pi_{L'}$
with $\alpha_u\in A^{(i)}_p$, $L=(i;z)$, $L'=(i;z')$ and $z,\,z'\in U(1)$, 
we obtain the unitary equivalence as follows:
\begin{alignat}{1}
&\mbox{Rep}^{(p)}[\,i\,]\circ\tau^{(i)} \cong \mbox{Rep}^{(p)}[\,i\,], \quad
\tau^{(i)}\in\mbox{Aut}^{(i)}_{U(2^p)}(\mbox{CAR}),
\end{alignat}
where Rep$^{(p)}[\,i\,]$ is defined by \eqref{Fock-like rep}.
In general, however, we have 
$\mbox{Rep}^{(p)}[j]\circ\tau^{(i)} \not\cong \mbox{Rep}^{(p)}[j]$ 
for $j\not=i$,
hence $\tau^{(i)}\in\mbox{Aut}^{(i)}_{U(2^p)}(\mbox{CAR})$ is generally 
an outer $\ast$-automorphism of the CAR algebra.
\par
Now, we consider the subgroup of $\mbox{Aut}_U(\mbox{CAR})$ defined by 
\begin{eqnarray}
&&  \mbox{Aut}^{(1)}_{U}(\mbox{CAR})
    \equiv \bigcup_{p\in\boldN}\mbox{Aut}^{(1)}_{U(2^p)}(\mbox{CAR}).
  \label{pi_s-auto}
\end{eqnarray}
Then, as seen from \eqref{CAR-1para-auto} with 
\eqref{CAR-1para-auto-by-O_{2^p}-1} and \eqref{CAR-1para-auto-by-O_{2^p}-2}, 
any one-parameter group of $\ast$-automorphisms corresponding to the time 
evolution of a (quasi-)free fermion system is contained in 
$\mbox{Aut}^{(1)}_{U}(\mbox{CAR})$. 
As for a generic one-parameter group of $\ast$-automorphism in 
\eqref{pi_s-auto}, because of its nonlinearity, it describes a time 
evolution in which the particle number changes generally. 
In the following, we show this property of \eqref{pi_s-auto} by using 
a few simple examples.
\vskip10pt
\noindent
{\bf Example 1.}\ Let $\{\alpha_t \mid t\!\in\!\boldR\}$ be a one-parameter 
group of $\ast$-automorphisms of $\calO_4$ defined by
\begin{alignat}{1}
&
\begin{cases}
\displaystyle
\alpha_t(s_i) = s_i, \quad i=1,2,\\[5pt]
\displaystyle
\alpha_t(s_3) = \cos\theta_t\,s_3 - \sin\theta_t\,s_4, \quad
\alpha_t(s_4) = \sin\theta_t\,s_3 + \cos\theta_t\,s_4, 
\quad\theta_t \equiv \mu t,
\end{cases} \label{O_4-auto}
\end{alignat}
where $\mu$ is a nonvanishing real constant. 
Then, the corresponding induced one-parameter group 
$\{\tau_t\mid t\in\boldR\}$ of $\ast$-automorphisms of the CAR algebra 
is obtained as follows:
\begin{alignat}{1}
\tau_t(a_{2m-1})
&= G_{m-1}
     \bigg[a_{2m-1} 
      \!- \sin\theta_t\Big(\!\sin\theta_t\,(a_{2m-1}+a_{2m-1}^*) 
      \!- \cos\theta_t\,W_{2m-1}\Big) a_{2m}^*a_{2m}\bigg], 
  \label{induced-CAR-auto-1}\\
\tau_t(a_{2m})
&= G_{m-1}
     \Big[\cos\theta_t\,a_{2m} 
     + \sin\theta_t\,W_{2(m-1)}\,(a_{2m-1}-a_{2m-1}^*)\,a_{2m}\Big], 
  \label{induced-CAR-auto-2}\\ 
G_n
&\equiv
\prod_{k=1}^n F_k, \quad G_0\equiv I,  \\
F_k
&\equiv
I - 2\sin\theta_t\Big(\sin\theta_t\,I 
  - \cos\theta_t\,W_{2(k-1)}\,
       (a_{2k-1} - a_{2k-1}^*) \Big)a_{2k}^*a_{2k}, 
  \label{induced-CAR-auto-3}\\
W_n
&\equiv  
\prod_{\ell=1}^{n}K_\ell, \quad W_0\equiv I,\quad
K_\ell\!\equiv\! a_\ell a_\ell^*\!-\!a_\ell^*a_\ell\!=
\exp(\sqrt{-1}\,\pi\,a^*_\ell a_\ell),\label{induced-CAR-auto-4}
\end{alignat}
where 
$m\in\boldN$, and $F_k$'s satisfy $[F_k,\,F_\ell]=0$.
Since the vacuum $e_1$ in the Fock representation at $t=0$, which is defined 
by \eqref{RFSvac} with \eqref{RFSfock}, satisfies
\begin{alignat}{1}
\tau_t(a_n)\,e_1 = 0,  \quad t\in\boldR,\quad n\in\boldN,
\end{alignat}
as it should be, we can adopt $e_1$ as the vacuum at any $t$.
Then, $\tau_t$ does {\it not\/} conserve the particle 
number due to the nonvanishing term proportional to $a_{2m-1}a_{2m}$ 
in $\tau_t(a_{2m})$ $(m\in\boldN)$. 
To show this in detail, we define the (formal) particle number operator 
at $t$, $N_t$, as follows:
\begin{alignat}{1}
N_t&\!\equiv\! \sum_{n=1}^{\infty} \tau_t(a_n^* a_n), \label{N_t-1}\\
\tau_t(a_{2m-1}^*a_{2m-1})
&\!=\!H_{m-1}\!\bigg[ a_{2m-1}^* a_{2m-1}
  \!+\! \sin\theta_t \Big(\!\sin\theta_t\,K_{2m-1}  
  \!+\! \cos\theta_t(a_{2m-1}\!+\!a_{2m-1}^*)\!\Big) a_{2m}^* a_{2m}\bigg]\!,
      \label{N_t-2}\\
\tau_t(a_{2m}^* a_{2m})
&\!=\! H_{m-1}\,a_{2m}^* a_{2m},  \label{N_t-3}\\
H_n
&\!\equiv\!
\prod_{k=1}^{n}\Big(I + \sin^2(2\theta_t)\,a_{2k}^* a_{2k}\Big),
  \label{N_t-4}
\quad H_0\equiv I.
\end{alignat}
Obviously, $N_t$ is explicitly dependent on $t$.
An eigenvector of $N_t$ with an eigenvalue $k$ is called a {\it $k$-particle
state vector\/} at $t$. 
A generic $k$-particle state vector at $t$ is, of course, a linear 
combination of vectors in the form of 
$\tau_t(a^*_{n_1}\cdots a^*_{n_k})\,e_1$ $(n_1<\cdots<n_k)$, and
is orthogonal to any $k'$-particle state vector at $t$ with $k'\not=k$.
Then, since, from \eqref{induced-CAR-auto-2}, we have
\begin{alignat}{1}
\tau_t(a_{2m}^*)\,e_1 
&= (\cos\theta_t \,I - \sin\theta_t\,a_{2m-1}^*)a_{2m}^*\,e_1, 
  \qquad m\in\boldN,
\end{alignat}
a one-particle state vector at $t\not=\pi c/\mu$ $(c\in\boldZ)$ 
given by $\tau_t(a_{2m}^*)\,e_1$
is {\it not\/} orthogonal to a two-particle state vector at $t=0$ given by 
$\tau_0(a_{2m-1}^*a_{2m}^*)\,e_1=a_{2m-1}^*a_{2m}^*\,e_1$ $(m\in\boldN)$:
\begin{alignat}{1}
\langle\,\tau_t(a_{2m}^*)\,e_1\mid\tau_0(a_{2m-1}^*a_{2m}^*)\,e_1\,\rangle
&=-\sin\theta_t.  \label{2p-1p-transition}
\end{alignat}
Therefore, the one-parameter group $\{\tau_t\}$ of $\ast$-automorphisms 
given by \eqref{induced-CAR-auto-1}--\eqref{induced-CAR-auto-3}
does not conserve the particle number.
\par
Next, we consider the expectation value of $N_t$ by 
a $k$-particle state at $t=0$, $v_k$, as follows:
\begin{alignat}{1}
\omega(N_t;\,v_k)&\equiv \langle\,v_k\mid N_t\,v_k\,\rangle,\quad
N_0\,v_k=k\,v_k, \quad k\in\boldN,
\end{alignat}
which may give the particle number of $v_k$ at $t$. 
From \eqref{N_t-1}--\eqref{N_t-4}, it is straightforward to calculate 
$\omega(N_t;\,v_k)$ for each $v_k$.
More precisely, we denote a $k$-particle state for a specific set of 
the creation operators by
\begin{alignat}{1}
v_{n_1,n_2,\ldots,n_k}
\equiv a_{n_1}^*a_{n_2}^*\cdots a_{n_k}^*\,e_1, \quad n_1<n_2<\cdots<n_k.
\end{alignat}
Then, for one-particle state vectors and two-particle state vectors, 
we obtain
\begin{alignat}{1}
\omega(N_t;\,v_{n_1})
&=\begin{cases}
   1 & \mbox{for $n_1\!=\!2m_1\!-\!1$,} \\[2pt] 
   1 + \sin^2\theta_t & \mbox{for $n_1\!=\!2m_1$,} 
  \end{cases}
\\
\omega(N_t;\,v_{n_1,n_2}) 
&=\begin{cases}
   2 & \mbox{for $n_1\!=\!2m_1\!-\!1,$\  
             $n_2\!=\!2m_2\!-\!1$,} \\[2pt] 
   2 - \sin^2\theta_t & 
    \mbox{for $n_1\!=\!2m_1\!-\!1,$\  $n_2\!=\!2m_1$,} \\[2pt] 
   2 + \sin^2\theta_t & 
    \mbox{for $n_1\!=\!2m_1\!-\!1,$\  $n_2\!=\!2m_2$,} \\[2pt] 
   2 + \sin^2\theta_t + \sin^2(2\theta_t) & 
    \mbox{for $n_1\!=\!2m_1,\phantom{-1}$\  $n_2\!=\!2m_2\!-\!1$,} \\[2pt] 
   (2+\sin^2(2\theta_t))(1+\sin^2\theta_t) & 
    \mbox{for $n_1\!=\!2m_1,\phantom{-1}$\  $n_2\!=\!2m_2$,} 
  \end{cases}
\end{alignat}
where $m_1,m_2\in\boldN\ (m_1<m_2)$. 
As for $k$-particle state vectors,
in particular case in which either
all of $n_1,\ldots,n_k$ $(n_1<\cdots<n_k)$ are odd or they are even,
it is easy to obtain the following:
\begin{alignat}{1}
\omega(N_t;\,v_{2m_1-1,\ldots,2m_k-1})
&=k,\\
\omega(N_t;\,v_{2m_1,\ldots,2m_k})
&=\frac{(1+\sin^2(2\theta_t))^k-1}{\sin^2(2\theta_t)}\,(1+\sin^2\theta_t).
\end{alignat} 
In this way, $\omega(N_t;\,v_k)$ is, in general, dependent on $t$ nontrivially.
\vskip10pt
\noindent
{\bf Example 2.}\ Let $\{\alpha_t\mid t\!\in\!\boldR\}$ be a one-parameter 
group of $\ast$-automorphisms of $\calO_8$ defined by
\begin{alignat}{1}
&
\begin{cases}
\displaystyle
\alpha_t(s_i) = s_i, \quad i=1,2,3,4,6,7,\\[5pt]
\displaystyle
\alpha_t(s_5) = \cos\theta_t\,s_5 - \sin\theta_t\,s_8, \quad
\alpha_t(s_8) = \sin\theta_t\,s_5 + \cos\theta_t\,s_8, \quad
\theta_t \equiv \mu t,
\end{cases} \label{O_8-auto}
\end{alignat}
where $\mu$ is a nonvanishing real constant.
Then, the corresponding induced one-parameter group $\{\tau_t\mid t\in\boldR\}$ 
of $\ast$-automorphisms of the CAR algebra is much simpler than 
\eqref{induced-CAR-auto-1}--\eqref{induced-CAR-auto-4},
since we have $\alpha_t(\zeta_3(X))=\zeta_3(\alpha_t(X))$ $(X\in\calO_8)$ 
for the recursive map $\zeta_3$ defined by \eqref{srfs_3},
and $\zeta_p(X_1\cdots X_n)=\zeta_p(X_1)\cdots\zeta_p(X_n)$ for an odd 
integer $n$ and $p\in\boldN$.
We obtain the following:
\begin{alignat}{1}
\tau_t(a_{3m-2})
=&\;a_{3m-2} + \big( (\cos\theta_t-1)a_{3m-2} 
 + \sin\theta_t\,a_{3m-1}^* \big) a_{3m}^*a_{3m},\label{O_8-induced-CAR-auto-1}\\
\tau_t(a_{3m-1})
=&\; a_{3m-1} + \big( - \sin\theta_t\,a_{3m-2}^* 
 + (\cos\theta_t-1)a_{3m-1} \big) a_{3m}^*a_{3m},\label{O_8-induced-CAR-auto-2}\\
\tau_t(a_{3m})
=&\; \Big[ \cos\theta_t\,I 
         + (1-\cos\theta_t)(a_{3m-2}^*a_{3m-2}-a_{3m-1}^*a_{3m-1})^2
        \nonumber\\
&\;\;    + \sin\theta_t(a_{3m-2}a_{3m-1}+a_{3m-2}^*a_{3m-1}^*)\Big]a_{3m}, 
        \label{O_8-induced-CAR-auto-3}
\end{alignat}
where $m\in\boldN$.
As seen from the term $a_{3m-2}a_{3m-2}a_{3m}$ in \eqref{O_8-induced-CAR-auto-3},
a one-particle state vector at $t\not=\pi c/\mu$ $(c\in\boldZ)$ 
given by $\tau_t(a_{3m}^*)\,e_1$ is 
{\it not\/} orthogonal to a three-particle state vector at $t=0$ given by $\tau_0(a_{3m-2}^*a_{3m-1}^*a_{3m}^*)\,e_1=a_{3m-2}^*a_{3m-1}^*a_{3m}^*\,e_1$
as follows:
\begin{alignat}{1}
\langle\,\tau_t(a_{3m}^*)\,e_1\mid
\tau_0(a_{3m-2}^*a_{3m-1}^*a_{3m}^*)\,e_1\,\rangle
&=-\sin\theta_t.
\end{alignat}
The particle number operator $N_t$ in the present case is given by
\begin{alignat}{1}
N_t&=\sum_{n=1}^\infty \tau_t(a_n^*a_n), \\
\tau_t(a_{3m-2}^*a_{3m-2})
&=
a_{3m-2}^*a_{3m-2} 
 - \sin^2\theta_t\,(a_{3m-2}^*a_{3m-2}+a_{3m-1}^*a_{3m-1}-I)a_{3m}^*a_{3m}
   \nonumber\\
&\qquad
 + \sin\theta_t\,\cos\theta_t\,(a_{3m-2}^*a_{3m-1}^*+a_{3m-1}a_{3m-2})a_{3m}^*a_{3m},\\
\tau_t(a_{3m-1}^*a_{3m-1})
&=
a_{3m-1}^*a_{3m-1} 
 - \sin^2\theta_t\,(a_{3m-2}^*a_{3m-2}+a_{3m-1}^*a_{3m-1}-I)a_{3m}^*a_{3m}
   \nonumber\\
&\qquad
 + \sin\theta_t\,\cos\theta_t\,(a_{3m-2}^*a_{3m-1}^*+a_{3m-1}a_{3m-2})a_{3m}^*a_{3m},\\
\tau_t(a_{3m}^*a_{3m})
&=
a_{3m}^*a_{3m},
\end{alignat}
where $m\in\boldN$.
It is easy to obtain the expectation values of $N_t$ by $k$-particle state 
vectors at $t=0$ as follows:
\begin{alignat}{1}
\omega(N_t;\,v_{n_1})
&=\begin{cases}
    1 &\mbox{for $n_1=3m-2,\,3m-1$,}\\[5pt]
    1+2\sin^2\theta_t &\mbox{for $n_1=3m$,}
  \end{cases} \\[5pt]
\omega(N_t;\,v_{n_1,n_2})
&=\omega(N_t;\,v_{n_1})+\omega(N_t;\,v_{n_2}),\\[5pt]
\omega(N_t;\,v_{n_1,n_2,n_3})
&=\begin{cases}
    \displaystyle
    3-2\sin^2\theta_t 
    &\mbox{for $n_1\!=\!3m\!-\!2,\,n_2\!=\!3m\!-\!1,\,n_3\!=\!3m$,}\\[5pt]
    \displaystyle
    \sum_{i=1}^3\omega(N_t;\,v_{n_i})
    &\mbox{otherwise,}
  \end{cases}
\end{alignat}
where $m\in\boldN$, and likewise in the case of $k\geqq4$.
\vskip10pt
\noindent
{\bf Example 3.}\ Let $\{\alpha_t\!\mid\!t\!\in\!\boldR\}$ be a one-parameter 
group of $\ast$-automorphisms of $\calO_{16}$ defined by
\begin{alignat}{1}
&
\begin{cases}
\multicolumn{2}{l}{\displaystyle
\alpha_t(s_i) = s_i, \quad i=1,4,6,7,10,11,13,16,}\\[5pt]
\displaystyle
\alpha_t(s_2) = \cos\theta_t\,s_2 - \sin\theta_t\,s_{15},
&\alpha_t(s_{15}) = \sin\theta_t\,s_2 + \cos\theta_t\,s_{15}, \\[5pt]
\displaystyle
\alpha_t(s_3) = \cos\theta_t\,s_3 - \sin\theta_t\,s_{14},
&\alpha_t(s_{14}) = \sin\theta_t\,s_3 + \cos\theta_t\,s_{14}, \\[5pt]
\displaystyle
\alpha_t(s_5) = \cos\theta_t\,s_5 - \sin\theta_t\,s_{12},
&\alpha_t(s_{12}) = \sin\theta_t\,s_5 + \cos\theta_t\,s_{12}, \\[5pt]
\displaystyle
\alpha_t(s_9) = \cos\theta_t\,s_9 - \sin\theta_t\,s_8,
&\alpha_t(s_8) = \sin\theta_t\,s_9 + \cos\theta_t\,s_8, \\[5pt]
\displaystyle\hspace*{21pt}\theta_t \equiv \mu t,
\end{cases} \label{O_{16}-auto}
\end{alignat}
where $\mu$ is a nonvanishing real constant.
In this case, it is also possible to introduce one to four mutually 
different constants $\mu$'s for the $SO(2)$ action on four pairs 
$(s_2,s_{15})$, $(s_3,s_{14})$, $(s_5,s_{12})$, and $(s_9,s_{8})$. 
If we have introduced more than one constants $\mu$'s whose 
ratios are irrational for at least one pair of them,
$\{\alpha_t\}$ would become nonperiodic in contrast with the
previous examples. 
Anyway, for simplicity, we have introduced only one constant $\mu$.
Then, the corresponding one-parameter group $\{\tau_t\mid t\in\boldR\}$ 
of $\ast$-automorphisms of the CAR algebra is obtained as follows:
\begin{alignat}{1}
\tau_t(a_{m,j_1}) 
=&\,  \cos\theta_t\, a_{m,j_1} + \sin\theta_t\, b_{m,j_1}, 
\quad  a_{m,j_1}\equiv a_{4(m-1)+j_1},  \label{O_{16}-induced-CAR-auto-1}\\
b_{m,j_1}
\equiv
&\,     a_{m,j_2}  \,a_{m,j_3}  \,a_{m,j_4} 
     +  a_{m,j_2}^*\,a_{m,j_3}^*\,a_{m,j_4}   
     +  a_{m,j_3}^*\,a_{m,j_4}^*\,a_{m,j_2}  
     -  a_{m,j_4}^*\,a_{m,j_2}^*\,a_{m,j_3},  \hspace*{-3pt}
     \label{O_{16}-induced-CAR-auto-2}
\end{alignat}
where $m\in\boldN$, $(j_1,j_2,j_3,j_4)$ is a cyclic permutation of 
$(1,2,3,4)$.
From the existence of 
$a_{m,j_2}\,a_{m,j_3}\,a_{m,j_4}$ in \eqref{O_{16}-induced-CAR-auto-1} 
with \eqref{O_{16}-induced-CAR-auto-2}, a one-particle state vector 
at $t\not=\pi c/\mu$ $(c\in\boldZ)$ is not orthogonal with a three-particle 
state vector at $t=0$ as in Example 2.
The particle number operator $N_t$ is given by
\begin{alignat}{1}
N_t\!=&\!\sum_{n=1}^\infty \tau_t(a_n^*a_n) \nonumber \\
   \!=&\!\sum_{m=1}^\infty \sum_{j_1=1}^4 \Big[ 
     (1+2\sin^2\theta_t)
     a_{m,j_1}^*\,a_{m,j_1}     
     \!+\! 2 \sin^2\!\theta_t \big( 
         2 a_{m,j_2}^*a_{m,j_2}\,a_{m,j_3}^*a_{m,j_3}\,a_{m,j_4}^*a_{m,j_4} 
  	    \nonumber\\
    &\hphantom{\sum_{m=1}^\infty \sum_{i=1}^4 \Big[}
     \!-\! a_{m,j_2}^*a_{m,j_2}\,a_{m,j_3}^*a_{m,j_3}
     \!-\! a_{m,j_3}^*a_{m,j_3}\,a_{m,j_4}^*a_{m,j_4}
     \!-\! a_{m,j_4}^*a_{m,j_4}\,a_{m,+j_2}^*a_{m,j_2}\big) 
      \nonumber\\
    &\hphantom{\sum_{m=1}^\infty \sum_{i=1}^4 \Big[}
     \!-\!2\sin\theta_t\,\cos\theta_t
       \big( a_{m,j_1}\,a_{m,j_2}^*\,
             a_{m,j_3}^*\,a_{m,j_4}^*    
     \!+\!a_{m,j_1}^*\,a_{m,j_2}\,a_{m,j_3}\,a_{m,j_4} \big) \Big],
\end{alignat}
where $(j_1,j_2,j_3,j_4)$ is a cyclic permutation of $(1,2,3,4)$.
The expectation values of $N_t$ by $k$-particle state vectors at $t=0$
are obtained as follows:
\begin{alignat}{1}
\omega(N_t;\,v_{n_1}) 
=&\, 1+2\sin^2\theta_t, \\ 
\omega(N_t;\,v_{n_1,n_2})
=&\,2+4(1-\delta_{m_1,m_2})\sin^2\theta_t, \\
\omega(N_t;\,v_{n_1,n_2,n_3})
=&\,3+\big(6+4(\delta_{m_1,m_2}\delta_{m_2,m_3}-\delta_{m_1,m_2}
            -\delta_{m_2,m_3}-\delta_{m_3,m_1})\big)\sin^2\theta_t, \\
\omega(N_t;\,v_{n_1,n_2,n_3,n_4})
=&\,4 
  + 4\big(2\!+\! \delta_{m_1,m_2}\delta_{m_2,m_3}
           \!+\! \delta_{m_2,m_3}\delta_{m_3,m_4}
           \!+\! \delta_{m_3,m_4}\delta_{m_4,m_1}
           \!+\! \delta_{m_4,m_1}\delta_{m_1,m_2} \nonumber\\
&     
	       \!-\! \delta_{m_1,m_2} \!-\! \delta_{m_1,m_3} 
	       \!-\! \delta_{m_1,m_4} \!-\! \delta_{m_2,m_3} 
	       \!-\! \delta_{m_2,m_4} \!-\! \delta_{m_3,m_4}\big)
                 \sin^2\theta_t,
\end{alignat}
where $n_i\equiv4(m_i-1)+j_i$, $m_i\in\boldN$, $j_i=1,2,3,4$
with $n_{i_1}<n_{i_2}$ for $i_1<i_2$, and likewise in the case of 
$k\geqq5$.
\vskip10pt
\par
It is straightforward to generalize the one-parameter group $\{\alpha_t\}$ 
of $\ast$-automorphisms in the above examples to the case of $\calO_{2^p}$. 
The corresponding induced one-parameter group $\{\tau_t\}$ of 
$\ast$-automorphisms of the CAR algebra may, in general, contain mixing of 
one-particle states and $r$-particle states with $r\leqq p$.
By composing such $\ast$-automorphisms which mutually commute,  
it is possible to make various one-parameter group of $\ast$-automorphisms 
of the CAR algebra, which change the particle number with keeping the Fock 
vacuum invariant.
%
%
%
\vskip40pt
\section{Discussion}
In the present paper, we have shown that it is possible to reveal some
nontrivial structures of the CAR algebra concretely in terms of generators 
without difficulties by transcribing various properties of the Cuntz 
algebra through our recursive fermion system.   
\par
As far as the permutation representations of the Cuntz algebra are 
concerned, the standard recursive fermion system yields irreducible 
representations only from those with central cycles of length 1. 
It is remarkable that, from any irreducible permutation representation 
with a central cycle of length greater than 1, a suitable nonstandard 
recursive fermion system, which is obtained from the standard one by 
using an inhomogeneous endomorphism, yields an irreducible one. 
According to the recent study in $C^*$-algebra,\cite{KOS} for any two 
pure states (or irreducible representations) in a rather wide class of 
$C^*$-algebras including the Cuntz algebra, there exists a 
$\ast$-automorphism connecting them. However, it seems still unknown 
how to construct such a $\ast$-automorphism explicitly in terms of 
generators.  If it becomes to possible to construct it, we may apply 
it in the recursive fermion system by replacing the above inhomogeneous 
endomorphisms.
\par
As shown in Sec.\,3-2, a set of permutation endomorphisms  
$\varphi_{\sigma_p}$ $(p\in\boldN)$, which induce $\ast$-homomorphisms 
of the CAR algebra to its even subalgebra, has interesting properties 
such as \eqref{p-branch-endo-1}, \eqref{p-branch-endo-2} and 
\eqref{branch-formula}, although it is not closed with respect to the 
composition.  It may be useful for a systematic study of such even-CAR 
endomorphisms to extend the set of $\varphi_{\sigma_p}$ $(p\in\boldN)$ 
so as to constitute an abelian  semigroup. 
Let $r$ be an odd positive integer and $P\equiv(p_1,p_2,\ldots,p_r)$ 
with $p_1,\ldots,p_r\in\boldN$ and $p_1<p_2<\cdots<p_r$. 
We define a family of the $(p_r+1)$-th order permutation endomorphisms 
$\varphi_{\sigma_P}$ of $\calO_2$ by \eqref{ext-perm-endo} 
with $\sigma_P\in \mathfrak{S}_{2^{p_r+1}}$ being given by
\begin{alignat}{1}
&\begin{cases}
\displaystyle
\sigma_P(1,j_1,\ldots,j_{p_r}) \equiv (1,j_1,\ldots,j_{p_r}), &\\
\displaystyle
\sigma_P(2,j_1,\ldots,j_{p_1-1},j_{p_1},\ldots,j_{p_2-1},j_{p_2},
     \ldots,j_{p_r-1},j_{p_r}) &\\
\qquad 
\equiv (2,j_1,\ldots,j_{p_1-1},\hatj_{p_1},\ldots,j_{p_2-1},
      \hatj_{p_2},\ldots,j_{p_r-1},\hatj_{p_r}) &
\end{cases}
\end{alignat}
with $\hatj\equiv 3-j$.
Then, $\varphi_{\sigma_P}$ is written as follows:
\begin{alignat}{1}
&
\begin{cases}
\displaystyle
\varphi_{\sigma_P}(s_1)=s_1,& \\
\displaystyle
\varphi_{\sigma_P}(s_2)=s_2\,\prod_{k=1}^r\rho^{p_k-1}(J), \quad 
  J\equiv s_{2;1}+s_{1;2},&
\end{cases}
\end{alignat}
where $\rho$ is the canonical endomorphism of $\calO_2$.
Let ${\cal P}$ be the whole set of $P$'s, each of which consists of 
an odd number of mutually different positive integers. 
Then, it is straightforward to show that 
$\{\varphi_{\sigma_P}\mid P\in{\cal P}\}$ constitutes an abelian 
semigroup with respect to the composition, that is, 
for any $P, Q\in{\cal P}$, there exists an element $R\in{\cal P}$ 
such that
\begin{alignat}{1}
\varphi_{\sigma_R}=\varphi_{\sigma_P}\circ\varphi_{\sigma_Q}
=\varphi_{\sigma_Q}\circ\varphi_{\sigma_P}.
\end{alignat}
\par
As for the KMS state of the CAR algebra given in Sec.\,7, we have 
constructed it from Rep$(1)$ of $\calO_2$ by using the above even-CAR 
endomorphism $\varphi_{\sigma_p}$ but not from a KMS state of the Cuntz 
algebra. Since it is known that the inverse temperature for the KMS 
state of the Cuntz algebra with respect to a certain type of one-parameter 
group of $\ast$-automorphisms is unique,\cite{OP,Evans,BJO} it does not 
seem desirable to construct a KMS state of the CAR algebra by restricting 
that of the Cuntz algebra with a unique inverse temperature.
For more study of KMS states of the CAR algebra in view of the Cuntz
algebra, it seems important to clarify in what class of one-parameter 
groups of $\ast$-automorphisms the uniqueness of the KMS state of the 
Cuntz algebra is valid. 
\par
In Sec.\,8, we have constructed nontrivial one-parameter groups $\{\tau_t\}$ 
of $\ast$-automorphisms of the CAR algebra which preserve the Fock vacuum 
invariant. For rather simple cases such as Examples 2 and 3, we can also
construct the generator of $\tau_t$, that is, the Hamiltonian.
From the unitary $u_t\in U(1,\calO_{2^p})$ associated with the 
$\ast$-automorphism $\alpha_t$ of $\calO_{2^p}$, which is defined by
\begin{alignat}{1}
u_t&\equiv \sum_{i=1}^{2^p}\alpha_t(s_i)\,s_i^*,
\end{alignat}
we obtain
\begin{alignat}{1}
\tau_t(a_j)&=\varPhi_{SR_p}^{-1}(u_t\,\bolda_j\,u_t^*), \quad j=1,\ldots,p,
 \label{unitary-for-seeds}
\end{alignat}
where $\bolda_j$'s are the seeds of the standard RFS$_p$.
In the case of Examples 2 and 3, since $\{a_{p(m-1)+j} \mid j=1,\ldots,p\}$ 
for a fixed $m\in\boldN$ transform in the same way as the seeds, 
by extrapolating the expression for $u_t$ in terms of the seeds to that of 
$\{a_{p(m-1)+j}\mid m\in\boldN,\ j=1,\ldots,p\}$,
it is straightforward construct the corresponding Hamitonians $H$ generating 
$\tau_t$ as follows:
\begin{alignat}{1}
 &\tau_t(a_n) = e^{\sqrt{-1}\,H\,t}\,a_n\,e^{-\sqrt{-1}\,H\,t}, \quad H^*=H,\\
\mbox{Example 2:}\ 
  H&=\sqrt{-1}\,\mu\sum_{m=1}^\infty\big(a_{3m-2}^*\,a_{3m-1}^*
                              -a_{3m-1}\,a_{3m-2}\big)a_{3m}^*a_{3m},\\
\mbox{Example 3:}\ 
  H&=\sqrt{-1}\,\mu\sum_{m=1}^\infty\sum_{j_1=1}^4\big(
        a_{4(m-1)+j_1}^*\,a_{4(m-1)+j_2}^*\,a_{4(m-1)+j_3}^*\,a_{4(m-1)+j_4}
	 \nonumber\\
&\hspace*{90pt}
       -a_{4(m-1)+j_4}^*\,a_{4(m-1)+j_3}\,a_{4(m-1)+j_2}\,a_{4(m-1)+j_1}\big),
\end{alignat}
where $(j_1,j_2,j_3,j_4)$ is a cyclic permutation of $(1,2,3,4)$.
Here, it should be noted that the above $H$'s are well-defined only in 
the Fock representation, or more precisely, in the representations induced 
from those of the Cuntz algebra which are kept invariant under the 
corresponding $\alpha_t$.
\par
It is interesting to consider expectation values of products of 
$\tau_t(a_m)$'s and $\tau_t(a_m^*)$'s with mutually different time 
variables by the vacuum $e_1$, which is written as
\begin{alignat}{1}
\omega\big(a^\sharp_{m_1}(t_1)\,a^\sharp_{m_2}(t_2)\,\cdots\,
            a^\sharp_{m_n}(t_n)\big)
&\equiv \langle\,e_1\mid a^\sharp_{m_1}(t_1)\,a^\sharp_{m_2}(t_2)\,
  \cdots\,a^\sharp_{m_n}(t_n) \,e_1\,\rangle,
\end{alignat}
where $a^\sharp_m(t)$ denotes $\tau_t(a_m)$ or $\tau_t(a^*_m)$. 
In contrast with those obeying linear transformations such as 
\eqref{CAR-1para-auto}, we found there are nontrivial truncated 
$n$-point functions, where truncation means subtraction of 
contributions from lower-point functions. 
As illustration, we consider those consisting only of $a_1$, $a_1^*$, 
$a_2$ and $a_2^*$ in the case of Example 1. 
Since all one-point functions vanish, there is no difference between 
truncated functions and untruncated ones in the case of $n=2,3$.
From \eqref{induced-CAR-auto-1} and \eqref{induced-CAR-auto-2}, 
we obtain the following two-point functions and three-point ones:
\begin{alignat}{1}
\omega\big(a_1(t_1)a_1^*(t_2)\big)&=1,\\
\omega\big(a_2(t_1)a_2^*(t_2)\big)
&=\cos(\theta_{1}-\theta_{2}), \\[5pt]
\omega\big(a_2(t_1)a_2^*(t_2)a_1^*(t_3)\big)
&=\omega\big(a_1(t_2)a_2(t_4)a_2^*(t_3)\big)^* 
=\sin(\theta_1-\theta_2),\\
\omega\big(a_2(t_1)a_1^*(t_2)a_2^*(t_3)\big)
&=\omega\big(a_2(t_3)a_1(t_2)a_2^*(t_1)\big)^*
=\sin(\theta_1-\theta_2)\cos(\theta_2-\theta_3), \label{a_2a_1^*a_2^*}
\end{alignat}
and others vanish, where $\theta_i\equiv\theta_{t_i}$.
Here, \eqref{a_2a_1^*a_2^*} reproduces \eqref{2p-1p-transition} with $n=1$ 
by setting $t_1=t$, $t_2=t_3=0$.
As for the truncated four-point functions, they are obtained as follows:
\begin{alignat}{1}
&\omega\big(a_1(t_1)a_2(t_2)a_1^*(t_3)a_2^*(t_4)\big)_{\mbox{\scriptsize T}}
=\omega\big(a_2(t_4)a_1(t_3)a_2^*(t_2)a_1^*(t_1)\big)_{\mbox{\scriptsize T}}^* 
  \nonumber\\
&\qquad = \omega\big(a_1(t_1)a_2(t_2)a_1^*(t_3)a_2^*(t_4)\big)
   + \omega\big(a_1(t_1)a_1^*(t_3)\big)\,\omega\big(a_2(t_2)a_2^*(t_4)\big)
  \nonumber\\
&\qquad =
 - \cos(\theta_3-\theta_2)\cos(\theta_3-\theta_4)
       + \cos(\theta_2-\theta_4)\nonumber\\
&\qquad =
  \sin(\theta_2-\theta_3)\sin(\theta_3-\theta_4), \\
&\omega\big(a_2(t_1)a_1(t_2)a_1^*(t_3)a_2^*(t_4)\big)_{\mbox{\scriptsize T}}
 \nonumber\\
&\qquad = \omega\big(a_2(t_1)a_1(t_2)a_1^*(t_3)a_2^*(t_4)\big)
  - \omega\big(a_1(t_2)a_1^*(t_3)\big)\,\omega\big(a_2(t_1)a_2^*(t_4)\big)
   \nonumber\\
&\qquad =
\cos(\theta_1\!-\!\theta_2)\cos(\theta_2\!-\!\theta_3)\cos(\theta_3\!-\!\theta_4) 
     - \cos(\theta_1\!-\!\theta_4)\nonumber\\ 
&\qquad =
 \sin(\theta_1\!-\!\theta_2)\sin(\theta_2\!-\!\theta_3)\cos(\theta_3\!-\!\theta_4)
      +\sin(\theta_1\!-\!\theta_3)\sin(\theta_3\!-\!\theta_4), \\
&\omega\big(a_2(t_1)a_1^*(t_2)a_1(t_3)a_2^*(t_4)\big)_{\mbox{\scriptsize T}}
= \sin(\theta_1-\theta_2)\cos(\theta_2-\theta_3)\sin(\theta_3-\theta_4), 
\end{alignat}
and others vanish, where a subscript T denotes truncation. Likewise, we can 
show there are nonvanishing truncated five-point functions and six-point ones.
To clarify the physical meaning of these results, it is necessary to study
in more detail the one-parameter group $\{\tau_t\}$ of $\ast$-automorphisms 
as time evolutions of quantum field theoretical dynamical systems.
\vskip40pt
%

%
%
\end{document}